\documentclass[a4paper,11pt]{article}
\usepackage{jheppub} 
\usepackage{lineno}
\nolinenumbers
\usepackage{mystuff}
\pgfplotsset{compat=1.18}
\usepackage{graphicx} 
\usepackage{xcolor}

\definecolor{amethyst}{rgb}{0.54, 0.17, 0.89}
\definecolor{coral}{rgb}{1.0, 0.3, 0.4}

\arxivnumber{2510.13941} 

\title{
Diffeomorphism invariant tensor networks for 3d gravity
}

\author[a,b,c]{Vijay Balasubramanian}
\author[a]{Charlie Cummings}
\affiliation[a]{David Rittenhouse Laboratory, University of Pennsylvania,  209 S.33rd Street, Philadelphia, Pennsylvania 19104, USA}
\affiliation[b]{Santa Fe Institute, 1399 Hyde Park Road, Santa Fe, NM 87501, USA}
\affiliation[c]{Theoretische Natuurkunde, Vrije Universiteit Brussel, Pleinlaan 2, B-1050 Brussels, Belgium}

\emailAdd{charlie5@sas.upenn.edu}
\emailAdd{vijay@physics.upenn.edu}

\abstract{
Tensor networks prepare states which share many features of states in quantum gravity. However, standard constructions are not diffeomorphism invariant and do not support an algebra of non-commuting area operators. Recently, analogues of both problems were addressed in a tensor network discretization of topological field theories (TFT) with finite or compact gauge groups. Here, we extend this work towards gravity by generalizing to  gauge groups that are discrete or continuous, compact or non-compact. Applied to $\SL(2,\R) \times \SL(2,\R)$ Chern-Simons theory, 
our construction can be interpreted as building states of three dimensional gravity with a negative cosmological constant. 
Our tensor networks prepare states which satisfy the constraints of Chern-Simons theory. In metric variables, this implies that the states we construct satisfy the Wheeler-DeWitt equation and momentum constraints, and so are diffeomorphism invariant.}

\begin{document}
\maketitle
\flushbottom

\section{Introduction}

Tensor networks are tools for constructing quantum states with a particular entanglement structure. They are built by contracting the ``in-plane legs'' of a collection of tensors while leaving some legs of the tensors free. This procedure defines a state in the Hilbert space of the free (or ``boundary'') legs. The particular state depends on the tensors we contract and the pattern of their contraction, which can be efficiently represented by a graph $\Lambda$ with vertices representing the tensors, and edges representing the contracted legs. This graph $\Lambda$ can be understood as  embedded in a static Cauchy slice $\Sigma$,  emphasizing that the associated tensor network does not capture  information about the dynamics of the theory. If we include matter (``out-of-plane'') legs, then the tensor network can instead be viewed as a map from the out-of-plane legs to the boundary legs.

Traditionally, the bulk legs of these tensor networks are taken to be featureless; the only data specifying them is their bond dimension, i.e., the dimension of the Hilbert space of each in-plane leg. This technique is often used in holography to make states whose entanglement structure matches the expected behavior of the Ryu-Takayanagi (RT) formula \cite{Ryu_2006,Swingle_2012}. To make this match, the geometry of the tensor network is taken to be a discretization of a static slice of an AdS spacetime, the tensors are taken to be Haar random \cite{Hayden_2015}, and the dimension of the Hilbert spaces associated to the bulk legs is taken to be large.\footnote{An alternative is to have each vertex represent a ``perfect tensor'', as in the HaPPY code \cite{Pastawski:2015qua}.}

Despite matching the RT formula,  tensor networks have other properties that  are not found in generic states of quantum gravity. For example, in precisely the same limit that gives the RT formula, tensor network states are R\'enyi flat \cite{Dong_2019}. In other words, if $\rho$ is the reduced density matrix for a subset of boundary legs, the family of R\'enyi entropies
    \begin{align}
        S_n(\rho) = \frac{1}{1-n} \log\left(\frac{\tr[\rho^n]}{(\tr[\rho])^n}\right)
    \end{align}
is independent of $n$ in tensor network states. In contrast, the R\'enyi entropy of generic semiclassical states of quantum gravity depends on $n$ \cite{Dong_2019}.

There is a gravitational interpretation of this phenomenon. R\'enyi flatness implies that the gravitational action of the $n$-replica spacetime is independent of $n$ \cite{Dong_2019}. Because the gravitational action is essentially the area of the Ryu-Takayanagi (RT) surface in the $n$-replica spacetime \cite{Lewkowycz_2013}, this essentially means that the path integral preparing $\rho$ has a delta function $\delta(A)$ in it, where $A$ is the area of the RT surface. So a tensor network state can be interpreted as an eigenstate of the area operator \cite{Dong_2019}. Alternatively, one could imagine canonically quantizing gravity with canonically conjugate variables $(h_{ij}, K_{ij})$, the spatial metric and extrinsic curvature of the Cauchy slice $\Sigma$. From this perspective, a tensor network can be thought of as an approximation of a spatial metric eigenstate $\ket{h_{ij}}$. In contrast, semiclassical states of gravity are instead given by coherent superpositions of such states; otherwise they would not have a well-defined extrinsic curvature  in the semi-classical limit. Thus, traditional tensor networks represent a certain class of non-semiclassical states in the quantum gravity Hilbert space.

To emphasize this point, consider an AdS spacetime. There is a classical operator (phase space function) $A(R)$ in canonical general relativity which measures the minimal area among all boundary-anchored surfaces homologous to a boundary subregion $R$. If $R_1, R_2$ are two such subregions such that $R_1 \cap R_2 \neq \emptyset$, then the Poisson bracket between their respective area operators fails to vanish:
    \begin{align}
        \{A(R_1),A(R_2)\} \neq 0 \,.
    \end{align}
By AdS/CFT, after quantizing the theory, this implies that $A(R_1)$ and $A(R_2)$ should fail to commute as operators on the boundary CFT Hilbert space. However, in a fixed area state, $A(R_1),A(R_2)$ will always commute, because they act as c-numbers on such a state. Thus, states prepared by traditional tensor networks fail to reproduce the expected canonical commutation relations of classical gravity, demonstrating how non-semiclassical they are. If we cure this problem by generalizing the standard random tensor network paradigm, the resulting states may retain more of the features expected in semi-classical gravity.

Tensor network states also differ from semiclassical states of gravity in lacking manifest time evolution. As explained above, tensor networks should be thought of as preparing states on a static Cauchy slice $\Sigma$. However, in gravity, bulk time evolution on $\Sigma$ generates the entire Wheeler-DeWitt patch that causally completes $\Sigma$. Because diffeomorphism invariance implies that gravity is a totally constrained system, this time evolution is implemented by ensuring that the states of $\Sigma$ satisfy the Wheeler-DeWitt (WDW) equation $H_{WDW} \ket{\Psi} = 0$, as well as the momentum constraints. Therefore, one might hope that building tensor network states which satisfy some version of the gravitational constraints could shed light on how to introduce dynamics into the tensor network paradigm \cite{Akers:2024ixq}. 

Three dimensional gravity is an especially fertile testing ground for this idea. In three dimensions, gravity is topological, as there are no gravitons. When the cosmological constant is negative, the action  in Lorentzian signature is equivalent to $\SL(2,\R) \times \SL(2,\R)$ Chern-Simons theory \cite{Witten:1988hc}.\footnote{The equivalence is at the level of the action, and the path integral measures in these two theories are different. However, see \cite{Collier_2023,Hartman:2025cyj,Chen:2024unp} for recent proposals concerning a TQFT with a possible quantum equivalence around saddle point geometries.} 
Likewise, Lorentzian de Sitter and Euclidean Anti-de Sitter gravity have the same action as $\SL(2,\C)$ Chern-Simons theory. When the cosmological constant vanishes, the action of 3D gravity matches  $\text{ISO}(1,2)$ (the Poincar\'e group) or $\text{ISO}(3)$ (the Euclidean group) Chern-Simons theory. Finally, the action of gravity in Euclidean de Sitter space is equivalent to a pair of $\SU(2)$ Chern-Simons fields. Thus, because Chern-Simons theory is a field theory, we might hope that the constraints are easier to implement by studying 3D gravity in its Chern-Simons variables. This would lead to tensor network states  which share more of the expected features of semiclassical gravity.\footnote{As pointed by Witten in work on the Chern-Simons/3D gravity correspondence \cite{Witten:1988hc}, this strategy is the starting point of loop quantum gravity. This suggests that the tools developed to study spin networks \cite{Rovelli:1995ac,Girelli:2005ii,Marzuoli:2004sk,Han2017:prd,Dupuis:2019yds,Alexandrov:2011ab} could shed some light on the dynamics of topological tensor networks. However, tensor networks and spin networks differ in important ways. Topological tensor networks are actually more closely related to ``string-nets'' \cite{Levin_2005}, which are similar to spin networks but to our knowledge have not been demonstrated to be equivalent. See \cite{Konopka_2008} for more details about the difference between these  approaches.}

Progress on this issue was recently made by Akers, Soni, and Wei (ASW) \cite{akers2024multipartite} and Dong, McBride and Weng  (DMW) \cite{Dong2024}, but also see \cite{Donnelly:2016qqt,Qi:2022lbd,Singh_2010,Colafranceschi:2020ern}, as well as \cite{Basteiro:2022xvu,Basteiro:2024cuh,Basteiro:2024crz,Basteiro:2022zur,Caputa:2020fbc,Frenkel:2024smt,Soni:2024aop,Singh:2017tet,Colafranceschi:2022dig,Chirco:2017wgl,Chirco:2021chk,Meusburger:2008bs,Kim:2015qoa} for other approaches constructing more realistic tensor networks for quantum gravity. These authors constructed a tensor network which hosts a discretized version of a gauge theory on each bulk leg, rather than a featureless Hilbert space. These kinds of models were first discovered in condensed matter theory by Kitaev \cite{Kitaev:1997wr}, as well as Levin and Wen \cite{Levin_2005}. The extra structure allowed them to impose a toy version of the gravitational constraints.  Furthermore, the states satisfying the constraints can be thought of as a particular superposition of more traditional random tensor network states, with the superposition designed to ensure the analog of the WDW equation is satisfied. In analogy with coherent states, this shows that the tensor networks in \cite{akers2024multipartite,Dong2024} are indeed ``closer'' to semiclassical states of gravity. 

The ASW model defines an analog of the area operator, and shows that such operators do not commute for overlapping boundary subregions \cite{akers2024multipartite}. Furthermore, the tensor networks in this model are topological, in a sense we will explain  below. So this  approach is suitable for constructing a model of 3D gravity, which is topological. We can think of the discrete graph defining a topological tensor network as discretizing the background manifold that the Chern-Simons connections live on, perhaps by a simplical decomposition of the background manifold $M$ that the connections propagate on. Because Chern-Simons theory is topological, this discretization is arbitrary, and so the diffeomorphism invariance of the theory remains unbroken. This is how the ASW networks are able to satisfy the analog of the Wheeler-DeWitt equation, even though the state is constructed from a discretized graph which naively seems to break diffeomorphism invariance. Furthermore, interpreted appropriately, topological tensor networks retain an RT-like formula computing von Neumann entropy, thus preserving the feature of standard tensor networks which makes them analogous to gravity in the first place. 

A similar construction was considered by Dong, McBride, and Weng (DMW) \cite{Dong2024}, but there were two major differences with the ASW model. While DMW consider states which satisfy the ``electric constraints'' (see below for definitions), ASW consider states which satisfy both electric and ``magnetic'' constraints. The magnetic constraints play a crucial role in simulating diffeomorphism invariance in gravity. Specifically, the diffeomorphism constraints in 3D gravity map onto both the electric and magnetic constraints of the $G_k \times G_{-k}$ Chern-Simons theory it is equivalent to \cite{akers2024multipartite,Levin_2005}. A second crucial difference  is that ASW consider a finite gauge group $G$, while DMW allow for any compact group (of which a finite group is a special case). These are both in contrast to the more physically relevant cases of $\SL(2,\R) \times \SL(2,\R)$, $\SL(2,\C)$, $\text{ISO}(1,2)$, or $\text{ISO}(3)$. 

In this paper, we will focus on the case of $\SL(2,\R) \times \SL(2,\R)$ Chern-Simons theory. The reason is that $G_k$-Levin-Wen models ($k$ is the level of the associated Chern-Simons theory) are always associated with ``doubled'' Chern-Simons theories $G_k \times \overline{G_{k}}$, where $\overline{G_{k}}$ denotes the orientation reversed version of $G_k$. One way to see this is that the states described by Levin-Wen models (topological tensor networks) are always time reversal symmetric \cite{Levin_2005}, but $G_k$ Chern-Simons theory maps to $\overline{G_{k}}$ Chern-Simons theory under time reversal.\footnote{In more technical terms, the $G_k$ Levin-Wen model is associated with the Turaev-Viro TQFT of the Drinfeld center $D(G_k) \cong G_k \times \overline{G_{k}}$, not just $G_k$. In the case $G = \SL(2,\R)$ and level $k = i\sigma$, $\overline{\SL(2,\R)_{k}} = \SL(2,\R)_{k} $ because the level is imaginary.} Another way to see this is that the action of Chern-Simons theories only has one time derivative, so the Chern-Simons path integral is a phase space path integral, not a configuration space path integral. Wave functions are functions of \emph{half} of the phase space coordinates, i.e., they are a function of the positions or momentum, but not both simultaneously. The definition of the Hilbert space thus requires a splitting of this phase space (more precisely, a choice of polarization) to separate the canonically conjugate variables \cite{Witten1989Jones,Witten1991ComplexCS,Witten:1988hc}. Thus, to capture the full Levin-Wen model, we need to double the degrees of freedom of the associated Chern-Simons theory. This is simplest to do in the case of $\SL(2,\R)_k$, where this doubling simply produces $\SL(2,\R)_k \times \overline{\SL(2,\R)_k}$. We will not pursue  generalizations to other gauge groups here. We also focus on constructing states of gravity in the $G_N \to 0$ limit. There are two reasons for focusing on this limit. First, the level $k = i\sigma$ of Chern-Simons theory is inversely proportional to Newton's constant, and the structure of $\SL(2,\R)_k$ simplifies to the representation theory of $\SL(2,\R)$ in the $k \to \infty$ limit \cite{Maldacena_2001}. Second, we do not consider the non-perturbative effects of topology change on the Hilbert space, so we need to take this limit for consistency anyways.

There are two technical challenges to generalizing the ASW/DMW models to the gauge groups  directly relevant for 3D gravity. The first is that the magnetic constraints are subtle when $G$ is not discrete: note that although DMW considered continuous gauge groups, they did not implement the magnetic constraints. In contrast, the electric constraints are subtle when $G$ is non-compact, which is the case for all the gravitational gauge groups mentioned above other than that of Euclidean deSitter space. If one could generalize the tensor networks of the ASW/DMW constructions to simultaneously satisfy the electric/magnetic constraints and support non-compact, continuous gauge groups, the resulting states could be interpreted as states of $\SL(2,\R) \times \SL(2,\R)$ Chern-Simons theory, and through a change of variables, 3D gravity. In this paper, we will do exactly that. 

Three sections and three appendices follow.  In Sec.~\ref{sec:themodel}, we explain how to construct topological tensor network states for a wide class of gauge groups. In Sec.~\ref{sec:Hphys}, we analyze some of the physical properties of these states, and verify that they produce a bulk-to-boundary map. We conclude the main text with a discussion in Sec.~\ref{sec:discussion}.  Appendix~\ref{sec:crashcourse} discusses methods of non-Abelian harmonic analysis that we will use, while Appendix~\ref{sec:move12} explains state-preserving moves on topological tensor networks.  Appendix~\ref{sec:DA} discusses the generalization of the quantum double algebra associated to a class of groups that we term ``transformable''.

\section{Topological tensor networks}\label{sec:themodel}

In this section, we will explain how to construct topological tensor network states with gauge group $G$. Akers, Soni, and Wei \cite{akers2024multipartite} took $G$ to be finite.  Finite groups are  tractable because they are discrete and compact (UV  and IR finite in physics jargon). In contrast, our analysis will apply to a wider class that we will call {\it transformable groups}. A transformable group can be continuous or discrete, compact or non-compact, and most importantly for our purposes, $G = \SL(2,\R)$ is transformable.


Below, we first define the notion of a transformable group, and discuss examples.  Then we  explain how to construct the pre-Hilbert space (ignoring the ``out-of-plane'' legs) that arises before  imposing the gauge constraints, which are equivalent to imposing the Wheeler-DeWitt equation in gravity. Next, we  explain the gauge constraints, and discuss the physical Hilbert space of gauge invariant states. Finally, we  discuss how to incorporate  out-of-plane legs.

\subsection{Transformable groups}
\label{sec:transformable}

To define topological tensor network states, we will have to perform non-Abelian Fourier transforms of functions of the gauge group $G$. For simplicity, we will mostly consider Lie groups.\footnote{If we dropped the assumption that $G$ is a Lie group, we would have to separately require that t $G$ be locally compact and separable as a topological space. Technically, this is more general, but we will simply assume $G$ is a Lie group as appropriate for the relation with 3D gravity.}
We also include discrete groups that can be viewed as comprising a zero-dimensional manifold with many disconnected components. In harmonic analysis, there are two conditions a Lie group $G$ must satisfy for the Fourier transform to exist: $G$ must be unimodular and type I (see below for definitions) \cite{Dixmier1977,Haar}.
We will refer to any group which satisfies these conditions as ``transformable''. 
Examples of transformable groups include semi-simple Lie groups (including $\SL(2,\R)$ or $\SL(2,\C)$), and compact groups.

First, recall that a left-invariant Haar measure $dg$ is not always right-invariant \cite{Haar}. Because $dg^{-1}$ is right-invariant, this means that $dg \neq dg^{-1}$ in general. However, for a wide class of groups, the left-invariant Haar measure \emph{is} also right-invariant, and for such groups $dg = dg^{-1}$. We will restrict our analysis to groups with this property, as we will use this inversion invariance to demonstrate the Hermiticity of various projectors within $L^2(G)$. Such groups are called unimodular groups. Examples include any compact, semi-simple, or connected reductive Lie group \cite{Haar}. 

As an aside, one could imagine a possibility for removing this constraint on $G$ as follows. Let $h\in G$ fixed. Because the left Haar measure $dg$ is unique up to an overall scaling, and $d(gh)$ is also a left Haar measure, it follows that $d(gh) = \Delta(h) dg$, where $\Delta(h)$ is a fixed, positive real number for fixed $h$. This defines a function $\Delta: G \to \R^+$ called the modular function of $G$ \cite{Haar}. 
One can use this function to show that 
the measure $d\mu(g) = \sqrt{\Delta(g^{-1})} dg$ \emph{is} invariant under inversions. On the other hand, $d\mu(g)$ is not translation invariant unless $\Delta(g) = 1$ for all $G$. This is where the name unimodular comes from. The authors of \cite{Marolf:2000iq} argued that when $G$ is non-unimodular,  $d\mu(g)$ is the appropriate measure for defining  constraints, not the Haar measure. Thus, our construction below may generalize to non-unimodular groups if we use $d\mu(g)$ instead of the Haar measure, but for simplicity we will restrict to the unimodular case.

Second, if our goal is to Fourier transform $L^2(G)$, then the ``momentum space'' that we transform into must be unique. This momentum space will be labeled by  irreducible unitary representations $\pi$ of $G$. We should be free to take the trace of operators in either the position or representation bases, so the trace within each irreducible representation $\pi$ should exist. This restricts the algebra of operators acting on $\pi$ to be of type I or type II  \cite{Dixmier1977,Sorce:2023fdx}. Any group with the property that all of its unitary irreducible representation (irreps) are type I algebras is called a type I group.\footnote{A type I algebra is just the usual algebra of matrix multiplication.} Any group which is not type I has both type II and type III algebras as unitary irreps \cite{Dixmier1977}, so it only makes sense to distinguish type I and non-type I groups. Thus, we will demand that $G$ is type I. Examples include any compact, semi-simple, or connected reductive Lie group \cite{Dixmier1977}.

To summarize, we restrict to type I, unimodular Lie groups, which we  refer to as ``transformable.''  Henceforth, we assume that the gauge group $G$ is transformable.

\subsection{The pre-Hilbert space}

\begin{figure}
    \centering
    \begin{tikzpicture}[scale=5]
        \def\nn{4}
        \foreach\xx in {1,...,\nn}{
            \draw ({\xx/(\nn+1)},0) -- ({\xx/(\nn+1)},1);
            \draw (0,{\xx/(\nn+1)}) -- (1,{\xx/(\nn+1)});
            
            \draw ({\xx/(\nn+1)},{0.5+sqrt(\xx/(\nn+1))*sqrt(1-\xx/(\nn+1) )}) circle (0.015);
            \draw ({\xx/(\nn+1)},{0.5-sqrt(\xx/(\nn+1))*sqrt(1-\xx/(\nn+1) )}) circle (0.015);
            \draw ({0.5+sqrt(\xx/(\nn+1))*sqrt(1-\xx/(\nn+1) )},{\xx/(\nn+1)}) circle (0.015);
            \draw ({0.5-sqrt(\xx/(\nn+1))*sqrt(1-\xx/(\nn+1) )},{\xx/(\nn+1)}) circle (0.015);
        \foreach\yy in {1,...,\nn}{
            \filldraw ({\xx/(\nn+1)},{\yy/(\nn+1)}) circle (0.015);
        }}
        \fill[white,even odd rule]  (0.5,0.5) circle (0.5)
        (0,0)--(1,0)--(1,1)--(0,1)--cycle;
         \draw (0.5,0.5) circle (0.5);
        \foreach\xx in {1,...,\nn}{
            \filldraw[fill=white] ({\xx/(\nn+1)},{0.5+sqrt(\xx/(\nn+1))*sqrt(1-\xx/(\nn+1) )}) circle (0.015);
            \filldraw[fill=white] ({\xx/(\nn+1)},{0.5-sqrt(\xx/(\nn+1))*sqrt(1-\xx/(\nn+1) )}) circle (0.015);
            \filldraw[fill=white] ({0.5+sqrt(\xx/(\nn+1))*sqrt(1-\xx/(\nn+1) )},{\xx/(\nn+1)}) circle (0.015);
            \filldraw[fill=white] ({0.5-sqrt(\xx/(\nn+1))*sqrt(1-\xx/(\nn+1) )},{\xx/(\nn+1)}) circle (0.015);
            }
    \end{tikzpicture}
    \caption{A lattice $\Lambda$ tessellating the disk $\Sigma$. The bulk vertices are in black, and the boundary vertices are in white.}
    \label{fig:tensornetwork}
\end{figure}
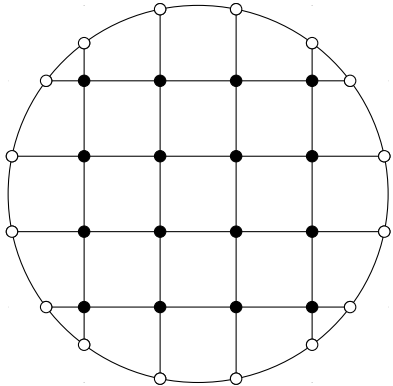

We will first consider the in-plane legs, and explain how to incorporate the out-of-plane legs in Sec.~\ref{sec:outofplane}. Consider an orientable, two-dimensional surface $\Sigma$, possibly with boundary, with a graph $\Lambda$ embedded into it (see  Fig.~\ref{fig:tensornetwork}). For concreteness, imagine $\Sigma$ is topologically a disc with boundary, but the generalization to other surfaces is straightforward. This choice of $\Sigma$ can be thought of as a Cauchy slice of AdS$_3$. The graph $\Lambda$ has vertices $v \in V$, oriented edges $\ell \in L$, and plaquettes (faces) $p \in P$. Additionally, an edge $\ell$ is said to be a boundary leg if one of its vertices is attached to $\partial \Sigma$, otherwise it is called a bulk leg. 

For every edge $\ell \in L$, associate a Hilbert space $\Ha_\ell := L^2(G)$, where $G$ is a transformable group (see Sec.~\ref{sec:transformable} for the definition). In the case of Lorentzian AdS gravity, we take $G = \SL(2,\R)$. Let us briefly discuss the structure of $\Ha_\ell$ (details in Appendix \ref{sec:crashcourse}).
Each Hilbert space $\Ha_\ell$ has a natural basis, called the group basis,
\begin{align}
    L^2(G) =  \text{span}\{\ket{g} \,|\, g \in G\}\,.
\end{align}
When $G$ is finite, the group basis has a natural inner product 
\begin{align}
    \braket{g}{h} = |G|\delta_{gh} \,.
\end{align}
When $G$ is continuous, the group basis is instead delta function normalized
\begin{align}
    \braket{g}{h} = \delta(g^{-1}h) \,.
\end{align}
We will generally use the continuum notation, but if we use the definition $\delta(g^{-1}h) := |G|\delta_{gh}$,  the same formulas will hold for finite $G$.\footnote{In the case when $G$ is discrete but non-compact (such as the integers), we should not include the $|G|$ in this definition.} Similarly, we will use the notations
\begin{align}
    \frac{1}{|G|}\sum_{g \in G} = \int_G dg \label{eqn:finiteGintegral}
\end{align}
interchangeably.

However, there is another basis, essentially the ``momentum basis'', that we will later need. To describe this basis, we use the Peter-Weyl theorem \cite{Tao2011PeterWeyl,peter1927weyl,Haar}. When $G$ is compact, the Peter-Weyl theorem says that
\begin{align}
    L^2(G) = \bigoplus_{\pi \in \widehat{G}} d_\pi \cdot [V_\pi \otimes V_\pi^*]\,,
\end{align}
where $\widehat{G}$ is the set of unitary representations of $G$, $V_\pi$ is the vector space of the unitary representation of $G$ labeled by $\pi$, $V_\pi^*$ is its dual space, and $d_\pi = \dim(V_\pi)$. For example, if $G = \SU(2)$, then $\widehat{G} = \frac{1}{2} \N_j$ is the set of spin quantum numbers, $\pi \sim j$ is a particular spin quantum number, $V_\pi = \C^{2j+1}$ is the vector space for the spin-$j$ representation, and $d_j= 2j+1$. The notation $d_\pi \cdot [\cdots]$ is shorthand for the dilatation of the inner product
\begin{align}
    \braket{A}{B}_{d_\pi \cdot [V_\pi \otimes V_\pi^*]} := d_\pi \cdot \braket{A}{B}_{V_\pi \otimes V_\pi^*}\,, \label{eqn:peterweylcompact}
\end{align}
where $\braket{A}{B}_{V_\pi \otimes V_\pi^*}$ is the Hilbert-Schmidt inner product on $V_\pi \otimes V_\pi^*$.\footnote{As explained in Appendix \ref{sec:crashcourse}, $V_\pi \otimes V_\pi^*$ can be thought of as a vector space of matrix elements, so $\ket{A}$ can be thought of as a matrix. The Hilbert-Schmidt inner product is the usual definition $\braket{A}{B}_{V_\pi \otimes V_\pi^*}=\tr[A^\dagger B]$.} With this notation, the Peter-Weyl theorem is not just an equality of vector spaces: it is an equality of Hilbert spaces, because the dilatation by $d_\pi $ ensures the inner products of the two sides agree, and therefore that the Fourier transform between the group basis and the representation basis is unitary. See Appendix~\ref{sec:crashcourse} for more details.

When $G$ is non-compact, we can still use a version of the Peter-Weyl theorem (which is  called the Plancherel decomposition in this case) \cite{HarishChandra1952,HarishChandra1954complex,HarishChandra1976,Mackey1976,segal1950,mautner1951} which says
\begin{align}
    L^2(G) = \int^{\oplus}_{\widehat{G}} d\mu(\pi) V_\pi \otimes V_\pi^* \,. \label{eqn:peterweylnoncompact}
\end{align}
The direct integral is analogous to the direct sum of vector spaces, but generalized to include both continuous and discrete families of representations. 

This decomposition is similar to the compact $G$ case, but with some key differences. $\widehat{G}$ is the set of irreducible unitary representations of $G$, also called the unitary dual of $G$. The unitary dual $\widehat{G}$ of a non-compact group is a topological space which has a much more complicated structure than for compact $G$.
For example, $\SL(2,\R)$ has both a discrete series (like spin $j$ of $\SU(2)$) and a continuous series (like momentum $k$ of $\R$) of representations, as well as a complementary series, limits of the discrete series, and the trivial representation \cite{langSL2R}. $\widehat{G}$ is generally not a manifold, even in the simplest case of $\SL(2,\R)$, and the explicit characterization of the topology of $\widehat{G}$ is sometimes not even known explicitly.\footnote{In principle, $\widehat{G}$ is always a topological space with the Fell topology \cite{fell1962topology,Dixmier1977} (weak convergence of matrix elements), but we mean that the explicit characterization of the Fell topology on $\widehat{G}$ is not always known.} Importantly, $d\mu(\pi)$ is the \emph{Plancherel measure} of $G$, which is a measure on $\widehat{G}$ which allows us to integrate over it, despite its complicated topological structure.  The reason we assumed $G$ is a transformable group (in particular, that it is type I) is so the topology of $\widehat{G}$ is tame enough for the Plancherel measure to exist.

The Plancherel measure has two important properties. First, $d\mu(\pi)$ does not have support on all of $\widehat{G}$: it assigns zero measure to some of the unitary representations $\pi \in \widehat{G}$. In the case of $\SL(2,\R)$, the Plancherel measure has support only on the principal and discrete series.
In particular, notice that this does not include the trivial representation, because the constant function is not square normalizable.\footnote{Technically, it is because the constant function is not normalizable in $L^{2+\epsilon}(G)$ for any $\epsilon > 0$ when $G$ is non-compact.} Thus, the ``uniform measure'' $d\pi$ of $\widehat{G}$ and the Plancherel measure do not even have the same support, so there is no coordinate transformation on $\widehat{G}$ which relates them. Because it is the Plancherel measure which appears in \eqref{eqn:peterweylnoncompact}, we will mostly use this measure instead of $d\pi$.

Second, the Plancherel measure rescales the inner product of each representation $V_\pi \otimes V_\pi^*$ by an appropriate dilatation factor. This rescaling ensures that \eqref{eqn:peterweylnoncompact} is an equality of Hilbert spaces, not just vector spaces. In other words, the Plancherel measure is the unique measure which makes the Fourier transform between the group basis of $L^2(G)$ and the RHS of \eqref{eqn:peterweylnoncompact} unitary. While $d\mu(\pi)$ is less natural than $d\pi$ in that it weighs different representations of $\widehat{G}$ non-uniformly, it is more natural because it is the measure which actually arises in the Plancherel decomposition of transformable groups.

We can unify \eqref{eqn:peterweylcompact} and \eqref{eqn:peterweylnoncompact} if we define $d\mu(\pi) = d_\pi$ when $G$ is compact.\footnote{When $G$ is compact, $\widehat{G}$ is discrete, so $\int_{\widehat{G}} d\mu(\pi) f(\pi) = \sum_{\pi \in \widehat{G}} d_\pi\, f(\pi)$.} If we restored the group volume dependence, we would find that $d\mu(\pi) = \frac{d_\pi}{\text{Vol}(G)}$. When $G$ is non-compact, both $d_\pi$ and $\text{Vol}(G)$ are infinite, but the Plancherel measure still exists, is finite, and is unique if $G$ is a transformable group \cite{Mackey1976}. However, its explicit form is more complicated (if known at all), but for the gauge groups that are relevant for gravity its explicit form is known \cite{HarishChandra1952,langSL2R,kitaev2018noteswidetildemathrmsl2mathbbrrepresentations,Haar,kleppner_lipsman_1972}.

Now that we understand the basics of $L^2(G)$, we can define the ``pre-Hilbert space'' associated to the graph $\Lambda$ as
\begin{align}
    \Ha(\Lambda) = \bigotimes_{\ell \in L} \Ha_\ell \,. \label{eqn:preH}
\end{align}
This Hilbert space has a ``local'' group basis $\{\ket{g_1,\cdots, g_{|L|}}\}$, i.e., a choice of group element for each leg of $\Lambda$. The pre-Hilbert space $\Ha(\Lambda)$ is much larger than the Hilbert space of physical states $\Ha_{phys}$ we will ultimately be interested in. This is because physical states must be gauge invariant, and therefore they must satisfy the gauge constraints (which we will define below). When $G$ is a finite group, the physical Hilbert space $\Ha_{phys}$ is precisely the subspace of $\Ha(\Lambda)$ that is annihilated by the gauge constraints. When $G$ is non-compact or continuous, this is essentially still true, but we have to be careful about precisely what we mean by a subspace.

In gravity, the gauge constraints impose diffeomorphism invariance, and arise directly from the equations of motion for the metric. In terms of the continuum Chern-Simons variables of 3D gravity, the equations of motion impose flatness of the gauge fields. In our discretized model, however, these constraints split into two types: the so-called electric and magnetic constraints \cite{Levin_2005}. The electric constraints are what impose Gauss' law at every vertex, and the magnetic constraints impose vanishing flux of the Chern-Simons gauge fields around each plaquette. We now discuss both constraints in more detail, and then use them to define the physical Hilbert space.

\subsection{Magnetic operators}

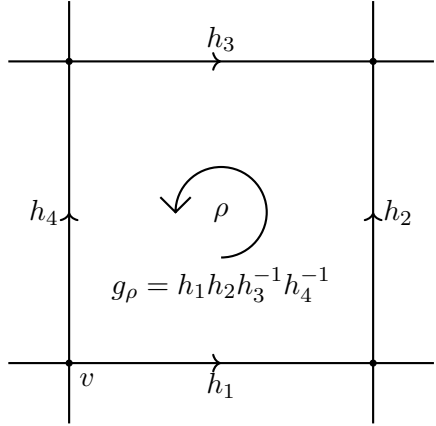
\begin{figure}
    \centering
    \begin{tikzpicture}[scale=4]
        \def\rr{0.15}
        \def\vv{0.05}
        \def\eps{0.2}
        
        \draw[thick,->-=0.5] (-\eps,0) -- (1+\eps,0);
        \draw[thick,->-=0.5] (1,-\eps) -- (1,1+\eps);
        \draw[thick,->-=0.5] (-\eps,1) -- (1+\eps,1);
        \draw[thick,->-=0.5] (0,-\eps) -- (0,1+\eps);
        
        \filldraw (0,0) circle (0.01);
        \filldraw (1,0) circle (0.01);
        \filldraw (0,1) circle (0.01);
        \filldraw (1,1) circle (0.01);
        \node[anchor=north west] at (0,0) {$v$};
        \node[anchor=north] at (0.5,0) {$h_1$};
        \node[anchor=west] at (1,0.5) {$h_2$};
        \node[anchor=south] at (0.5,1) {$h_3$};
        \node[anchor=east] at (0,0.5) {$h_4$};
        
        \draw[thick] (0.5,{0.5-\rr}) arc (-90:180:\rr)
        ({0.5-\rr - \vv},{0.5 + \vv}) --  ({0.5-\rr},{0.5}) --  ({0.5-\rr + \vv},{0.5 + \vv}); 
        \node at (0.5,0.5) {$\rho$};
        \node[anchor=north] at (0.5,{0.5-\rr}) {$g_\rho = h_1 h_2 h_3^{-1} h_4^{-1}$};
    \end{tikzpicture}
    \caption{An example of a plaquette that $B_{(v,p)}(h)$ acts on in \eqref{eqn:Bdef}. }
     \label{fig:Bdef}
\end{figure}

Consider a vector $\ket{g_1,\cdots,g_{|L|}} \in \Ha(\Lambda)$, where $|L|$ is the number of legs. The set of such states forms a complete basis for $\Ha(\Lambda)$. Next, let $(v,p)$ be a choice of vertex and plaquette, such that $v \in \partial p$. We will call such a pair a site. Let $S$ denote a choice of one site per plaquette. The choice of vertex for each site is arbitrary.
Then, define $\rho$ to be the counter-clockwise path around $\partial p$ which begins and ends at the vertex $v$. Letting $h \in G$, we can define ``magnetic'' operators $B_{(v,p)}(h)$ via
\begin{align}
    B_{(v,p)}(h)\ket{g_1,\cdots,g_{|L|}} = \delta(h^{-1}g_\rho)\ket{g_1,\cdots,g_{|L|}}\,. \label{eqn:Bdef}
\end{align}
Here, $g_\rho$ is the product of the group elements along the links in the orientation of $\rho$, and $\delta(g)$ is the delta function on $G$ which integrates to one on any open set of $G$ containing the identity. For an example, see Fig.~\ref{fig:Bdef}. Note that $B_{(v,p)}(h)$ is Hermitian for any $h$.

We can see from its definition that $B_{(v,p)}(h)$ annihilates any state for which the flux around the plaquette $p$ is not $h$. We can also see that the definition of $B_{(v,p)}(h)$ is subtle when $G$ is continuous, for $\delta(e)$ diverges in this case. Nevertheless, the action of $B_{(v,p)}(h)$ is well-defined.
If instead we considered a more general state $\ket{\psi}$ with a wave function over group elements given by 
\begin{align}
    \braket{g_1,\cdots,g_{|L|}}{\psi} = \psi(g_1,\cdots,g_{|L|})\,,
\end{align}
then by inserting a resolution of the identity, $B_{(v,p)}(h)$ acts as
\begin{align}
    B_{(v,p)}(h)\ket{\psi} = \int dg_1 \cdots dg_{|L|}\delta(h^{-1}g_\rho) \psi(g_1,\cdots,g_{|L|}) \ket{g_1,\cdots,g_{|L|}}\,.
\end{align}
Finally, note that if $f$ is any bounded function of $G$, and $dg$ is the Haar measure, we can define a more general class of operators
\begin{align}
    B_{(v,p)}[f] := \int dg \,f(g) B_{(v,p)}(g)\,.
\end{align}
We can think of $B_{(v,p)}(g)$ as forming a basis of more general magnetic operators $B_{(v,p)}[f]$. Magnetic operators of this form are bounded operators, which means they are valid observables on $\Ha(\Lambda)$. To see that they are bounded, let $||f||_\infty = \sup_{g \in G} |f(g)|$. Then by the triangle inequality for integration, we can compute
\begin{align}
    ||B_{(v,p)}[f] \ket{\psi}|| &= \left|\left| \int dg_1 \cdots dg_{|L|}f(g_\rho) \psi(g_1,\cdots,g_{|L|}) \ket{g_1,\cdots,g_{|L|}} \right|\right|
    \\&\leq ||f||_\infty \cdot||\ket{\psi}|| \,.
\end{align}
We will use this perspective in Sec.~\ref{sec:Hphys} (and Appendix \ref{sec:DA}) to understand the algebra of electric and magnetic operators in more detail.

We said above that the operators $B_{(v,p)}(h)$ measure the flux of the gauge field around the plaquette $p$. The equations of motion require that the flux vanishes: therefore, the physical states must lie in the image of $B_{(v,p)}(e)$ for each site, where $e$ is the identity element of $G$. Furthermore, operators $B_{(v,p)}(e)$ at different sites commute with each other, as one can check from the definition, so we can impose this condition independently at each plaquette.
Thus, the magnetic constraint operator will be defined as
\begin{align}
    \Pi_B = \bigotimes_{(v,p) \in S} 
    B_{(v,p)}(e) \,.
\end{align}
The key property of $\Pi_B$ is that it annihilates states which do not satisfy the constraints, regardless of whether $G$ is discrete or continuous. The normalization of $\Pi_B$ is more delicate. One can check that 
\begin{align}
    B_{(v,p)}(e)B_{(v,p)}(e) = \delta(e) B_{(v,p)}(e)\,,
\end{align}
so up to the normalization of the constant $\delta(e)$, $ B_{(v,p)}(e)$ is indeed the projector onto the states of $\Ha(\Lambda)$ which satisfy the magnetic constraint at a particular site. When $G$ is discrete, we are done, for $\delta(e)$ is finite and we can divide $B_{(v,p)}(e)$ by $\delta(e)$ to obtain a  projection operator $\Pi_B$ acting on $\Ha(\Lambda)$. When $G$ is continuous, however, we must be more careful, because $\delta(e)$ diverges. Note that while \cite{Dong2024} allowed for continuous gauge groups, they did not impose the magnetic constraint on their tensor networks. Thus, this normalization issue did not arise for them.

To understand how to proceed for continuous $G$, let us analyze the discrete case more carefully. The Hilbert space satisfying the constraints is simply
\begin{align}
    \Pi_B \Ha(\Lambda) = \text{span}\{\delta(e)^{-|S|/2}\Pi_B \ket{\psi} \text{ such that }\ket{\psi} \in \Ha(\Lambda)\} \,. \label{eqn:PiBHeasy}
\end{align}
Here, $|S|$ is the number of sites, and the $\delta(e)^{-|S|/2}$ is for later comparison to continuous groups. For discrete groups, this factor will not affect the definition of the Hilbert space at all. The resulting formulas, however, will continue to be meaningful when $G$ is continuous after all the $\delta(e)$'s have canceled out. The inner product between the states $\delta(e)^{-|S|/2}\Pi_B \ket{\psi}$ and $\delta(e)^{-|S|/2}\Pi_B \ket{\sigma}$ is given by
\begin{align}
   \frac{1}{\delta(e)^{|S|}}\bra{\sigma}\Pi_B^\dagger \Pi_B \ket{\psi} = \bra{\sigma} \Pi_B \ket{\psi} \,. \label{eqn:PiBHinnerproducteasy}
\end{align}

Crucially, however, this parameterization of $\Pi_B\Ha(\Lambda)$ is highly redundant. To see this, let $\ket{\chi}$ be any state such that $\Pi_B \ket{\chi} = 0$. Call the subspace of all such states $\Ha_{null}$. Then for any physical state $\ket\psi$, and any $\ket{\chi} \in \Ha_{null}$, the states
\begin{align}
    \Pi_B \ket{\psi} = \Pi_B (\ket{\psi} + \ket{\chi}) 
\end{align}
map onto precisely the same state of $\Pi_B \Ha(\Lambda)$. To remove this redundancy, we can instead consider states to be labeled by the formal set of equivalence classes
\begin{align}
    [\ket{\psi} \sim \ket{\psi} + \ket{\chi}] & \, \text{ for all }  \ket{\psi} \in \Ha(\Lambda) \,\text{ and } \, \ket{\chi} \in \Ha_{null}\,. \label{eqn:magnullstates}
\end{align}
We will denote the equivalence class containing a state $\ket{\psi}$ as $\ket{\psi}\rangle$. Next, it will be convenient to define $\Ha(\Lambda)_\infty$ as the subspace of $\Ha(\Lambda)$ of bounded $L^2$ wave functions, or equivalently, the intersection of $L^2$ and $L^\infty$.

This subspace is not a Hilbert space because it is not complete, but it is dense in $\Ha(\Lambda)$. 
In quantum mechanics we are used to requiring that wavefunctions are square integrable so that they have a probabilistic interpretation, namely that they are in $L^2(\R)$.   A wave function can have a singularity which is locally of the form $|x|^{-\frac{p}{2}}$ for $0 < p < 1$ and still be normalizable in the $L^2$ inner product.
As we will see below, it will turn out for us that wave functions must be in $L^2 \cap L^\infty$ to be normalizable after we impose the magnetic constraint, as we will see below. Note that when $G$ is discrete, all $L^2$ functions are bounded, which explains why this subtlety did not arise in \cite{akers2024multipartite}, which considered finite gauge groups. When $G$ is continuous, we have to impose the additional $L^\infty$ condition by hand.

The reason this restriction normally doesn't arise in quantum mechanics is because Hilbert spaces are required to be complete. The subspace $L^2 \cap L^\infty$ is not complete with respect to the $L^2$ norm, and so completing this subspace to a full Hilbert space leads us back to all of $L^2$. In contrast, this vector space \emph{will} be complete with respect to an alternative definition of the inner product. This alternative definition will agree with the usual inner product when $G$ is discrete, but continues to be meaningful when $G$ is continuous, and so is well-motivated.

With this in mind, we can define the vector space
\begin{align}
    (\Pi_B \Ha(\Lambda))_{pre} = \Ha(\Lambda)_\infty / \Ha_{null} \,,
\end{align}
and define the inner product on this vector space to be
\begin{align}
    \langle \braket{\sigma}{\psi}\rangle := \bra{\sigma} \Pi_B \ket{\psi} \,. \label{eqn:PiBHinnerproducthard}
\end{align}
Note that this definition of inner product for $(\Pi_B \Ha(\Lambda))_{pre}$ does not depend on a choice of representative for $\ket{\psi}\rangle$ or $\ket{\sigma}\rangle$. 
Finally, we define the Hilbert space $\Pi_B \Ha(\Lambda)$ as the completion of $(\Pi_B \Ha(\Lambda))_{pre}$ with respect to the inner product \eqref{eqn:PiBHinnerproducthard}:
\begin{align}
    \Pi_B \Ha(\Lambda) = \overline{\Pi_B \Ha(\Lambda))_{pre}}\,. \label{eqn:PiBHhard}
\end{align}
When $G$ is discrete, the resulting Hilbert space \eqref{eqn:PiBHhard} with inner product \eqref{eqn:PiBHinnerproducthard} is isomorphic to \eqref{eqn:PiBHeasy} and \eqref{eqn:PiBHinnerproducteasy}, so there is no physical difference between them.
In this case, the equivalence class definition \eqref{eqn:PiBHhard} is not necessary, and the simpler (but completely equivalent) definition \eqref{eqn:PiBHeasy} may be preferred. 

\begin{figure}
    \centering
    \begin{tikzpicture}[scale=2]
        \filldraw (0,-1) circle (0.015);
        \filldraw[fill opacity=0.1,thick,->-=0] (0,0) circle (1);
        \node[anchor=north] at (0,-1) {$v$};
        \node[anchor=west] at (1.1,0) {$\ell = \partial p$};
    \end{tikzpicture}
    \caption{The most basic topological tensor network $\Lambda_1$. There is a single vertex, plaquette $p$ (shaded in grey), and leg $\ell = \partial p$.}
    \label{fig:basicplaquette}
\end{figure}
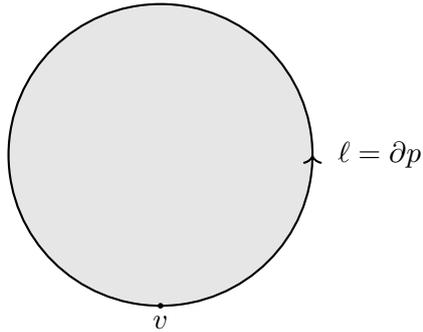

However, when $G$ is continuous, the divergence of $\delta(e)$ means we can not normalize $\Pi_B$ to define the analog of \eqref{eqn:PiBHeasy}. But the set of equivalence classes \eqref{eqn:PiBHhard} still exists when $G$ is continuous, and is complete with respect to the inner product \eqref{eqn:PiBHinnerproducthard}, so it really is a Hilbert space. The completeness of $\Pi_B \Ha(\Lambda)$ is subtle. For simplicity, consider the case when $\Lambda_1$ has a single plaquette (Fig.~\ref{fig:basicplaquette}), so $\Pi_B = B_{(v,p)}(e)$. Assume that the wavefunctions of $\ket{\psi},\ket{\sigma} \in \Ha(\Lambda)_\infty$ are bounded as well as square integrable. Again, not all wave functions in $\Ha(\Lambda_1)$ are of this form, but the set of such wavefunctions is dense in $\Ha(\Lambda_1)$.
We can then compute
\begin{align}
    \langle \braket{\sigma}{\psi}\rangle &= \int d[g,h] \,\sigma(g)^* \psi(h) \bra{g}B_{(v,p)}(e)\ket{h} 
    \\&= \int d[g,h] \,\sigma(g)^* \psi(h) \delta(h)\braket{g}{h}
    \\&= \sigma(e)^* \psi(e) \,. \label{eqn:Lambda1innerproduct}
\end{align}
Since $\psi,\sigma$ are bounded functions, this inner product converges. In this case, the Cauchy-Schwartz inequality implies that the Hilbert space completion $\Pi_B\Ha(\Lambda_1)$ is one dimensional, because all normalized vectors have maximal overlap. 

Because the magnetic operators $B_{(v,p)}(e)$ commute for disjoint plaquettes, a similar analysis applies to more general graphs $\Lambda$ as well. In that case, the $\delta(g)$ in the inner product above is replaced with $\delta(g_\rho)$, as explained in the definition of $B_{(v,p)}(e)$.
We can see from the definition that all wave functions for which $\psi(g_\rho = e)=0$ are null states in this inner product, and must be modded out in the definition of $\Pi_B\Ha(\Lambda)$. In this sense, the resulting Hilbert space has support only on states in the image of $\Pi_B$, but remains normalizable in the new inner product. Thus, we will take this to be the proper definition of $\Pi_B\Ha(\Lambda)$ from the start, as it is compatible with both discrete and continuous groups.

As explained above, to make $(\Pi_B \Ha(\Lambda))_{pre}$ into the Hilbert space $\Pi_B\Ha(\Lambda)$, we must complete it with respect to the inner product \eqref{eqn:PiBHinnerproducthard}. 
Actually, $(\Pi_B \Ha(\Lambda))_{pre}$ is \emph{already} complete with respect to the inner product \eqref{eqn:PiBHinnerproducthard}, but is not complete with respect to the inner product on $\Ha(\Lambda)$.  
For example, there are unbounded wave functions on $\Ha(\Lambda)$ for which \eqref{eqn:PiBHinnerproducthard} doesn't converge, so they do not correspond to normalizable states in $\Pi_B \Ha(\Lambda)$. Conversely, while every equivalence class $\ket{\psi}\rangle$ has a representative which is normalizable in the $\Ha(\Lambda)$ inner product, there are representatives $\psi(x)$ which are normalizable in $\Pi_B\Ha(\Lambda)$ but not in $\Ha(\Lambda)$---these additional representatives are elements of equivalence classes which already exist, so they do not contribute new states to $\Pi_B\Ha(\Lambda)$. But sometimes, these additional representatives have a clearer physical interpretation as we will see below.

One  way to phrase the difference between the $\Ha(\Lambda)$ and $\Pi_B\Ha(\Lambda)$ completions is that unit norm states in $\Ha(\Lambda)$ no longer have unit norm with respect to the $\Pi_B \Ha(\Lambda)$ inner product, even when $G$ is discrete. This implies that a Cauchy sequence of vectors $\ket{\psi_n}\rangle \in \Pi_B\Ha(\Lambda)$ may not have a Cauchy sequence of representatives $\ket{\psi_n} \in \Ha(\Lambda)$.
To be concrete, let us again consider the example graph $\Lambda_1$ of Fig.~\ref{fig:basicplaquette}. Consider the sequence of vectors $\ket{\psi_n}$ in the pre-Hilbert space $\Ha(\Lambda)$ with wave functions
\begin{align}
    \psi_n(g) = (N_n)^{-\frac{1}{2}} \Theta_n(g)\,.
\end{align}
Here, $\Theta_n(g)$ is the indicator function for a ball $V_n(e)$ of the identity which has Haar volume $1/n$ and minimal surface area, i.e., $\Theta_n(g) = 1$ if $g \in V_n(e)$, and is zero otherwise.\footnote{The minimal surface area condition ensures that $V_n(e)$ is approximately spherical as $n \to \infty$, which ensures regularity in this limit. We should also demand that $U_{n+1}(e) \subset U_{n}(e)$ and that $g \in V_n(e) \implies g^{-1} \in V_n(e)$ for all $n$.} 
$N_n$ is a normalization constant that depends on the inner product $\psi_n$ is normalized with respect to. On the one hand, if we demand $\braket{\psi_n} = 1$, then $N_n = n$. Fixing the ratio $n/m$, we can use this to show that there is a positive constant $C$ (that depends on this ratio) such that
\begin{align}
    \lim_{n,m\to\infty}||\ket{\psi_n} - \ket{\psi_m}||_{\Ha(\Lambda_1)} > C \,.
\end{align}
This implies that $\lim_{n\to\infty} \ket{\psi_n}$ is not a Cauchy sequence, and so does not converge to a vector in $\Ha(\Lambda_1)$. 

On the other hand, if we demand $\langle \braket{\psi_n}\rangle = 1$, then $N_n = 1$ for all $n$. Furthermore, one can show directly from \eqref{eqn:Lambda1innerproduct} that regardless of $n,m$, we have that 
\begin{align}
    ||\ket{\psi_n}\rangle - \ket{\psi_m}\rangle||_{\Pi_B \Ha(\Lambda_1)} = 0\,.
\end{align}
This implies that regardless of $n$, the states $\ket{\psi_n}$ are in the same equivalence class of $\Pi_B\Ha(\Lambda_1)$. Therefore, the limit of these representatives $\lim_{n\to\infty}\ket{\psi_n}$ \emph{does} converge to a vector in $\Pi_B \Ha(\Lambda_1)$, which we can call $\ket{\psi_\infty}$. The wave function of $\ket{\psi_\infty}$ is the indicator function of the identity element. 
In fact, $\ket{\psi_\infty}\rangle = \ket{\psi_1}\rangle$, so this ``completion'' which introduced $\ket{\psi_\infty}$
did not actually introduce new states into $\Pi_B \Ha(\Lambda_1)$.
This shows that the actual support of $\ket{\psi_1}\rangle$ is only on $g=e$, because the two representatives $\ket{\psi_1}$ and $\ket{\psi_\infty}$ only differ by a null state. Even though many representatives of $\ket{\psi_1}\rangle$ have support on group elements $g \neq e$, which naively violates the magnetic constraint, these representatives are as valid as $\ket{\psi_\infty}$, if not more so.


Let us compare this procedure to a less rigorous (but perhaps more intuitive) approach.\footnote{For discrete groups, this procedure is perfectly well-defined.} Let $\Theta_n(g)$ be as above.
We now define
\begin{align}
    \widetilde{B}_{(v,p)}(e) := \lim_{n \to \infty} B_{(v,p)} [\Theta_n]\,.
\end{align}
This has the effect of averaging $B_{(v,p)}$ over a small neighborhood $V_n(e)$ of the identity, rather than evaluating it at the identity itself. We must keep $n$ finite at all steps in a calculation, and take the limit that $n \to \infty$ at the end. The reason this approach is less rigorous than the formal procedure of changing the inner product is that there could be issues with order of limits. Furthermore, we will not be explicit about how the wave functions that $\widetilde{B}_{(v,p)}(e)$ acts on are allowed to depend on $n$. This definition of $\widetilde{B}_{(v,p)}(e)$ should therefore be thought of as a heuristic version of the above procedure of redefining the inner product. 
With this definition, we can see that
\begin{align}
   \widetilde{B}_{(v,p)}(e)\widetilde{B}_{(v,p)}(e)  
    &= \lim_{n \to \infty} \int_{V_n(e)} \int_{V_n(e)} dg dh B_s(g)B_s(h)
    \\&=  \lim_{n \to \infty} \int_{V_n(e)} \int_{V_n(e)} dg dh \delta(g^{-1}h) B_s(g)
    \\&=  \lim_{n \to \infty} \int_{V_n(e)} dg  B_s(g) \left(\int_{V_n(g)} dh  \delta(h)\right)
    \\&=  \lim_{n \to \infty} \int_{V_n(e)} dg  B_s(g) = \widetilde{B}_{(v,p)}(e)
\end{align}
In the third line, we used that $g \in V_n(e)$, so $e \in V_n(g)$ (this is part of the definition of $V_n$).
Furthermore, one can show that $\widetilde{B}_{(v,p)}(e)^\dagger \widetilde{B}_{(v,p)}(e) = \widetilde{B}_{(v,p)}(e)$,  and so $\widetilde{B}_{(v,p)}(e)$ is a projector
onto states of $\Ha(\Lambda)$ with zero flux around the $p$ plaquette.

As an example, consider the case of a single plaquette from above. Then
\begin{align}
    \widetilde{B}_{(v,p)}(e) \ket{\psi} = \lim_{n \to \infty} \int_{V_n(e)} dg\, \psi(g) \ket{g}.
\end{align}
The norm of this state is
\begin{align}
    \bra{\psi} \widetilde{B}_{(v,p)}(e)^\dagger \widetilde{B}_{(v,p)}(e) \ket{\psi} = \lim_{n \to \infty}  \int_{V_n(e)} dg\, \psi^*(g)\psi(g).
\end{align}
The only states which survive this projection are those where\footnote{Technically, such a state would not be a member of $L^2(G)$, but we could fix this by allowing $\psi$ itself depend on $\sigma$, so that it limits to \eqref{eqn:psideltanormalization} in the limit $n \to \infty$. For example, $\psi$ could be a Gaussian with variance $n^{-2}$. }
\begin{align}
    |\psi(g)|^2 = |\widetilde{\psi}(e)|^2 \delta(g) + \cdots \label{eqn:psideltanormalization}
\end{align}
which, upon inspection, is essentially the same class of states that survive \eqref{eqn:PiBHhard}. The difference is that while the proper method of defining the inner product keeps the wavefunctions finite, the ``large-$n$  limit'' approach instead formally moves an infinite constant into the wave function itself to cancel the infinite $\delta(e)$.

So from now on, we will take \eqref{eqn:PiBHhard} and \eqref{eqn:PiBHinnerproducthard} to be our definition of $\Pi_B\Ha(\Lambda)$. We will wait to discuss this Hilbert space in more detail until after we define and apply the electric constraints.

\subsection{Electric operators} \label{sec:electricops}

\begin{figure}
    \centering
    \begin{tikzpicture}
    \def\sep{0.5}
    \def\xx{0.6}
      \node at (-2-\sep,0) {\begin{tikzpicture}[scale=2]
          \draw[thick,->-=0.25,->-=0.75] (0,-1) -- (0,1);
          \draw[thick,->-=0.25,->-=0.75]  (-1,0) -- (1,0);
          \filldraw (0,0) circle (0.025);
          \node[anchor = north west] at (0,0) {$v$};
          \node[anchor=west] at (0,-\xx) {$g_1$};
          \node[anchor=west] at (0,\xx) {$g_2$};
          \node[anchor=south] at (-\xx,0) {$g_3$};
          \node[anchor=south] at (\xx,0) {$g_4$};
      \end{tikzpicture}};
      \draw[thick,->] (-\sep/2,0) -- (\sep/2,0);
      \node[anchor=north] at (0,0) {$A_v(h)$};
      \node at (2+\sep,0) {\begin{tikzpicture}[scale=2]
          \draw[thick,->-=0.25,->-=0.75] (0,-1) -- (0,1);
          \draw[thick,->-=0.25,->-=0.75]  (-1,0) -- (1,0);
          \filldraw (0,0) circle (0.025);
          \node[anchor = north west] at (0,0) {$v$};
          \node[anchor=west] at (0,-\xx) {$g_1 h^{-1}$};
          \node[anchor=west] at (0,\xx) {$hg_2$};
          \node[anchor=south] at (-\xx,0) {$g_3 h^{-1}$};
          \node[anchor=south] at (\xx,0) {$hg_4$};
      \end{tikzpicture}};
    \end{tikzpicture}
    \caption{The action of $A_v(h)$ on an example vertex}
    \label{fig:Adef}
\end{figure}
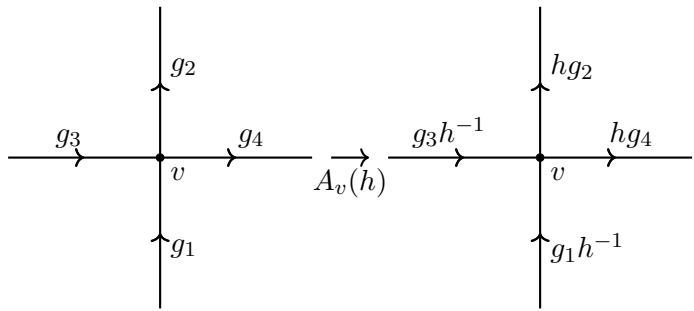

Let $v \in V$ be a vertex, and $\ket{g_1,\cdots,g_{|L|}}$ as above. Then for legs $\ell_i$ which are attached to $v$, we can define an operator $A_v(h)$ by left multiplying outflowing $\ell_i$ by $g_i \to h g_i$, and inflowing $\ell_i$ by $g_i \to g_i h^{-1}$. See Fig.~\ref{fig:Adef} for an example.

Similar to the magnetic operators, for any $f \in L^1(G)$, we can think of $A_v(h)$ as forming a basis for the more general class of electric operators
\begin{align}
    A_v[f] = \int dh \,f(h) A_v(h) \,,
\end{align}
where $dh$ is the Haar measure.
One can show from a direct computation that these smeared operators are bounded. Despite the fact that the constant function is not $L^1$, we will be particularly interested in the operator
\begin{align}
    A_v[1] = \int dh\, A_v(h)\,,
\end{align}
because this operator has the property that
\begin{align}
    A_v(g) A_v[1] &= \int dh\, A_v(g) A_v(h) \label{eqn:Av1invariant}
    \\&= \int dh \,A_v(gh)
    \\&= A_v[1]\,.
\end{align}
We used the left invariance of the Haar measure in the third line. Note that $A_v[1]^\dagger = A_v[1]$ because $dg = dg^{-1}$. Because the constant function is not $L^1$, $A_v[1]$ is not a bounded operator, similar to $B_{(v,p)}(e)$. Nevertheless, it will be a useful mathematical object to construct the physical Hilbert space.

We can also think of $A_v(h)$ as enacting a gauge transformation on the lattice, so the states on which $A_v(h)$ acts trivially will be gauge invariant. Now observe that the gauge invariant states will lie in the image of $A_v[1]$, because $A_v(h) A_v[1] = A_v[1]$ for any $h \in G$.  In other words, if we first act with $A_v[1]$, then a subsequent gauge transformation enacted by any $A_v(h)$ produces no change; so the action of $A_v[1]$ produces invariant states. Furthermore, $A_v[1]$  commutes with $A_{v'}[1]$ for any $v,v' \in V$, so we can impose this constraint on each vertex simultaneously. Thus, the electric constraint operator can be defined as
\begin{align}
    \Pi_A = \bigotimes_{v \in V} A_v[1]\,.
\end{align}
Any state in the image of this operator will be gauge invariant.

As an aside, if we wrote the action of $A_v[1]$ in the representation basis of $L^2(G)$ instead, one would find that only configurations which fuse to the trivial representation at $v$ survive the action of $A_v[1]$. In other words, $A_v[1]$ enforces charge conservation at each vertex. This is another way to see why the definition of $A_v[1]$ is more subtle in the case of non-compact $G$: the trivial representation actually does not appear in the Plancherel decomposition of $L^2(G)$ in \eqref{eqn:peterweylnoncompact} when $G$ is non-compact \cite{Haar}, so defining a projection onto charge-neutral states requires some extra mathematical machinery.

When $G$ is compact, $A_v[1]$ is proportional to a projector onto the gauge invariant subspace of $\Ha(\Lambda)$, because $A_v[1] A_v[1] = \text{Vol}(G) \cdot A_v[1]$, which we can see by integrating both sides of \eqref{eqn:Av1invariant} with respect to $g$. So after rescaling $\Pi_A$ by an appropriate power of the group volume, $\Pi_A$ will be a projector when $G$ is compact.  Thus the gauge invariant Hilbert space for compact groups can be constructed as
\begin{align}
    \Pi_A \Ha(\Lambda) = \span\{\text{Vol}(G)^{-|L|/2} \Pi_A \ket{\psi} \text{ such that } \ket{\psi} \in \Ha(\Lambda)\} \,.\label{eqn:PiAHeasy}
\end{align}
When $G$ is non-compact, we must be more careful, since the volume of non-compact groups diverges, and we can not normalize $\Pi_A$ in this way. Note that this is  analogous to the issues we faced in implementing the magnetic constraints for continuous groups. We will deal with this challenge analogously.

The solution to this problem is that while $A_v[1]$ does not define a projector on $\Ha(\Lambda)$ when $G$ is non-compact, we can still define $\Ha_{null} \subset \Ha(\Lambda)$ to be set of states which $\Pi_A$ annihilates.\footnote{It is easy to construct such states: for any $\ket{\psi} \in \Ha(\Lambda)$ and any $g \in G$, $(A_v(g)-1)\ket{\psi}$ is a null state. More generally, any linear combination of such states is null.}
The set of equivalence classes
\begin{align}
    [\ket{\psi} \sim \ket{\psi} + \ket{\chi}] & \, \text{ for all }  \ket{\psi} \in \Ha(\Lambda) \,\text{ and } \, \ket{\chi} \in \Ha_{null}\,. \label{eqn:electricnullstates}
\end{align}
forms a vector space, and we will denote the equivalence class containing $\ket{\psi}$ by $\ket{\psi}\rangle$. Next, it will be convenient to define $\Ha(\Lambda)_1$ as the subspace of $\Ha(\Lambda)$ whose wave functions are $L^1$, in addition to being $L^2$.\footnote{A function $f(g)$ is in $L^1(G)$ if $\int_G dg |f(g)|$ converges.} $\Ha(\Lambda)_1$ is dense in $\Ha(\Lambda)$, but is not a Hilbert subspace, for it is not complete with respect to the $\Ha(\Lambda)$ inner product. Not all wave functions in $\Ha(\Lambda)$ are in $\Ha(\Lambda)_1$: for example, in the case of $L^2(\R)$, there are wave functions with tails that go as $x^{-1}$ as $x \to \infty$, which are $L^2$ but not $L^1$. It turns out that a wave function must be in $L^1 \cap L^2$ to be normalizable after we impose the electric constraint. Note that when $G$ is compact, all $L^2$ functions are $L^1$. This explains why this subtlety did not arise in \cite{akers2024multipartite,Dong2024}, who considered compact gauge groups.

The reason this restriction  does not usually arise in quantum mechanics is because Hilbert spaces must be complete, and if we completed $\Ha(\Lambda)_1$ with respect to the usual $L^2$ inner product, we would get back all of $\Ha(\Lambda)$. We will see below that similarly to the magnetic constraints, we can define a new inner product where $\Ha(\Lambda)_1 / \Ha_{null}$ \emph{is} complete, and this alternative definition of the inner product agrees with the usual definition when $G$ is compact. We can now define the vector space
\begin{align}
    (\Pi_A \Ha(\Lambda))_{pre} = \Ha(\Lambda)_1 / \Ha_{null}
\end{align}
equipped with the inner product
\begin{align}
    \langle \braket{\sigma}{\psi}\rangle = \bra{\sigma} \Pi_A \ket{\psi} \,. \label{eqn:PiAinnerproducthard}
\end{align}
Note that this definition does not depend on a choice of representative.
This defines a Hilbert space we will call $\Pi_A\Ha(\Lambda)$, which analogously to the magnetic case, is defined as the completion
\begin{align}
    \Pi_A \Ha(\Lambda) = \overline{(\Pi_A \Ha(\Lambda))_{pre}}
\end{align}
with respect to the inner product \eqref{eqn:PiAinnerproducthard}. If $G$ is compact, this procedure leads to the same Hilbert space as the simpler definition \eqref{eqn:PiAHeasy}. But when $G$ is non-compact, this definition of the inner product still produces a gauge invariant Hilbert space. 

To see this, consider the simple case when $\Lambda_1$ contains a single leg as in Fig.~\ref{fig:basicplaquette}, so $\Ha(\Lambda_1) = L^2(G)$. Let $\ket{\psi},\ket{\sigma}$ be $L^1$ functions as well as $L^2$. Then \eqref{eqn:PiAinnerproducthard} reduces to
\begin{align}
     \langle \braket{\sigma}{\psi}\rangle 
     &= \int d[g,h,k] \,\sigma^*(g) \psi(h) \bra{g} A_v(k)\ket{h}
     \\&= \int d[g,h,k] \,\sigma^*(g) \psi(h) \braket{g}{kh}
     \\&= \int d[h,k] \,\sigma^*(kh) \psi(h) 
     \\&= \left(\int dk \,\sigma(k)\right)^*\left(\int dh \,\psi(h) \right)\,. \label{eqn:L1}
\end{align}
Because $\psi,\sigma$ are $L^1$ as well as $L^2$ functions, this inner product converges.  Similar to the magnetic case, the Cauchy-Schwartz inequality implies that the completed Hilbert space $\Pi_A \Ha(\Lambda_1)$ is one dimensional because all normalized vectors have maximal overlap. Finally, because $A_v(g)$ commutes with $A_{v'}(h)$ for any vertices $v \neq v'$, a similar analysis can be applied to any graph $\Lambda$ by imposing $A_v[1]$ separately for each vertex.

As a concrete example, let $G=\R$, and $\Lambda_1$ be as above. Roughly, the gauge invariant state satisfying the Gauss constraint is the constant function. The problem is that there is no constant function in $L^2(\R)$, for the constant function is not normalizable. However, the gauge invariant inner product is given by
\begin{align}
    \langle\braket{\sigma}{\psi}\rangle = \left(\int dx\, \sigma(x) \right)^* \left(\int dx \,\psi(x) \right) \,.
\end{align}
We can see that any state whose wave function integrates to zero is a null state in this inner product. Quotienting out the null states, which are spanned by all the non-zero Fourier modes, we are left with a one dimensional Hilbert space which is spanned by the constant function, even though the constant function is not  a valid state of $L^2(\R)$ by itself. Any wave function which integrates to one is a valid representative of the constant function, as it will differ from the constant function by purely non-zero Fourier modes (null states).

The resulting Hilbert space $\Pi_A \Ha(\Lambda)$ defined by \eqref{eqn:PiBHhard} is called the Hilbert space of $G$ co-invariants, while the simpler definition \eqref{eqn:PiAHeasy} is called the Hilbert space of $G$ invariants \cite{AMHJprelim,DeVuyst:2024fxc,Held:2025mai}.\footnote{We thank Elba Alonso-Monsalve for many helpful discussions about the construction of the co-invariant Hilbert space.} This technique also goes by the name of group averaging \cite{AMHJprelim,Ashtekar:1995zh,Marolf:2000iq,Marolf_2009,Giulini:1998kf} and is related to BRST-BV quantization \cite{Henneaux:1992ig,Henneaux:2025ocw,Penington:2023dql}. The reader may  be familiar with BRST quantization in the case of continuum quantum field theory, so it is useful to compare why the method we are using serves the same purpose. The operator $\Pi_A$ has a $\text{Vol}(G)^{|L|}$ divergence. In the continuum, this is divergent because $|L|\to\infty$. In our case, although $|L|$ is finite, $\text{Vol}(G)$ is infinite. Thus, in both cases, the role of BRST quantization is to provide a formal way to eliminate  extra factors of the gauge group volume as necessary to obtain finite answers.

In Appendix \ref{sec:DA}, we show that $[\Pi_A, \Pi_B] = 0$,\footnote{Strictly speaking, this requires a definition of the action of $\Pi_{A,B}$ on the separate Hilbert spaces $\Pi_{B,A}\Ha(\Lambda)$, respectively. What we really mean is that the natural definition of these actions commute.} so there is an unambiguous, simultaneously invariant Hilbert space, which is isomorphic to the image
\begin{align}
    \Ha_{phys}(\Sigma) := \Pi_A \Pi_B \Ha(\Lambda) \,.
\end{align}
This is the physical Hilbert space, which we will analyze in more detail in Sec.~\ref{sec:Hphys}. Recall that $\Sigma$ is the two-dimensional surface that $\Lambda$ tesselates. We choose the notation $\Ha_{phys}(\Sigma)$, as opposed to $\Ha_{phys}(\Lambda)$, for reasons that will become clear in Sec.~\ref{sec:Hphys}.

\subsection{Out-of-plane legs} \label{sec:outofplane}

In conventional tensor networks, local matter degrees of freedom can be incorporated by including ``out-of-plane'' legs for the bulk vertices. These are legs of the tensor network which are not contracted with a second bulk vertex, and may live in a different Hilbert space than the in-plane bulk legs. The motivation behind this convention is that one  imagines that each bulk vertex represents some spatial subregion of the bulk, and the matter leg accounts for the possible matter states within that subregion. 
We will also refer to the out-of-plane legs as ``matter legs'', but we are really thinking about the out-of-plane legs as describing the local physics in a patch (perhaps a Hubble volume) of spacetime. This is consistent with the perspective of out-of-plane legs in traditional tensor networks.

In topological tensor networks, we will adopt a similar procedure. However, instead of only associating out-of-plane legs with Hilbert spaces living at vertices, we will need to also associate them to plaquettes \cite{akers2024multipartite}. Physically, one way to think about this is that matter could in principle contain both electric and magnetic charges, and thus needs to be sensitive to the physics both at vertices and plaquettes. Mathematically, this is because both the electric and magnetic constraints must be satisfied by the matter degrees of freedom if the Wheeler-DeWitt equation is satisfied in the continuum. This means there should be some action of both the electric operators $A_v$ and magnetic operators $B_{(v,p)}$ on the matter Hilbert space. This would not be possible if we simply attached the matter Hilbert spaces to each vertex. 

Recall that we called each pair $(v,p)$ a site, with the set of all sites being denoted $S$.  To each site, we can associate a matter Hilbert space.\footnote{If there are not enough vertices or plaquettes in some lattice $\Lambda$ for each matter Hilbert space to be associated to disjoint sites, we can use moves 1 and 2 defined below to find a new lattice $\Lambda'$ where each matter site is disjoint \cite{akers2024multipartite}.} 
With this in mind, we will include the matter degrees of freedom in the pre-Hilbert space as \cite{akers2024multipartite}
\begin{align}
    \Ha(\Lambda) = \bigotimes_{\ell \in L} \Ha_\ell \bigotimes_{(v,p) \in S} \Ha^{matt}_{(v,p)} \,. \label{eqn:preHmatter}
\end{align}
Furthermore, the electric and magnetic operators should incorporate the gauge transformations of the matter degrees of freedom. Let $A_v^{matt}, B_{(v,p)}^{matt}$ denote, respectively, the electric and magnetic action of the gauge group on $\Ha^{matt}_{(v,p)}$. Then compared to the matter-free case, we must make the following substitutions \cite{akers2024multipartite} in the definition of the constraint operators:
\begin{align}
    A_v(h) &\to A_v(h) A_v^{matt}(h)\,, \label{eqn:Aconstraintwithmatter}\\
    B_{(v,p)}(h) &\to \int dg B_{(v,p)}(hg^{-1})B^{matt}_{(v,p)}(g) \,.
\end{align} 
After these substitutions, the rest of the analysis of the constraint operators is unchanged. The matter-free case is equivalent to taking $\Ha^{matt}_{(v,p)} = \C$, because vector spaces $V$ and $V \otimes \C$ are isomorphic.
Note that this is consistent with the fact that in pure 3D gravity, there are no local degrees of freedom, so physics in a local region of space requires additional fields to be nontrivial.

\section{The physical Hilbert space}\label{sec:Hphys}

After applying both the electric and magnetic constraints, the inner product on $\Ha_{phys}(\Sigma)$ takes the form
\begin{align}
    \langle \braket{\sigma}{\psi}\rangle_{phys} &= \bra{\sigma} \Pi_A \Pi_B \ket{\psi} \\&:= \int \prod_{v \in V} dg_v \bra{\sigma} \left[\prod_{v \in V} A_v(g_v) \right]\left[ \prod_{(v,p) \in S} B_{(v,p)}(e) \right] \ket{\psi}\,,
\end{align}
where $\ket{\psi}\rangle$ is an equivalence class of wave functions in $\Ha(\Lambda)$, identified up to states $\ket{\chi}$ such that $\Pi_A \Pi_B \ket{\chi} = 0$. If $\Pi_{A,B}$ were projectors in the usual sense, this would agree with the usual definition of projection onto the gauge invariant subspace. But as explained above, $\Pi_A$ is not quite a projector when $G$ is non-compact ($\Pi_A^2 \sim \text{Vol}(G) \Pi_A$), and $\Pi_B$ is not quite a projector when $G$ is continuous $(\Pi_B^2 \sim \delta(e) \Pi_B$), so the more careful definitions explained in Sec.~\ref{sec:themodel} are needed to make sense of the physical Hilbert space in general. These more careful definitions simply remove  additional divergent factors which arise from the $\Pi^2$ in the naive definition, rendering the inner product between states finite. Because all  the operators $A_v(g)$ and $B_{(v,p)}(e)$ commute, this inner product is well-defined.

\subsection{Comparison with traditional tensor networks}

At this point, we have constructed the Hilbert space $\Ha_{phys}(\Sigma)$ of a topological tensor network. At first, this class of states seems quite different than traditional tensor network states, so it is illuminating to compare the two more concretely.

A traditional tensor network is constructed as follows. Let $v$ be a vertex with $n$ in-plane legs attached to it, and define the Hilbert space
\begin{align}
    \Ha_v = \Ha_v^{(1)} \otimes \cdots \otimes \Ha_v^{(n)} \otimes \Ha^{matt}_v
\end{align}
associated with this vertex. $\Ha^{matt}_v$ is the Hilbert space associated to the out-of-plane leg located at the vertex $v$. If there is no matter leg at $v$, we can think of $\Ha^{matt}_v = \C$ as the one dimensional Hilbert space.

A state $\ket{\psi_v} \in \Ha_v$ can be thought of as a tensor with $n$ legs which has been placed at this vertex. To construct a traditional tensor network, we must contract the legs of these tensors according to the graph $\Lambda$ which defines the tensor network. Suppose we want to contract the $i$th leg of a vertex $v$ with the $j$th leg of a vertex $v'$. To do so, we define another Hilbert space $\Ha_{\ell}$ associated with this leg. In terms of the decomposition of $\Ha_v, \Ha_{v'}$, we can think of $\Ha_\ell = \Ha_v^{(i)} \otimes \Ha_{v'}^{(j)} $. Note that for this contraction to be consistent, we need $\Ha_v^{(i)} \cong \Ha_{v'}^{(j)}$. If $\ket{\chi, ij}$ is the maximally entangled Bell pair of $\Ha_\ell$, then the state
\begin{align}
    \ket{\Psi} = \left(\bigotimes_{\langle ij\rangle} \bra{\chi,ij}\right) \bigotimes_{v \in V} \ket{\psi_v} \label{eqn:tradTNverts}
\end{align}
represents a state with support on the boundary vertices $\Ha_{\partial}$ and the matter legs $\Ha_{matt}$. The notation $\langle ij\rangle$ indicates that we should project with Bell pairs on all the tensor legs which are connected according to the graph $\Lambda$. The Bell pairs ensure that the tensors of each vertex are contracted in the usual way: a Bell pair will first project the $i$th leg of $v$ and the $j$th leg of $v'$ onto the same state, and sum over a basis of all possible states in a uniform way. The only legs which remain uncontracted are the boundary legs  and the matter legs. Finally, if we flip all the matter legs from kets to bras, then we can equivalently think of $\ket{\Psi}$ as a map $\Psi: \Ha_{matt} \to \Ha_\partial$. So a tensor network is a map from the bulk to the boundary Hilbert spaces.

However, there is a ``dual'' perspective we can take on this state which is equally valid. Instead of constructing the tensor network by first building $\Ha_v$ and contracting these tensors using the Bell pairs at each leg, we could instead build the tensor network by placing a Bell pair on each leg and contracting the ends of these Bell pairs according to particular tensors at each vertex. In other words, the same tensor network state $\ket{\Psi}$ can be thought of as 
\begin{align}
    \ket{\Psi} = \left(\bigotimes_{v \in V} \Psi_v(\ell_1,\cdots,\ell_n)\right) \bigotimes_{\ell \in L} \ket{\chi,\ell} \,.\label{eqn:tradTNlegs}
\end{align}
Here, $\ket{\chi,\ell}$ is a Bell pair associated with each leg of the tensor network. $\Psi_v(\ell_1,\cdots,\ell_n)$ is a tensor which contracts the half of the Bell pairs at $\ell_1, \cdots, \ell_n$ attached to the vertex $v$, and produces a state in the matter Hilbert space $\Ha_{matt}$. In terms of the states $\ket{\psi_v}$ of \eqref{eqn:tradTNverts}, $\Psi_v$ is just $\ket{\psi_v}$ with the in-plane legs viewed as bras instead of kets. This is the same tensor because all we did was swap the bras and kets in the contractions of \eqref{eqn:tradTNverts}. If we then contracted a state $\ket{\psi_{matt}} \in \Ha_{matt}$ onto the matter legs of $\ket{\Psi}$, then we get a state in $\Ha_\partial$. So $\ket{\Psi}$ is still a bulk-to-boundary map.

A topological tensor network has the same basic structure as this legs-first perspective, with some differences. In the definition of $\Ha(\Lambda)$ in \eqref{eqn:preH} or \eqref{eqn:preHmatter}, we associated a copy of $L^2(G)$ with each in-plane leg. For a fixed representation $\pi$, this has the same tensor factorization $V_\pi \otimes V_\pi^*$ as in a traditional tensor network. This suggests we should think of the vector space $V_\pi$ of \eqref{eqn:peterweylnoncompact} as associated with one vertex of $\ell$, and $V_\pi^*$ as associated with the other vertex. Because we then sum over all possible representations, we can think of a topological tensor network as preparing a superposition of tensor networks. In fact, because there are an infinite number of representations $\pi \in \widehat{G}$, a topological tensor network can be thought of as a superposition of infinitely many traditional tensor networks. 

In traditional tensor networks, the structure of the state is determined by two pieces of data: the choice of tensors $\Psi_v$ at each vertex and the state on each leg $\ket{\chi,\ell}$. In our case, the electric constraints impose the contractions with the tensor $\Psi_v$.
As explained in Sec.~\ref{sec:themodel}, the electric constraint $A_v[1]$ forces the representations meeting at a vertex $v$ to fuse to the trivial representation. The operator which performs this fusion is called an intertwiner, and is explained in Appendix \ref{sec:crashcourse}. Satisfying the electric constraint forces the $\Psi_v$ tensors to be interwiners. The magnetic constraints are what force the states of the bulk legs to be Bell pairs $\ket{\chi,\ell}$. To see this, note that an example of a state in the pre-Hilbert space which satisfies the magnetic constraint is the product $\ket{e,\cdots,e}_{bulk}\ket{\psi}_{\partial}$ of delta functions on the identity element on every bulk leg, and an arbitrary state on the boundary legs. If we Fourier transform the state $\ket{e}$ on a single bulk leg (see Appendix \ref{sec:crashcourse}), we find that
\begin{align}
    \ket{e} = \int d\mu(\pi) \sum_{m,n} \pi(e)_{mn}\ket{\pi,mn} = \int d\mu(\pi) \sum_{m} \ket{\pi,mm}=  \int d\mu(\pi) \ket{\chi_\pi} \,,
\end{align}
where we used that $\pi(e) = \Id_{V_\pi}$. When $G$ is non-compact, the state
\begin{align}
    \ket{\chi_\pi} = \sum_{m} \ket{\pi,mm} 
\end{align}
is defined by an infinite sum, so we need to be careful about convergence issues. However, it turns out that the wave function of $\ket{\chi_\pi}$ does converge to an $L^1$ function called the character function of the representation $\pi$, regardless of the compactness of $G$. These character functions are foundational to the structure of topological tensor networks: we explain them in more detail in Appendix \ref{sec:crashcourse}. From the form of the wave function, we can see that the character function can be thought of as the Bell pair for a particular representation. Thus, the magnetic constraint forces the initial state of the bulk legs of the tensor network to be maximally entangled within each sector $\pi$. Although we only showed this for the specific example $\ket{e,\cdots,e} \ket{\psi}$, it turns out that after also applying the electric constraints, this is the most general case. In gravitational variables, the momentum constraint determines the entanglement structure on the legs of the topological tensor network, and the Wheeler-DeWitt equation fixes the tensors we use to contract these legs with.

Thus, for topological tensor networks, after contracting the matter legs onto a state $\ket{\psi_{matt}}$, the only independent degrees of freedom are on the boundary legs. It is interesting that the equations of motion of gravity are the mechanism which imposes this. 
We will analyze this property in more detail in Sec.~\ref{sec:bulktoboundary} after explaining some other features of the physical Hilbert space which will be useful in this analysis.

\subsection{Lattice deformations}

As explained in Sec.~\ref{sec:themodel}, the pre-Hilbert space $\Ha(\Lambda)$ has many null states in the inner product for $\Ha_{phys}(\Sigma)$ which must be quotiented out. One implication is that we can add extra states to $\Ha(\Lambda)$ without changing $\Ha_{phys}(\Sigma)$, as long as these additional states are annihilated by either $\Pi_A$ or $\Pi_B$. We can also remove states from $\Ha(\Lambda)$ without affecting $\Ha_{phys}(\Sigma)$, as long as these states are annihilated by either $\Pi_A$ or $\Pi_B$.
Using this freedom, it turns out that different graphs $\Lambda$ and $\Lambda'$ that generate different pre-Hilbert spaces $\Ha(\Lambda)$ and $\Ha(\Lambda')$ can lead to the same physical Hilbert space $\Ha_{phys}(\Sigma)$. This was shown for finite groups in \cite{akers2024multipartite}, and we prove this for transformable groups in Appendix~\ref{sec:move12}. Two graphs $\Lambda,\Lambda'$ will lead to the same physical Hilbert space if and only if they 1) have the same Euler characteristic, i.e., if they tesselate the same surface $\Sigma$, and 2) have the same number of boundary legs. This is our reason for the notation $\Ha_{phys}(\Sigma)$, as opposed to $\Ha_{phys}(\Lambda)$. 

There are two families of ``moves'' which can transform a lattice $\Lambda \to \Lambda'$ with the same Euler character and boundary legs. Graphically, these moves consist of geometrically changing the graph $\Lambda$ by adding or removing adjacent vertex/leg or leg/plaquette pairs. Quantum mechanically, each move has an associated isometry\footnote{They are isometries when restricted to states in $\Ha_{phys}(\Sigma)$, thought of as subspaces of $\Ha(\Lambda)$ or $\Ha(\Lambda')$. For non-gauge-invariant states, $\Delta_{1}$ and $\Delta_2$ need not be isometries. } $\Delta_1$, $\Delta_2$ which sends states from $\Ha(\Lambda) \to \Ha(\Lambda')$. These moves are a direct consequence of the fact that states in the physical Hilbert space satisfy the constraints. We briefly review these moves here, and refer the reader to Appendix \ref{sec:move12} for more details.

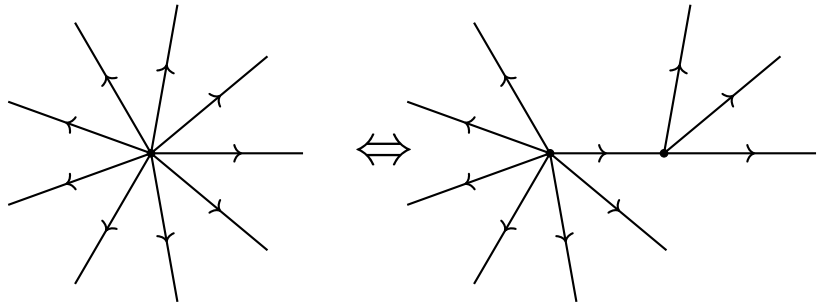
\begin{figure}
    \centering
    \begin{tikzpicture}
        \def\sep{3}
        \def\nn{9}
        \def\scale{2}
        \def\hh{0.75}
        \node at (-\sep,0) {\begin{tikzpicture}[scale=\scale]
        
        \filldraw (0,0) circle (0.025);
        \foreach \xx in {1,...,\nn}{
        \draw[thick,->-=0.6] (0,0) -- ({cos(360*\xx/\nn)},{sin(360*\xx/\nn)});
        }
        \end{tikzpicture}};
        \node[scale=2] at (0,0) {$\Leftrightarrow $};
        \node at (\sep,0) {\begin{tikzpicture}[scale=\scale]
        \filldraw (0,0) circle (0.025);
        \filldraw (\hh,0) circle (0.025);
        \foreach \xx in {1,...,3}{
        \draw[thick,->-=0.6] (\hh,0) -- ({\hh+cos(360*(\xx-1)/\nn)},{sin(360*(\xx-1)/\nn)});
        }
        \foreach \xx in {4,...,\nn}{
        \draw[thick,->-=0.6] (0,0) -- ({cos(360*(\xx-1)/\nn)},{sin(360*(\xx-1)/\nn)});
        }
        \draw[thick,->-=0.5] (0,0) -- (\hh,0);
        \end{tikzpicture}};
       
    \end{tikzpicture}
    \caption{An example of move 1, which can be performed in either direction to add or remove a bulk vertex/leg from the graph $\Lambda$.}
    \label{fig:move1}
\end{figure}

\paragraph{Move 1:} The first move, permitted by the electric constraint $\Pi_A$,  allows us to add or remove a vertex/leg pair from $\Lambda$ (Fig.~\ref{fig:move1}). We can do this move in either direction.
This move is possible because the states in the physical Hilbert space are gauge invariant. Physically, one way to see this is that if the flow of charge is conserved on the LHS of Fig.~\ref{fig:move1}, then it will also be conserved on the RHS. Mathematically, this move works because the fusion of unitary representations is associative, so the globally charge-neutral states of the LHS will be isomorphic to the globally charge-neutral states of the RHS.

\begin{figure}
    \centering
    \begin{tikzpicture}
        \def\sep{4}
        \def\nn{6}
        \def\rr{1.3}
        \def\scale{2}
        \def\hh{0.75}
        \node at (-\sep,0) {\begin{tikzpicture}[scale=\scale]
        
        \foreach \xx in {1,...,\nn}{
        \filldraw ({cos(360*(\xx-1)/\nn)},{sin(360*(\xx-1)/\nn)}) circle (0.025);
        \draw[thick,->-=0.6] ({cos(360*(\xx-1)/\nn)},{sin(360*(\xx-1)/\nn)}) -- ({\rr*cos(360*(\xx-1)/\nn)},{\rr*sin(360*(\xx-1)/\nn)});
        \draw[thick,->-=0.6] ({cos(360*(\xx-1)/\nn)},{sin(360*(\xx-1)/\nn)}) -- ({cos(360*\xx/\nn)},{sin(360*\xx/\nn)});
        }
        \end{tikzpicture}};
        \node[scale=2] at (0,0) {$\Leftrightarrow $};
        \node at (\sep,0)  {\begin{tikzpicture}[scale=\scale]

        \draw[thick,->-=0.5] (1,0) -- ({cos(360*(2)/\nn)},{sin(360*(2)/\nn)});
        
        \foreach \xx in {1,...,\nn}{
        \filldraw ({cos(360*(\xx-1)/\nn)},{sin(360*(\xx-1)/\nn)}) circle (0.025);
        \draw[thick,->-=0.6] ({cos(360*(\xx-1)/\nn)},{sin(360*(\xx-1)/\nn)}) -- ({\rr*cos(360*(\xx-1)/\nn)},{\rr*sin(360*(\xx-1)/\nn)});
        \draw[thick,->-=0.6] ({cos(360*(\xx-1)/\nn)},{sin(360*(\xx-1)/\nn)}) -- ({cos(360*\xx/\nn)},{sin(360*\xx/\nn)});
        }
        \end{tikzpicture}};
       
    \end{tikzpicture}
    \caption{An example of move 2, which can be performed in either direction to add or remove a leg/plaquette from the graph $\Lambda$.}
    \label{fig:move2}
\end{figure}
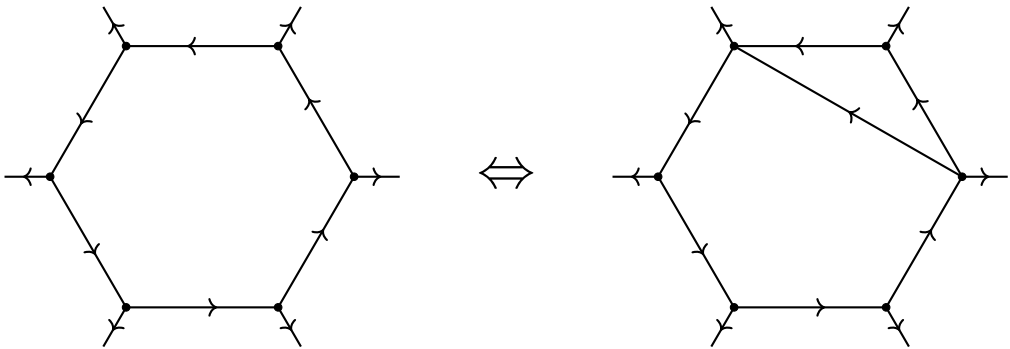

\paragraph{Move 2:} The second move allows us to add or remove a leg/plaquette pair from $\Lambda$ (Fig.~\ref{fig:move2}). Like move 1, we can perform this move in either direction. The magnetic constraint $\Pi_B$ allows this move, because if we bisect a plaquette with zero flux, the resulting plaquettes will continue to have zero flux. Physically, this move is possible because Chern-Simons theory is topological.

\begin{figure}
    \centering
    \begin{tikzpicture}[scale=3]
        \def\nn{6}
        \def\mm{3}
        \def\rr{1.1}
        
        \foreach\xx in {1,...,\nn}{
            \draw[thick,->-=0.6] (0,0) -- ({cos(360*(\xx)/(\nn+\mm))},{sin(360*(\xx)/(\nn+\mm))});
            \filldraw[thick,fill=white] ({cos(360*(\xx)/(\nn+\mm))},{sin(360*(\xx)/(\nn+\mm))}) circle (0.025);
            }
        \foreach\xx in {1,...,\mm}{
            \draw[thick,->-=0.6] (0,0) -- ({cos(360*(\xx+\nn)/(\nn+\mm))},{sin(360*(\xx+\nn)/(\nn+\mm))});

            \draw[blue,thick] ({cos(360*(\xx+\nn)/(\nn+\mm))},{sin(360*(\xx+\nn)/(\nn+\mm))}) -- ({\rr*cos(360*(\xx+\nn)/(\nn+\mm))},{\rr*sin(360*(\xx+\nn)/(\nn+\mm))});

            \fill[blue] ({\rr*cos(360*(\xx+\nn)/(\nn+\mm))},{\rr*sin(360*(\xx+\nn)/(\nn+\mm))}) circle (0.025);
            
            \filldraw ({cos(360*(\xx+\nn)/(\nn+\mm))},{sin(360*(\xx+\nn)/(\nn+\mm))}) circle (0.025);

            \draw[thick] ({\rr*cos(360*(\xx+\nn)/(\nn+\mm))},{\rr*sin(360*(\xx+\nn)/(\nn+\mm))}) circle ({\rr-1});

            }
        \filldraw (0,0) circle (0.025);
    \end{tikzpicture}
    \caption{Reduced lattice when matter is included. The boundary vertices are shown in white, and the out-of-plane legs are shown in blue. The matter legs are attached to a bulk vertex and plaquette, which we call a ``lollipop''. This lollipop is connected to the central vertex of the reduced lattice through another bulk leg. }
    \label{fig:lollipop}
\end{figure}
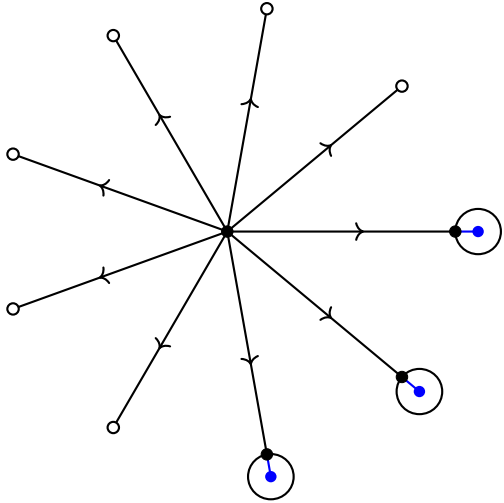

\paragraph{The reduced lattice:} Any two graphs $\Lambda, \Lambda'$ related by a sequence of these moves will define the same physical Hilbert space after the constraints are enforced. The presentation of $\Ha_{phys}(\Sigma)$ may depend on $\Lambda$, but all such presentations are isomorphic and therefore physically equivalent. When $\Sigma$ is a disk,\footnote{When $\Sigma$ is not a disk, other canonical representative lattices exist, but will contain non-trivial cycles.} there is a preferred representative $\Lambda_r$ called the reduced lattice which is particularly easy to work with (Fig.~\ref{fig:lollipop}).  Each matter Hilbert space is associated with a site, and therefore we will need to retain at least one site per matter degree of freedom in the reduced lattice. This leads to, in addition to boundary legs, an additional kind of leg called a ``lollipop factor'' which consists of a single vertex, a pair of legs, and a single plaquette, as well as the bulk matter leg itself. See \cite{akers2024multipartite} for more details. The reduced lattice consists of \emph{only} boundary legs and lollipop factors, connected with a single bulk vertex and no plaquettes. 

\subsection{Physical operators}

Let $\mathcal{O}$ be a bounded operator on $\Ha(\Lambda)$. We say $\mathcal{O}$ is physical  if it commutes with the constraint operator $\Pi_A \Pi_B$. This is a reasonable definition because if $\mathcal{O}$ is a physical operator and $\ket{\chi}$ is a null state, then
\begin{align}
    \Pi_A \Pi_B \mathcal{O} \ket{\chi} =  \mathcal{O}  \Pi_A \Pi_B \ket{\chi} = 0 \,.
\end{align}
Thus, $\mathcal{O}$ sends null states to null states.
This implies that we can define an operator $\widetilde{\mathcal{O}}$ on $\Ha_{phys}(\Sigma)$ by its action on a representative. In other words, we can define $\widetilde{\mathcal{O}}$ by
\begin{align}
    \widetilde{\mathcal{O}} \ket{\psi}\rangle = \ket{\mathcal{O} \cdot \psi}\rangle = [\mathcal{O}\ket{\psi} \sim \mathcal{O}\ket{\psi} + \ket{\chi}] \text{ for any}  \, \ket{\chi} \in \Ha_{null}\,.
\end{align}
If $\mathcal{O}$ did not map null states to null states, then $\widetilde{\mathcal{O}}$ would not be well-defined. 

Notice that we did \emph{not} define physical operators as $\widetilde{\mathcal{O}} = \Pi_A \Pi_B \mathcal{O}  \Pi_A \Pi_B$. When $G$ is a finite group so that $\Pi_A \Pi_B$ is a projection operator, these definitions are equivalent. But in the Hilbert space of co-invariants,  normalization issues arose  when we tried to square the constraint operators $\Pi_A \Pi_B$. The definition of physical operators using the commutation relation $[\Pi_A \Pi_B, \mathcal{O}]$ is linear in the constraint operators, and so is well defined for arbitrary transformable groups. 

Because physical operators have a representative $\widetilde{\mathcal{O}}$ on $\Ha_{phys}(\Sigma)$, we can use moves 1 and 2 above to relate the representative $\mathcal{O}$ on $\Ha(\Lambda)$ to another operator $\mathcal{O}'$ on $\Ha(\Lambda')$ if $\Lambda, \Lambda'$ lead to the same physical Hilbert space. Thus, even if we define a physical operator $\mathcal{O}$ on a particular representative $\Pi_A \Pi_B \Ha(\Lambda)$ of $\Ha_{phys}(\Sigma)$, we know it still exists as an operator on any other representative as well.

\subsubsection{Ribbon operators}

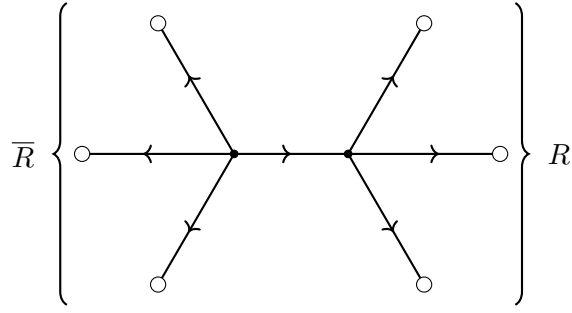
\begin{figure}
    \centering
    \begin{tikzpicture}[scale=2]
        \def\sep{3}
        \def\nn{6}
        \def\hh{0.75}
       
        \filldraw (0,0) circle (0.025);
        \filldraw (\hh,0) circle (0.025);
        \foreach \xx in {1,...,3}{
            \draw[thick,->-=0.6] (\hh,0) -- ({\hh+cos(360*(\xx-2)/\nn)},{sin(360*(\xx-2)/\nn)});
            \filldraw[fill=white] ({\hh+cos(360*(\xx-2)/\nn)},{sin(360*(\xx-2)/\nn)}) circle (0.05);
            }
        \foreach \xx in {4,...,\nn}{
        \draw[thick,->-=0.6] (0,0) -- ({cos(360*(\xx-2)/\nn)},{sin(360*(\xx-2)/\nn)});
        \filldraw[fill=white] ({cos(360*(\xx-2)/\nn)},{sin(360*(\xx-2)/\nn)}) circle (0.05);
        }
        \draw[thick,->-=0.5] (0,0) -- (\hh,0);

        \draw [thick,decorate,decoration={brace,amplitude=5pt}]
        (-1-0.1,-1) -- (-1-0.1,1) node[midway,xshift=-1.5em]{$\overline{R}$};

        \draw [thick,decorate,decoration={brace,amplitude=5pt,mirror}]
        (1+\hh+0.1,-1) -- (1+\hh+0.1,1) node[midway,xshift=1.5em]{$R$};
       
    \end{tikzpicture}
    \caption{The bowtie lattice between the boundary legs $R$ and $\overline{R}$.}
    \label{fig:bowtie}
\end{figure}

Next, we will construct an interesting explicit example of a physical operator. 
Let $R$ be a subset of the boundary legs of $\Lambda$. We think of this subset as specifying a subregion of the boundary of $\Sigma$, the surface that $\Lambda$ tessellates. Denote the complementary set of boundary legs by $\overline{R}$. Like the reduced lattice, there will be another representative graph $\Lambda_b$ which will be  convenient to work with. To define $\Lambda_b$, start with the reduced lattice $\Lambda_r$ and use move 1 to add a single leg separating the legs $R$ and $\overline{R}$ in the bulk. We call this the bowtie lattice. While there is a single reduced lattice, there is a different bowtie lattice for every choice of bipartition of boundary legs. Physically, this additional leg represents an infinitesimal thickening of the boundary separating $R,\overline{R}$. We will provide a stronger justification for this interpretation after we define the area operator for $R$, and see that it has support on this additional leg. 
We call this graph $\Lambda_b$ the bowtie separating $R$ and $\overline{R}$, and refer to the additional leg as the corner leg (see Fig.~\ref{fig:bowtie}).

Consider a gauge invariant state $\ket{\psi}\rangle$ in $\Ha_{phys}(\Sigma) = \Pi_A \Pi_B \Ha(\Lambda_b)$,  the physical Hilbert space. Such a state has a basis expansion
\begin{align}
    \ket{\psi}\rangle = \int d[\vec{g}_R,\vec{g}_{\overline{R}},h] \,\psi(\vec{g}_R, \vec{g}_{\overline{R}}, h) |\vec{g}_{\overline{R}},h,\vec{g}_R\rangle\rangle \,,
\end{align}
where $|\vec{g}_{\overline{R}},h,\vec{g}_R\rangle\rangle$ is the image of the group basis state $\ket{\vec{g}_{\overline{R}},h,\vec{g}_R}$ of the pre-Hilbert space $\Ha(\Lambda_b)$. Here, $\vec{g}_R, \vec{g}_{\overline{R}}$ refer to the collection of group elements on each of the boundary legs, and $h$ is the group element on the additional bulk leg (the corner leg). This resolution of the state follows from the linearity of the constraints $\Pi_A \Pi_B$ and a resolution of the identity on $\ket{\psi}$. Note that integrating over all the group elements of $|\vec{g}_{\overline{R}},h,\vec{g}_R\rangle\rangle$ is redundant due to the quotient by null states: this expression, however, is still well-defined because of the boundedness conditions of the wave function we explained in Sec.~\ref{sec:themodel}. For example, consider a graph $\Lambda_2$ with two boundary legs that are both oriented inwards and meet at a single vertex. The wave function of a state on $\Pi_A \Pi_B \Ha(\Lambda_2)$ will be
\begin{align}
    \ket{\psi}\rangle &= \int d[g,h]\,\psi(g,h) \ket{g,h}\rangle \label{eqn:Lambda2easystate} \\&= \int d[g,h]\,\psi(g,h) |h^{-1}g,e\rangle\rangle \\&= \int d[g,h]\,\psi(hg,h) \ket{g,e}\rangle\,.
\end{align}
In the first line, we used the fact that $\ket{g,h}$ and $\ket{h^{-1}g,e}$ are related by null states, and in the second line we used the left invariance of the Haar measure. If we define the reduced wave function
\begin{align}
    \widetilde{\psi}(g) = \int dh\,\psi(hg,h) \,, \label{eqn:reducedwavefn}
\end{align}
then we can see that 
\begin{align}
    \ket{\psi}\rangle = \int dg \,\widetilde{\psi}(g)\ket{g,e}\rangle \,.
\end{align}
The reason the reduced wave function $\widetilde{\psi}(g)$ is well-defined is because the representative $\psi(g,h)$ $L^1$ integrable, so \eqref{eqn:reducedwavefn} converges to another $L^1$ function.\footnote{To see this, compute the $L^1$ norm of $\widetilde{\psi}(g)$ and do a left multiplication $g \to hg$. This reduces to the $L^1$ norm of $\psi$, which is finite. } We can think of the integration \eqref{eqn:reducedwavefn} as integrating out the gauge-dependent degrees of freedom. But this is equivalent to the state \eqref{eqn:Lambda2easystate}, so we can use either presentation of the state if we wish.

Now, let $\chi_\pi(g)$ be a character function of an irreducible representation $\pi \in \widehat{G}$. When $G$ is compact, $\chi_\pi(g) = \tr[\pi(g)]$. When $G$ is non-compact, this is essentially still true, but the character function must first be appropriately normalized  to be well-defined. This makes the existence of $\chi_\pi(g)$ more subtle, and is explained in more detail in Appendix \ref{sec:crashcourse}. 
Nevertheless, irreducible unitary representations of non-compact groups still have character functions \cite{HarishChandra1976}. Importantly, though, $\chi_\pi(g)$ is always an $L^1$ function, which means it decays fast enough at infinity to be integrable with respect to the Haar measure \cite{Haar}. We will use these character functions to define the operator 
\begin{align}
    F_{\eth R}(\pi) \ket{\psi}\rangle &= \int d[\vec{g}_R,\vec{g}_{\overline{R}},h,k] \, \chi_\pi(k)\psi(\vec{g}_R, \vec{g}_{\overline{R}}, k^{-1}h) |\vec{g}_{\overline{R}},h,\vec{g}_R\rangle\rangle \,.
\end{align}
Notice that $F_{\eth R}(\pi)$ only acts on the corner leg of $\Lambda_b$, and not any of the boundary legs. This suggests we can think of $F_{\eth R}(\pi)$ as acting ``between'' the bulk subregions associated with the boundary legs $R, \overline{R}$. But if $F_{\eth R}(\pi)$ is a physical operator, then it also has an independent definition on $\Ha_{phys}(\Sigma)$, independent of the bowtie lattice $\Lambda_b$. This presentation of $F_{\eth R}(\pi)$ is simply a convenient choice to define this operator.

$F_{\eth R}(\pi)$ is only physically meaningful if it commutes with gauge transformations at both of the vertices of $\Lambda_b$. To see that $F_{\eth R}(\pi)$ commutes with gauge transformations, let $\ket{\psi}$ be a state in $\Ha(\Lambda_b)$, and recall that gauge transformations at the $R$ or $\overline{R}$ vertices act as
\begin{align}
    A_R(\ell) \ket{\vec{g}_{\overline{R}},h,\vec{g}_R} & = \ket{\vec{g}_{\overline{R}},h\ell^{-1},\ell \cdot\vec{g}_R}\,,\\
    A_{\overline{R}}(\ell) \ket{\vec{g}_{\overline{R}},h,\vec{g}_R} & = \ket{\ell\cdot\vec{g}_{\overline{R}},\ell h ,\vec{g}_R}\,.
\end{align}
By $\ell \cdot\vec{g}_R$, we mean a shorthand for left multiplication by $\ell$ on the outflowing legs of $R$, and right multiplication by $\ell^{-1}$ on the inflowing legs of $R$, as explained in Sec.~\ref{sec:themodel}. Then we can see that
\begin{align}
    A_R(\ell) F_{\eth R}(\pi)\ket{\vec{g}_{\overline{R}},h,\vec{g}_R} \rangle
    &= \int dk \chi_\pi(k) |\vec{g}_{\overline{R}},k^{-1}h\ell^{-1},\ell\cdot\vec{g}_R\rangle \rangle
    \\&= \int dk \chi_\pi(k) |\vec{g}_{\overline{R}},k^{-1}h,\vec{g}_R\rangle \rangle
    \\&=F_{\eth R}(\pi)|\vec{g}_{\overline{R}},h,\vec{g}_R\rangle \rangle\,,
\end{align}
where we used the gauge invariance of $|\vec{g}_{\overline{R}},kh,\vec{g}_R\rangle \rangle$ in the second line, and 
\begin{align}
    A_{\overline{R}}(\ell) F_{\eth R} (\pi)|\vec{g}_{\overline{R}},h,\vec{g}_R\rangle \rangle\,,
    &= \int dk \,\chi_\pi(k) |\ell\cdot \vec{g}_{\overline{R}},\ell k^{-1}h,\vec{g}_R\rangle\rangle\,,
    \\&= \int dk\, \chi_\pi(\ell^{-1}k\ell) |\ell\cdot \vec{g}_{\overline{R}}, k^{-1} \ell h,\vec{g}_R\rangle \rangle
      \\&= \int dk\, \chi_\pi(k) |\ell\cdot \vec{g}_{\overline{R}}, k \ell h,\vec{g}_R\rangle\rangle\,,
    \\&= F_{\eth R} (\pi) |\ell\cdot \vec{g}_{\overline{R}}, \ell h,\vec{g}_R \rangle\rangle \,,
    \\&= F_{\eth R} (\pi)|\vec{g}_{\overline{R}},h,\vec{g}_R\rangle \rangle \,.
\end{align}
We used the fact $\chi_\pi(g) = \chi_\pi(hgh^{-1})$ (cyclicity of the trace) in the third line, and the gauge invariance of $|\vec{g}_{\overline{R}},h,\vec{g}_R\rangle \rangle$ in the fifth line. Because we have shown that $F_{\eth R}(\pi)$ commutes with gauge transformations for a basis of $\Ha_{phys}(\Sigma)$, by linearity it is a physical operator on all of $\Ha_{phys}(\Sigma)$, despite the fact we defined it on the specific presentation $\Pi_A \Pi_B \Ha(\Lambda_b)$.

One can show from the definitions that
\begin{align}
    F_{\eth R} (\pi)F_{\eth R} (\omega) = \delta(\pi,\omega) F_{\eth R} (\pi)\,,
\end{align}
where $\delta(\pi,\omega)$ is the delta function with respect to the Plancherel measure $d\mu(\pi)$. This relies on equation \eqref{eqn:characterdeltafn} from Appendix \ref{sec:crashcourse}. Note that this implies $F_{\eth R} (\pi)$ and $F_{\eth R} (\omega)$ commute. If $f: \widehat{G}\to \C$ is a smearing function, then
\begin{align}
    F_{\eth R}(f) = \int d\mu(\pi) f(\pi) F_{\eth R}(\pi)
\end{align}
is also gauge invariant, so we can think of $ F_{\eth R}(\pi)$ as forming a basis of an (abelian) operator algebra $\mathcal{A}_{\eth R}$. Following  \cite{akers2024multipartite,Dong2024}, we can define an area operator ${\rm Area}_R \in \mathcal{A}_{\eth R}$ which measures the area of a bulk surface which is homologous to $R$ as a member of this algebra: 
\begin{align}
    {\rm Area}_R = \int d\mu(\pi) \log\left(\frac{d\mu(\pi)}{d\pi}\right) F_{\eth R}(\pi) \,.
\end{align}
Here, $\frac{d\mu(\pi)}{d\pi}$ is the measure-theoretic derivative (the Radon–Nikodym derivative) of the Plancherel measure with respect to the uniform measure of $\widehat{G}$, restricted to the support of the Plancherel measure. We can think of this as being the numerical coefficient of the Plancherel measure relative to the uniform measure $d\pi$ of $\widehat{G}$.\footnote{Really, the uniform measure restricted to the support of the Plancherel measure.} For example, $\frac{d\mu(\pi)}{d\pi} = \frac{d_\pi}{\text{Vol}(G)}$ when $G$ is compact. Indeed, \cite{akers2024multipartite,Dong2024} define the area operator as\footnote{Actually, these authors chose the normalization of the Haar measure with $\text{Vol}(G)=1$, but we restore the group volume dependence for comparison with non-compact groups.}
\begin{align}
    {\rm Area}_R = \int d\mu(\pi) \log\left(\frac{d_\pi}{\text{Vol}(G)}\right) F_{\eth R}(\pi) \,.
\end{align}
So our definition is the natural generalization of theirs. We will confirm that this is the correct definition in  \cite{upcoming} by showing that this operator contributes universally to the entanglement entropy of reduced states on a subset $R$ of the boundary legs of $\Lambda$.

However, if $R_1$ and $R_2$ are two overlapping boundary regions, then $[{\rm Area}_{R_1}, {\rm Area}_{R_2}] \neq 0$. As explained in the introduction, this is the expected behavior in semi-classical gravity. This was first demonstrated when $G$ is a finite group in \cite{akers2024multipartite}, and continues to be the case when $G$ is a transformable group with essentially the same proof. The reason is that because $F_{\eth R}(\pi)$ is a physical operator, we can determine its action on a physical state $\ket{\psi}\rangle$ by first acting $F_{\eth R}(\pi)$ on a representative $\ket{\psi}$ without loss of generality. Then, the analysis of \cite{akers2024multipartite} still applies, because we can quotient by null states after we perform the same manipulations. 

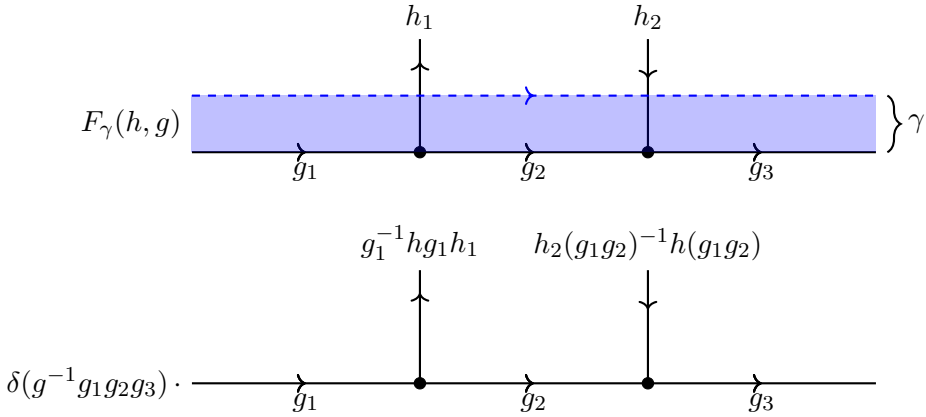
\begin{figure}
    \centering
    \begin{tikzpicture}
        \def\vv{1.5}
        \node at (0,\vv) {\begin{tikzpicture}[scale=3]
            \draw[thick,->-=0.1666,->-=0.5,->-=0.8333] (0,0) -- (3,0);
            \draw[thick,->-=0.8] (1,0) -- (1,0.5);
            \draw[thick,->-=0.35] (2,0.5) -- (2,0);
            \filldraw (1,0) circle (0.025);
            \filldraw (2,0) circle (0.025);
            
            \fill[blue,opacity=0.25] (0,0) -- (3,0) -- (3,0.25) -- (0,0.25) -- cycle;

            \draw[thick,dashed,blue,->-=0.5] (0,0.25) -- (3,0.25);

            \node[anchor=north] at (0.5,0) {$g_1$};
            \node[anchor=north] at (1.5,0) {$g_2$};
            \node[anchor=north] at (2.5,0) {$g_3$};

            \node[anchor=south] at (1,0.5) {$h_1$};
            \node[anchor=south] at (2,0.5) {$h_2$};

            \node[anchor=east] at (0,0.125) {$F_\gamma(h,g)$} ;
            \draw [thick,decorate,decoration={brace,amplitude=5pt,mirror}] (3.05,0) -- (3.05,0.25)
            node[midway,xshift=1em]{$\gamma$};
            \node at (-1,0) { };
            \node at (4,0) { };
        \end{tikzpicture}};

        \node at (0,-\vv) {\begin{tikzpicture}[scale=3]
            \draw[thick,->-=0.1666,->-=0.5,->-=0.8333] (0,0) -- (3,0);
            \draw[thick,->-=0.8] (1,0) -- (1,0.5);
            \draw[thick,->-=0.35] (2,0.5) -- (2,0);
            \filldraw (1,0) circle (0.025);
            \filldraw (2,0) circle (0.025);
            

            \node[anchor=north] at (0.5,0) {$g_1$};
            \node[anchor=north] at (1.5,0) {$g_2$};
            \node[anchor=north] at (2.5,0) {$g_3$};

            \node[anchor=south] at (1,0.5) {$g_1^{-1} h g_1 h_1$};
            \node[anchor=south] at (2,0.5) {$h_2 (g_1 g_2)^{-1} h (g_1 g_2)$};

            \node[anchor=east] at (0,0) {$\delta(g^{-1}g_1 g_2 g_3) \,\cdot$} ;
            \node at (-1,0) { };
            \node at (4,0) { };
        \end{tikzpicture}};
    \end{tikzpicture}
    \caption{An example of a ribbon operator with support on a ribbon $\gamma$. The ribbon operator measures the partial flux $g$ along the spine of the ribbon $\gamma$, and also enacts a parallel-transported group multiplication by $h$ on the spokes of $\gamma$. Figure adapted from (2.17) of \cite{akers2024multipartite}.}
    \label{fig:ribbon}
\end{figure}

The operator $F_{\eth R}(\pi)$ is a special case of a more general class of physical operators called ribbon operators. A ribbon operator is defined on a pair of adjacent paths (this pair is called a ribbon) through the graph $\Lambda$ (called the spine) and the dual graph (called the spokes). A ribbon is allowed to end on boundary vertices or matter legs, or form a closed loop, and is gauge invariant except at the endpoints of the ribbon. We can think of a ribbon operator as a generalization of a Wilson line. 
Given group elements $g,h \in G$ and a ribbon $\gamma$, the ribbon operator $F_\gamma(h,g)$ acts on $\Ha(\Lambda)$ as in Fig.~\ref{fig:ribbon}. These ribbon operators are gauge invariant away from the endpoints of the ribbon. When $G$ is a finite group, ribbon operators have a useful property: they are topological away from the matter legs. In other words, if $\gamma$ and $\gamma'$ are two ribbons which enclose a subgraph of $\Lambda$ which contains no matter legs, then $F_\gamma(h,g) = F_{\gamma'}(h,g)$ \cite{Bombin_2008}. We conjecture that this continues to hold for arbitrary transformable groups, but we will not present a formal proof here. However, we note that with the substitutions $\frac{1}{|G|}\sum_{g \in G} \to \int dg$ and $\sum_{\pi \in \widehat{G}} d_\pi \to \int d\mu(\pi)$, the same proof of this property for finite groups in \cite{Bombin_2008} seems to continue to hold.\footnote{Another useful substitution seems to be $\frac{1}{|C|} \sum_{[h] \in C} \to \sum_T\int_T dt \Delta(t)$, where $C$ is the set of conjugacy classes of $G$, $T$ labels the Cartan subgroups of $G$, and $\Delta(t)$ is the Weyl denominator formula (see Appendix \ref{sec:crashcourse}).} Furthermore, anyons (Wilson lines) in continuum Chern-Simons theories are indeed topological, so if we identify the ribbon operators with these excitations, that would imply that $F_\gamma$ are topological. It would be interesting to confirm this more precisely.

\paragraph{The quantum double algebra: } However, just as with $F_{\eth R}(\pi)$, we must smear ribbon operators with smearing functions $f(g),f'(h)$ to ensure that they are operators are bounded. The $g$ action is related to the magnetic operators because it measures the partial flux along the spine of the ribbon. The $h$ action is related to the electric operators because it enacts a (parallel transported) gauge transformation along the spokes of the ribbon. So, for $F_\gamma$ to be a bounded operator, we must demand that $f(g)$ is a bounded function, and $f'(h)$ is an $L^1$ function. The completion of these smeared operators is the operator algebra of ribbon operators. For more information about the full algebra of electric and magnetic operators on $\Ha(\Lambda)$, see Appendix \ref{sec:DA}. 

\subsection{Topological tensor networks as a bulk-to-boundary map} \label{sec:bulktoboundary}

The physical Hilbert space $\Ha_{phys}(\Sigma)$ that we defined in Sec.~\ref{sec:themodel} has an interesting structure which we will explore in greater detail in  \cite{upcoming}. But in this section, we will briefly sketch the proof that the physical Hilbert space of a topological tensor network with no out-of-plane legs should be thought of as a boundary Hilbert space, even though it is  constructed from a bulk lattice $\Lambda$. One of the reasons why quantum gravity researchers have been interested in traditional tensor networks is because they function as a toy model of some aspects of the bulk-to-boundary map which translates states between the two sides of a holographic duality \cite{Hayden_2015,Pastawski:2015qua}. Our result will imply that topological tensor networks can be also interpreted in this way. This was already demonstrated for finite groups in \cite{akers2024multipartite}, and we will show that the relationship continues to hold for all transformable groups.

For the moment, assume $G$ is compact, and neglect the out-of-plane legs. Consider the physical Hilbert space in the reduced lattice presentation, $\Ha_{phys}(\Sigma) = \Pi_A \Pi_B \Ha(\Lambda_r)$. Assume $\Lambda_r$ has $n$ boundary legs. For simplicity, let $\ket{\pi,ab;\vec{g}}$ be a basis for the pre-Hilbert space $\Ha(\Lambda_r)$, which consists of a state in the representation basis  $\ket{\pi,ab}$ for one leg, and, for notational convenience, states in the group basis $\ket{g_i}$ for the remaining legs. We take the orientation of the leg in the representation basis to be inflowing, so for $\ket{\pi,ab}$, the $a$ index (associated with $V_\pi$) lives ``on the boundary'' and the $b$ index (associated with $V_\pi^*$) is ``in the bulk''. A gauge transformation at the central vertex takes the form 
\begin{align}
    A_R(k) \ket{\pi,ab} \ket{\vec{g}\,} = \sum_c \pi(k)_{cb}\ket{\pi,ac} \ket{k\cdot\vec{g}\,} \,.
\end{align}
We can infer this from the fact the $b$ index is contracted into the bulk vertex, while the $a$ index remains free.
This implies that the equivalence class $\ket{\pi,ab ; \vec{g}\,}\rangle$ that defines a vector in the physical Hilbert space always has a fixed representation $\pi$ and index $a$, even after the quotient by null states.

We can then understand the structure of the physical Hilbert space by Fourier transforming the remaining legs of the reduced lattice, so they are all in the representation basis. With the compact notation
\begin{align}
    d\mu(\vec{\pi})& = d\mu(\pi_1) \cdots d\mu(\pi_n) \,,
    \\ V_{\vec{\pi}} & = \bigotimes_{\pi_i \in \vec{\pi}} V_{\pi_i}\,,
\end{align}
this Hilbert space can be decomposed as
\begin{align}
    \Ha_{phys}(\Sigma) = \int_{\widehat{G}}^\oplus d\mu(\vec{\pi}) V_{\vec{\pi}} \otimes \Pi_A[V_{\vec{\pi}}^*]\,. \label{eqn:Hphysboundary}
\end{align}

The states in the $\Pi_A[V_{\vec{\pi}}^*]$ subspace of $\Ha_{phys}(\Sigma)$ are completely determined by gauge invariance, and have a basis of intertwiners (see Appendix \ref{sec:crashcourse} for details). Intertwiners can be thought of as the generalization of Clebsch-Gordan coefficients of $\SU(2)$, and describe how different irreducible representations of $G$ ``fuse'' together to form other representations. Intertwiners are completely fixed by the representation theory of $G$. Therefore, the remaining data required to define a state in $\Ha_{phys}(\Sigma)$ only depends on the specification of the state in $V_\pi$ subspaces. Recall that $V_\pi$ are precisely the degrees of freedom associated with the boundary vertices. Thus, $\Ha_{phys}(\Sigma)$ itself is the boundary Hilbert space, despite its presentation based on the pre-Hilbert spaces $\Ha(\Lambda)$ with support on bulk legs. Interestingly, it is the Wheeler-DeWitt equation and momentum constraint (in Chern-Simons variables) that is directly responsible for for pushing the independent degrees of freedom to the boundary.

Recall that for a fixed choice of representations $\vec{\pi}$, $\Pi_A[V_{\vec{\pi}}^*]$ has $n$ tensor factors, and the resulting state on this subspace is highly entangled with respect to a product basis for $V_{\vec{\pi}}^*$. This entanglement structure is determined by the distinct ways that $V_{\vec{\pi}}^*$ can fuse to the trivial representation of $G$ (see Appendix \ref{sec:crashcourse}).  In particular, these degrees of freedom are not just bipartite entangled: they carry a rich multipartite entanglement structure that they inherit from the intertwiners of $G$. These are the multipartite edge modes of \cite{akers2024multipartite}. This multiparty entanglement structure is inherited by the boundary vertices $V_{\vec{\pi}}$ through the coupling in \eqref{eqn:Hphysboundary}. The precise entanglement structure is crucial for the non-commutativity of the area operators for overlapping boundary subregions. 



There are some obstacles to working with $\Ha_{phys}(\Sigma)$ directly in its boundary presentation. First, the explicit states in $\Pi_A[V_{\vec{\pi}}^*]$
can become difficult to compute for arbitrary compact Lie groups. For non-compact groups, the required representation theory data (the $6j$ symbols and the Clebsch-Gordan coefficients) is not even known in general, though this data exists in principle if $G$ is transformable (in particular, if it is type I). Second, in this presentation it is clear from \eqref{eqn:Hphysboundary} that $\Ha_{phys}(\Sigma)$ does not factorize across boundary vertices because of the integral over representations. In contrast, we expect the complete boundary Hilbert space to factorize when placed on the lattice of boundary vertices, because this is what occurs in AdS/CFT after UV regulating the CFT. Our interpretation of this is that $\Ha_{phys}(\Sigma)$ is a non-factorizing subspace of the complete boundary Hilbert space. Indeed, $\Ha_{phys}(\Sigma)$ only contains states with a fixed spatial topology $\Sigma$. Presumably, the true boundary Hilbert space should at least contain
\begin{align}
    \Ha_{phys} = \bigoplus_{\Sigma} \Ha_{phys}(\Sigma) \,.
\end{align}
where the sum is over tensor networks discretizing different topologies that can fill in the same boundary.  It is possible that such a sum would lead to factorized Hilbert space. Indeed, this similar to the mechanism for factorization of the two boundary Hilbert space into a tensor product of single boundary Hilbert spaces in \cite{Balasubramanian:2025zey,Balasubramanian:2024yxk}. We will explore some features of this factori\~{s}ation puzzle \cite{upcoming}.\footnote{We use the term ``factori\~{s}ation'' to avoid confusion with the terms ``factorisation'' and ``factorization'' which have been used in the recent literature to refer to conceptually distinct puzzles \cite{Boruch:2024kvv,Penington:2023dql}.
}

Now, consider the reduced lattice $\Lambda^{(m)}_{r}$, with $n$ boundary legs and $m$ lollipop factors (out-of-plane legs). $\Ha_{phys}(\Sigma)$ has support on both the boundary legs and the out-of-plane legs. Denote the set of lollipop factors by $\mathfrak{l}$, and the associated Hilbert space as $\Ha(\mathfrak{l})$. Based on the discussion above, define the boundary Hilbert space $\Ha_\partial = \Pi_A \Pi_B \Ha(\Lambda_r)$ as the physical Hilbert space of the reduced lattice with no lollipop factors, so it is the same as in \eqref{eqn:Hphysboundary}. Then given a state $\ket{T} \in \Ha(\mathfrak{l})$ and a state $\ket{\Psi}$ on $\Ha_{phys}(\Sigma) = \Pi_A \Pi_B \Ha(\Lambda_r^{(m)})$, the state
\begin{align}
    \ket{\psi} = (\Id_\partial \otimes \bra{T})\ket{\Psi} 
\end{align}
has support only on the boundary legs, so $\ket{\psi} \in \Ha_\partial$. Thus, the topological tensor network state can still be thought of as a bulk-to-boundary map
\begin{align}
    \ket{\Psi}: \Ha(\mathfrak{l}) \to \Ha_\partial \,.
\end{align}
Note that because of moves 1 and 2, this actually holds regardless of if we chose to define $\Ha_{phys}(\Sigma)$ using the reduced lattice or not. So really, a topological tensor network is a \emph{family} of bulk-to-boundary maps related by the isometries of moves 1,2 which take us from one lattice presentation of the physical Hilbert space to another. 

\subsection{Coset constructions} \label{sec:gravity}

At this point, we have constructed topological tensor networks which prepare states in $G \times \overline{G}$ Chern-Simons theories. When $G = \SL(2,\R)$, these  can be interpreted as states of 3D gravity.  That said, we should be cautious because
 the measure of $\SL(2,\R) \times \SL(2,\R)$ Chern-Simons theories integrates over non-invertible metrics. Thus, we expect some topological tensor networks with $G=\SL(2,\R)$ to have no geometric interpretation. While we do not work out all the details here, in this section, we will propose how one might cure this issue. 

The CFT dual of $\SL(2,\R) \times \SL(2,\R)$ Chern-Simons theory is a $\SL(2,\R)$ Wess-Zumino-Witten (WZW) model that lives on the boundary of the spacetime on which the Chern-Simons theory propagates \cite{Coussaert_1995}. In the large level limit, the Hilbert space of the $\SL(2,\R)$ WZW model on a circle is isomorphic to $L^2(\SL(2,\R))$. Indeed, the $G_k$-WZW model can be thought of as a worldsheet theory of a string propagating in the target space $G$ \cite{GEPNER1986493}, and in the classical limit $k \to \infty$, this reduces to the Hilbert space $L^2(G)$ of a point particle on this group manifold.

On the other hand, the CFT dual of the Virasoro TQFT \cite{Collier_2023}, which is thought to properly account for the measure of 3D gravity in AdS spacetimes, is the chiral Liouville CFT, which can be thought of as a coset construction $\SL(2,\R) / \text{U}(1)$ of the $\SL(2,\R)$ WZW model. The central charge of the Liouville CFT is related to the level of the WZW theory as $c = 6k$ \cite{Brown1986}. Here, we quotient by the diagonal $\text{U}(1)$  $g \to z g z^{-1}$, for $g \in \SL(2,\R)$ and $z \in \text{U}(1)$.
This suggests that  topological tensor networks for the Virasoro TQFT might be constructed by replacing the copy of $L^2(\SL(2,\R))$ at each leg in the pre-Hilbert space with a copy of
\begin{align}
    \Ha_\ell = L^2(\SL(2,\R) / \text{U}(1)) \,.
\end{align}
Actually, this is the Hilbert space of both the left and right movers of the Liouville CFT, so the bulk will be described by two copies of the Virasoro TQFT. In other words, we can think of each copy of the Virasoro TQFT as generalizing a single factor of $\SL(2,\R)$ in the Chern-Simons theory. In \cite{Hartman:2025cyj}, it was argued that this doubled-Virasoro TQFT is related to a Turaev-Viro theory for the conformal group (CTV) by a modular S transformation of the boundary conditions of the bulk fields. A closely related approach was also proposed in \cite{Chen:2024unp}, which utilizes boundary conformal field theory (BCFT) techniques to relate the boundary partition function of Liouville CFT to the bulk path integral of a TFT which reduces to Einstein gravity in the large level limit.\footnote{See \cite{Hung:2019bnq,Chen:2022wvy,Cheng:2023kxh,Hung:2024gma} for more details about this approach.} Depending on the boundary conditions of the bulk fields, we expect either the CTV, BCFT, or Virasoro TQFT approach will be the theory described by the $\SL(2,\R)/\text{U}(1)$ topological tensor network states in the large $k$ limit. 


How does this quotient affect the Plancherel decomposition of $L^2(\SL(2,\R))$? For a fixed representation $V_\pi \otimes V_\pi^*$, we should take the quotient space (viewing $\text{U}(1) \subset \SL(2,\R)$)
\begin{align}
    V_\pi \otimes V_\pi^* &\mapsto \int_{\text{U}(1)} d\theta\, \pi(\theta)  V_\pi \otimes V_\pi^* \pi(\theta)^\dagger 
    \\&= \int_{\text{U}(1)} d\theta\, (\pi(\theta)  V_\pi) \otimes (\pi(\theta)V_\pi)^* \,.
\end{align}
This implies that the only representations which survive the quotient are the ones with zero charge under this $\text{U}(1)$ action. As shown in \cite{langSL2R}, and explained in Appendix \ref{sec:sl2Rexample}, it is precisely the principal series that survives this quotient. In other words, we have the modified Plancherel decomposition
\begin{align}
    L^2(\SL(2,\R)/\text{U}(1)) = \int_{\text{principal}} d\mu(\lambda) V_\lambda \otimes V_\lambda^* \,.
\end{align}
Thus, we conjecture that the pre-Hilbert space for topological tensor networks of the Virasoro TQFT (or CTV), in the large level limit, can be constructed using this coset Hilbert space at each leg, instead of all of $L^2(\SL(2,\R))$. In other words, we should use the same constructions via intertwiners of $\SL(2,\R)$ at each vertex, but restricted to just the principal series. Indeed, in the large level limit, the continuous parameter $\lambda$ of the principal series is proportional to the Liouville momentum $P$ \cite{Jackson_2015}. The Plancherel measure $d\mu(\lambda)$ also matches the Cardy density of states in this limit \cite{Jackson_2015}. Actually, intertwiners have a natural generalization to finite level as well: the chiral vertex operators of Liouville CFT \cite{Jackson_2015}. So we expect that this coset construction can be generalized to finite level. 

In fact, the BCFT methods of \cite{Chen:2024unp} may be relevant for this generalization. However, \cite{Chen:2024unp} are mostly interested in tessellating the asymptotic boundary of three dimensional Euclidean manifolds, while our tensor networks are interpreted as tessellating a Cauchy slice. Additionally, their triangulation scheme constructs the dual graph $\widetilde{\Lambda}$ of a graph $\Lambda$ in our conventions, because their graphs are constructed by gluing legs in parallel, rather than end-to-end. Furthermore, while the electric constraints seem straightforward to implement in the generalization to cosets proposed above, the magnetic constraints are more subtle. Indeed, because the coset $\SL(2,\R)/\text{U}(1)$ is not a group, it is not straightforward  to impose the vanishing flux condition using the above coset construction, even in the large level limit. For finite gauge groups, the magnetic operator can be recast as a sum over central ribbon operators \cite{Buerschaper:2009zlr}, and we expect a suitable generalization to exist for non-compact groups as well.
Because the magnetic constraint is responsible for move 2 (see Appendix \ref{sec:move12}), which allows us to relate graphs $\Lambda$ tessellating the same surface $\Sigma$, we expect this constraint to be equivalent to some of the corresponding moves in CTV: see \cite{Chen:2024unp} and Sec.~3.1 of \cite{Hartman:2025cyj}. It would be interesting to see if this possible connection can be made precise.

\section{Discussion}\label{sec:discussion}

\begin{figure}
    \centering
    \begin{tikzpicture}
        \def\ss{4.75}
        \def\aa{0.05}
        \def\eps{0.002}
        \def\nn{5}
        \def\vv{0.9}

        \node at (0,0) {\begin{tikzpicture}[scale=\ss]
        
        \draw[thick] (0,-\eps) -- (0,1+\eps) (1,-\eps) -- (1,1+\eps);
        \draw[dashed,thick] (0,0) -- (1,1) (1,0) -- (0,1);
        \draw[thick,decorate,decoration={zigzag,segment length=10pt,amplitude=2.5pt}] (0,1) -- (1,1);
        \draw[thick,decorate,decoration={zigzag,segment length=10pt,amplitude=-2.5pt}] (0,0) -- (1,0);

        \draw[thick] (0,0.5) -- (1,0.5);
        \node[anchor=north] at (0.5,-\aa) {$\ket{\Psi}$};

        \foreach \xx in {1,...,\nn}{
        \filldraw ({\xx/(\nn+1)}, 0.5) circle (0.015);
       
        }
        \filldraw[thick,fill=white] (0,0.5) circle (0.015);
        \filldraw[thick,fill=white] (1,0.5) circle (0.015);
        \end{tikzpicture}};

        \node at (1.5*\ss,0) {\begin{tikzpicture}[scale=\ss]
        
        \draw[thick] (0,-\eps) -- (0,1+\eps) (1,-\eps) -- (1,1+\eps);
        \draw[dashed,thick] (0,0) -- (1,1) (1,0) -- (0,1);
        \draw[thick,decorate,decoration={zigzag,segment length=10pt,amplitude=2.5pt}] (0,1) -- (1,1);
        \draw[thick,decorate,decoration={zigzag,segment length=10pt,amplitude=-2.5pt}] (0,0) -- (1,0);

        \node[anchor=north] at (0.5,-\aa) {$A(\vec{g})\ket{\Psi} = \ket{\Psi}$};
    
        \draw[thick] plot[smooth,domain=0:1] (\x,{0.5+0.5*\vv*sin(180*\x)});
        \foreach \xx in {1,...,\nn}{
        \filldraw ({\xx/(\nn+1)}, {0.5+0.5*\vv*sin(180*\xx/(\nn+1))}) circle (0.015);
        }
        \filldraw[thick,fill=white] (0,0.5) circle (0.015);
        \filldraw[thick,fill=white] (1,0.5) circle (0.015);
        \end{tikzpicture}};

        \node[scale=2] at ({1.5*\ss/2},0) {$\Rightarrow $};

    \end{tikzpicture}
    \caption{A topological tensor network $\ket{\Psi}$ for a BTZ black hole (left). After a diffeomorphism $A(\vec{g}) = \bigotimes_{v} A_v(g_v)$, the tensor network $\Lambda$ is moved within the spacetime. Because the state $\ket{\Psi}$ satisfies the constraints, the state is unchanged.
    }
    \label{fig:BTZ}
\end{figure}
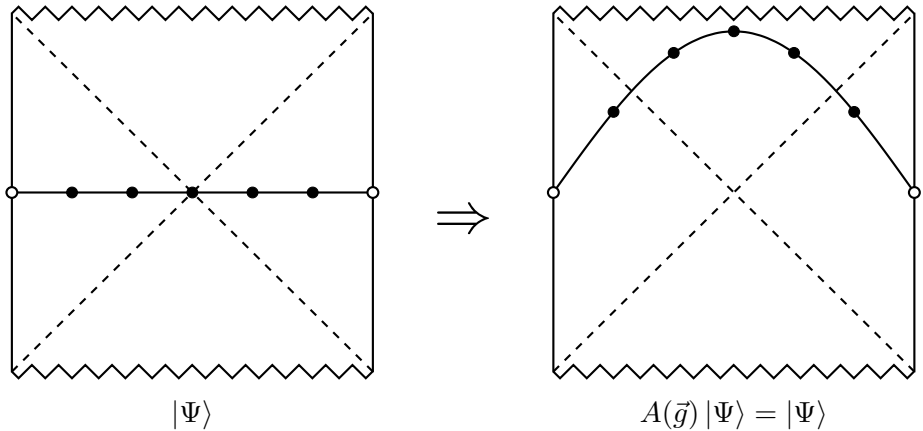

The main result of this paper is to construct semi-classical states of non-chiral Chern-Simons theories with transformable gauge groups. In metric variables when $G = \SL(2,\R)$, these states satisfy the Hamiltonian and momentum constraints of gravity, so they are diffeomorphism invariant. Because of this, if we imagine embedding the graph $\Lambda$ defining the tensor network $\ket{\Psi}$ in a spacetime, a diffeomorphism of the spacetime does not change $\ket{\Psi}$. So if topological tensor networks can be interpreted in this way, then they prepare diffeomorphism invariant states in the bulk. It would be particularly interesting to construct such toplogical tensor  networks representing the BTZ black hole. If we embed the  network on the time-reflection symmetric slice far from the singularity, then using only Wheeler-DeWitt time evolution, the center of the Cauchy slice that $\Lambda$ is embedded in can traverse beyond the horizon, and even come arbitrarily close to the singularity. It would be interesting to construct explicit topological tensor network states which can be interpreted in this way to better understand how the interior of a black hole is represented in the asymptotic state, i.e., in the state at the boundary of the topological tensor network.

In our construction, a copy of $L^2(G)$ is associated with each (oriented) leg $\ell$ of our tensor networks. If we use \eqref{eqn:peterweylnoncompact} to decompose one such copy of $L^2(G)$, we can think of the $V_\pi$ degrees of freedom as living at the outflowing vertex of the leg $\ell$, and the $V_\pi^*$ degrees of freedom as living at the inflowing vertex. A state in the $V_\pi \otimes V_\pi^*$ subspace of $L^2(G)$, then, represents a state that is not entangled between the endpoints of $\ell$. In this case we might as well drop the leg $\ell$ from the network, as there is no correlation between the state at the vertices of $\ell$ that indicates their geometric connection. In contrast, a group-basis eigenstate $\ket{g}$ has support on every $V_\pi \otimes V_\pi^*$, so it represents a state that is highly entangled  between the vertices. From this perspective, we could  think of the legs of the tensor network themselves as being ``generated'' by  entanglement between the degrees of freedom at the vertices. Furthermore, after imposing the gauge constraints, the only independent degrees of freedom have support on the boundary vertices (as explained in Sec.~\ref{sec:Hphys}). Thus, the bulk presentation of the Hilbert space can be thought of as being generated by entanglement of states of the boundary Hilbert space. This is reminiscent of the relationship between entanglement and geometry in holography \cite{VanRaamsdonk2010,Maldacena2013}. It would be interesting to make this connection more precise.

The Hilbert space $\Ha_{phys}(\Sigma)$ comprises states of certain Chern-Simons theories. A Chern-Simons theory, however, has two pieces of defining data: the gauge group $G$, and the level $t=k+i\sigma$. In gravity, we take $k=0$ and $\sigma$ to be inversely proportional to Newton's constant $G_N$ \cite{Witten:1988hc}. In our construction of the states, we did not specify the level of the Chern-Simons theory. The reason is because we have implicitly taken the large level limit $\sigma \to \infty$, or equivalently, $G_N \to 0$. To see this, consider an analogy with $\SU(2)$ Chern-Simons theory at level $k \in \Z$. This theory is defined by its collection of anyons (Wilson lines), which are in turn labeled by the integrable representations of $\SU(2)$ at level $k$. We can think of the collection of representations $\SU(2)_k$ as a ``cut-off'' version of the complete unitary dual $\widehat{\SU}(2)$. In fact, in the large level limit, $\lim_{k \to \infty} \SU(2)_k =  \widehat{\SU}(2)$. Similarly, for non-compact gauge groups like $G = \SL(2,\R)$, the large level limit of $G_\sigma$ is $\lim_{\sigma \to \infty} G_\sigma \to \widehat{G}$ \cite{Maldacena_2001}.\footnote{Really, the large level limit does not lead to all of $\widehat{G}$, but precisely the representations of $\widehat{G}$ in the support of the Plancherel measure.} So we have implicitly been working in the semi-classical limit $G_N \to 0$ by using the complete unitary dual $\widehat{G}$ instead of the quantum group $G_\sigma$. 

This also explains why we were able to construct states with a fixed spatial topology $\Sigma$: in the $G_N \to 0$ limit, states of 3D gravity reduce to their saddlepoints, for which the topology of $\Sigma$ does not fluctuate. In fact, 3D gravity is only equivalent to Chern-Simons theory at the level of the action. The topological field theory which properly accounts for the measure of gravity is two copies of the Virasoro TQFT \cite{Collier_2023}.\footnote{The need for two copies is because a single copy of the Virasoro TQFT is related to $\SL(2,\R)$ Chern-Simons theory, and the gauge group of gravity is $\SL(2,\R) \times \SL(2,\R)$.} We discussed a possible direction for generalization to build topological tensor networks for the Virasoro TQFT in Sec.~\ref{sec:gravity}. If this generalization holds, then we expect our states to agree with the Virasoro TQFT in the semiclassical limit, and that the Virasoro TQFT is the quantum theory which controls the loop corrections (finite level effects) in each chiral half of the bulk theory. 

Topological tensor networks (called string-nets in condensed matter theory) are believed to the the canonically quantized version of the $G_\sigma \times \overline{G_\sigma}$ Turaev-Viro topological quantum field theory \cite{kirillov2011stringnet,kirillov2010corners,Hu:2011pf,Wang2010,Turaev:1994xb,Kadar2009-aw,Buerschaper:2010vwd}, which in our case is the large level limit of a Chern-Simons theory. Therefore, topological tensor networks are \emph{not} simply discrete approximations to semi-classical states of 3D gravity: they are a tool to construct the exact continuum states in the large level $(G_N \to 0)$ limit. Note that recent work by Hartman has shown that the exact path integral of 3D gravity can be computed via triangulation of hyperbolic manifolds using conformal Turaev-Viro theory, a Turaev-Viro theory of the Virasoro group \cite{Hartman:2025ula,Hartman:2025cyj}.\footnote{This is not quite the same as the (doubled) Virasoro TQFT, but is related by S-duality of the boundary conditions defining the path integrals.} Thus, if the relationship between string-nets and Turaev-Viro theories continues to hold, then our work can be interpreted as a canonical quantization of these exact path integrals.\footnote{A similar model was also proposed in \cite{Chen:2024unp}, but with a focus on reproducing Liouville CFT exactly, rather than matching the exact partition function of 3D gravity in the bulk. We expect these models to be closely related.}
It would be very interesting if, in some sense, the triangulation of these hyperbolic manifolds could be restricted to a Cauchy slice $\Sigma$ and produce topological tensor networks like the ones we studied in a direct way after canonical quantization. 

Furthermore, in this paper, we considered topological tensor networks associated with doubled Chern-Simons theories $G_k \times \overline{G_k}$, and focused on $\SL(2,\R) \times \SL(2,\R)$. It would be interesting to construct similar models for the other gauge groups relevant for gravity, such as $\SL(2,\C)$, $\text{ISO}(1,2)$, and $\text{ISO}(3)$. The obstruction is that these groups do not immediately factorize into chiral halves, so their relationship with string-nets is not obvious. Perhaps some coset construction, analogous to that outlined in Sec.~\ref{sec:gravity}, will allow for the appropriate generalization. 

We did not consider how non-perturbative effects like topology change affect the constraint equations $\Pi_A, \Pi_B$. Thus, the topological tensor networks we consider should be thought of as describing spacetimes $M$ without dynamical wormholes: for all times $t$ (defined by the coordinate conjugate to the Wheeler-DeWitt Hamiltonian), the spatial topology of $\Sigma$ is constant. It would be interesting to understand how summing over spacetime topologies in the path integral affects the structure of the physical Hilbert space. To gain intuition, we could imagine taking Chern-Simons theory with compact gauge group and summing over bulk topologies as a toy model. If we work in Euclidean signature so that $\partial M = \partial \Sigma \times S^1$, then this restricts the spacetime manifolds $M$ to have topology $\Sigma \times_f S_t^1$, where $f: \Sigma \to \Sigma$ is a diffeomorphism, possibly large, which twists $\Sigma$ around the Euclidean time circle $S_t^1$. In Chern-Simons theory, the path integral on such manifolds prepares a fibered link state \cite{Balasubramanian:2025kaf}. In \cite{Cummings:2025zfe}, it was shown how to explicitly sum over the bulk topology of fibered link states (with a fixed number of boundary components) in Chern-Simons theories with compact gauge groups. In this work, it was demonstrated that the corresponding boundary states have a rich multipartite entanglement structure between the asymptotic boundary Hilbert spaces, which parallels the structure we found in Sec.~\ref{sec:Hphys}, following  \cite{akers2024multipartite}. In the case of topological tensor networks, this multipartite entanglement structure was a direct consequence of the local constraint equations from gravity. In \cite{Cummings:2025zfe}, this entanglement structure arose directly from the sum over bulk topologies.

It is important to note that the sources of these multipartite entanglement patterns is not quite the same. In topological tensor networks the multipartite entanglement structure is related to the fusion rules of the gauge group $G$. In contrast, in \cite{Cummings:2025zfe}, the multipartite entanglement is related to fusion rules of the mapping class group of $M$ (see \cite{Collier_2023,Balasubramanian:2025kaf,Cummings:2025zfe} for more details). However, we note that the gauge group $G$ (e.g., $\SL(2,\R) \times \SL(2,\R)$) of the Virasoro TQFT/Chern-Simons is essentially the connected component of the diffeomorphism group in 3D, while the mapping class group captures the global structure of the diffeomorphism group of the spacetime. So, perhaps, these multipartite entanglement structures could be combined appropriately to generate an entanglement structure corresponding to the entire diffeomorphism group of the spacetime. We leave this for future work.

\paragraph{Acknowledgments:} We thank Wayne Weng, Chris Akers, Jon Sorce, Elba Alonso-Monsalve, Leo Shaposhnik, Alexander Jahn, Wissam Chemissany for helpful discussions.
CC was supported by the National Science Foundation Graduate Research Fellowship under Grant No. DGE-2236662.   VB was supported in part by the DOE through DE-SC0013528 and QuantISED grant DE-SC0020360, and in part by the Eastman Professorship at Balliol College, University of Oxford. 

\appendix

\section{A crash course in non-Abelian harmonic analysis} \label{sec:crashcourse}

In this appendix, we will  review aspects of non-Abelian harmonic analysis which we will use in this paper.  \cite{Dixmier1977,Haar,knapp2002} for a more complete treatment. We  restrict the discussion to semi-simple groups, so our treatment will not apply to $\text{ISO}(1,2)$ or $\text{ISO}(3)$. However, after understanding the semi-simple case, representations of the latter are well understood using Mackey's machine \cite{Varadarajan2007}, which is  the generalization of the use of little groups to understand representations of the Poincar\'e group.

\subsection{Compact groups}

Consider a free particle on a circle. The states of this system live in the Hilbert space of square integrable functions on $S^1$, which the group manifold of $\text{U}(1)$.  This Hilbert space, $L^2(\text{U}(1))$, has two obvious bases: the position basis $\{\ket{\theta} \, | \, \theta \in [0,2\pi)\}$, and the momentum basis $\{\ket{n} \, | \, n \in \Z \}$, which are related by the Fourier transform $\braket{\theta}{n} = e^{2\pi i n \theta}$. If we denote the momentum space of $\text{U}(1)$ by $\widehat{U}(1) = \Z$ and
\begin{align}
    E_n = \text{span}\{\ket{n}\}\,,
\end{align}
then we have the decomposition
\begin{align}
    L^2(\text{U}(1)) = \bigoplus_{n \in \widehat{U}(1)} E_n \,.
\end{align}
Here, the Fourier transform is a unitary map from a function $f(\theta)$ of elements of $\text{U}(1)$ and another function $\widehat{f}(n)$ of $\widehat{U}(1)$. As we will see below, it turns out that there is an analogous correspondence between functions on transformable groups $G$ and functions on a different topological space $\widehat{G}$, which we will call the unitary dual of $G$. $\widehat{G}$ can be thought of as the ``momentum space'' of $G$. 

Now consider $\SU(2)$, the simplest example of a non-Abelian Lie group. This case is representative of the general case of compact groups, and is qualitatively similar to the case of $\text{U}(1)$. The major differences that arise are because $\SU(2)$ is non-Abelian, while $\text{U}(1)$ is Abelian.
For $\SU(2)$, the irreducible unitary representations are labeled by $\widehat{\SU}(2) = \frac{1}{2}\N$, where $j \in \frac{1}{2}\N$ labels the spin quantum number of a particular representation. The spin-$j$ representation of $\SU(2)$ is defined on the vector space $V_j = \C^{2j+1}$, which has dimension $d_j = 2j+1$. The group action of $\SU(2)$ on $V_j$ is defined by the Wigner D-matrix elements $D^j_{mn}(g)$, where $g \in \SU(2)$. Notice that for a fixed pair of indices $m,n$, each Wigner $D$-matrix element defines a function $D^j_{mn}: \SU(2) \to \C$. Because $\SU(2)$ is compact, this implies that we can define a vector of $L^2(\SU(2))$ by the wave function
\begin{align}
    \braket{g}{j,mn} = D^j_{mn}(g) \,.
\end{align}
Define the space
\begin{align}
    E_j = \text{span}\{\ket{j,mn} \,|\, m,n \in 1,\cdots,2j+1 \}\,.
\end{align}
which is a vector subspace of $L^2(\SU(2))$. Notice that $E_j = V_j \otimes V_j^*$, where the $m$ index labels a basis of $V_j$, and the $n$ index labels a basis of $V_j^*$. Furthermore, these matrix elements satisfy the orthogonality relation
\begin{align}
    \braket{j,mn}{\ell,pq} = \int_{\SU(2)} dg \,(D^j_{mn} (g))^* D^\ell_{pq}(g) = \left(\frac{\text{Vol}(\SU(2))}{d_j} \delta_{j\ell} \right)\delta_{mp} \delta_{nq} \,, \label{eqn:SU2orthog}
\end{align}
where $\text{Vol}(\SU(2)) = \int 1 \,dg$ is the group volume of $SU(2)$ in the Haar measure, and $d_j = \dim(V_j)$. We can prove this using Schur's lemma. Because $D^\ell(g)$ is a unitary representation of $G$,  $D^\ell_{mn}(g)^* = D^\ell_{nm}(g^{-1}) $. Therefore,
\begin{align}
    \braket{j,mn}{\ell,pq} &= \int_{\SU(2)} dg \, \bra{n}D^j (g^{-1}) \ketbra{m}{p} D^\ell(g) \ket{q}
    \\&= \, \bra{n} \left(\int_{\SU(2)} dg D^j (g^{-1}) \ketbra{m}{p} D^\ell(g)\right) \ket{q}
    \\&= \, \bra{n} O^{j\ell}_{mp}\ket{q} \,, \label{eqn:SU2orthogmiddle}
\end{align}
for some operator $O^{j\ell}_{mp}$. This operator is not arbitrary: using the right invariance of $dg$, we can show that for any $g \in \SU(2)$,
\begin{align}
   D^j(g) O^{jl}_{mp} = O^{jl}_{mp} D^\ell(g) \,.
\end{align}
When $j = \ell$, this says that $O^{jl}_{mp}$ commutes with the action of $\SU(2)$: it must be proportional to the identity. When $j \neq \ell$, no such operator exists because $V_j,V_\ell$ are irreducible, so $O^{jl}_{mp} = 0$ in that case. Together, this implies that
\begin{align}
    O^{j\ell}_{mp} = \delta_{j\ell} c^j_{mp} \frac{\Id_{V_j}}{d_j} 
\end{align}
for some constant $c^j_{mp}$, where $\Id$ is the identity operator. This constant can be determined by taking the trace
\begin{align}
    c^j_{mp} = \tr(O^{jj}_{mp}) = \int_{\SU(2)} dg \,\bra{p} D^j(g) D^j (g^{-1}) \ket{m} = \text{Vol}(\SU(2)) \braket{p}{m} \,.
\end{align}
Plugging this into \eqref{eqn:SU2orthogmiddle}, we obtain \eqref{eqn:SU2orthog}.

Equation \eqref{eqn:SU2orthog} shows that the representation basis $\ket{j,mn}$ is \emph{almost} orthonormal. We can make it orthonormal in two ways.  The first is to redefine  $\ket{j,mn} \to \left(\frac{d_j}{\text{Vol}(\SU(2))}\right)^{-1/2} \ket{j,mn}$ so that the basis is orthonormal. The second  is to define a measure on $\widehat{\SU}(2)$ by
\begin{align}
    \mu(j) = \frac{d_j}{\text{Vol}(\SU(2))}
\end{align}
and rescale the inner product on each $E_j$ by
\begin{align}
     \braket{j,mn}{j,pq}_{\mu(j) \cdot E_j} := \mu(j) \cdot \braket{j,mn}{j,pq}_{E_j} \,.
\end{align}
With this definition  \eqref{eqn:SU2orthog} becomes
\begin{align}
    \braket{j,mn}{\ell,pq} = \delta(j,\ell) \delta_{mp} \delta_{nq} \,,
\end{align}
where $\delta(j,\ell)$ is the delta function with respect to $\mu(j)$, viewed as a measure on $\widehat{\SU}(2)$. In other words, the basis $\ket{j,mn}$ \emph{is} orthonormal, but with respect to the $\mu(j)$-weighted inner product on $\widehat{G}$. The Peter-Weyl theorem then states that
\begin{align}
    L^2(\SU(2)) = \bigoplus_{j \in \widehat{SU}(2)} \mu(j) \cdot E_j \,. \label{eqn:SU2pw}
\end{align}
The fact that $ L^2(\SU(2)) \supset \bigoplus_{j \in \widehat{SU}(2)} \mu(j) \cdot E_j$ is not surprising, as we already explained that each $E_j$ is a subspace of $L^2(\SU(2))$. The reverse inclusion $\subset$ is initially surprising: it says that \emph{any} square integrable function of $\SU(2)$ can be expanded as a linear combination of Wigner D-matrix elements. But even this is familiar: we can think of the Wigner D-matrix elements as the ``spherical harmonics'' for $S^3$, the group manifold of $\SU(2)$), so the reverse inclusion is the statement that these generalized spherical harmonics span $L^2(\SU(2))$.

The reason we have chosen to state the Peter-Weyl theorem using the measure $\mu(j)$, called the \emph{Plancherel measure} of $\SU(2)$, is that it makes \eqref{eqn:SU2pw} an equality of Hilbert spaces, not just vector spaces, where the vectors $\ket{j,mn}$ are orthonormal in the modified inner product. In other words, this definition of the inner product makes the Fourier transform from the group basis $\ket{g}$ to the representation basis $\ket{j,mn}$ unitary.

For compact groups $G$, the same basic story always holds. There is always a discrete space of points $\pi \in \widehat{G}$ which labels the irreducible unitary representations of $G$. The Plancherel measure $\mu(\pi) = \frac{\dim(V_\pi)}{\text{Vol}(G)}$ is proportional to the dimension of the vector space the representation $\pi$ acts on. There is a subspace $E_\pi$ of $L^2(G)$ spanned by the matrix elements
\begin{align}
    \braket{g}{\pi,ij} = \pi_{ij}(g)
\end{align}
of the unitary representation $\pi$, which have an overlap
\begin{align}
    \braket{\pi,ij}{\omega,mn} = \delta(\pi,\omega) \delta_{im}\delta_{nj}\,.
\end{align}
Just like the $\SU(2)$ case, $\delta(\pi,\omega)$ is the delta distribution on $\widehat{G}$ with respect to the Plancherel measure. Finally, the Peter-Weyl theorem says that
\begin{align}
    L^2(G) = \bigoplus_{\pi \in \widehat{G}} \mu(\pi) \cdot E_\pi 
\end{align}
is a unitary equivalence of vector spaces.

\subsection{Representation theory of non-compact groups}

Now suppose $G$ is a non-compact transformable group (see Sec.~\ref{sec:transformable} for the definition of a transformable group). Based on the analogy with compact groups, to determine the Plancherel decomposition of $L^2(G)$, we must first construct its unitary dual $\widehat{G}$. Then, we must determine the Plancherel measure $d\mu(\pi)$. For a general group, neither of these tasks is straightforward, and in some cases the results not known. However, it has been shown that both $\widehat{G}$ and $d\mu(\pi)$ exist if $G$ is transformable \cite{Dixmier1977,Haar}. 

\subsubsection{The $KAN$ decomposition of $G$}

We will focus for concreteness on $G=\SL(2,\R)$, but the same basic picture holds for arbitrary real semi-simple Lie groups. Our discussion largely follows \cite{langSL2R,knapp2002}. $\SL(2,\R)$ is a real, three dimensional group which can be given coordinates $x_k\in [0,2\pi)$, $x_a\in \R^+$ and $x_n \in \R^{>0}$, such that
\begin{align}
    g(\theta,x_a,x_n) = \begin{pmatrix}
        \cos(x_k) & \sin(x_k) \\ -\sin(x_k) & \cos(x_k)
    \end{pmatrix}\begin{pmatrix}
        x_a & 0 \\ 0 & x_a^{-1}
    \end{pmatrix}
    \begin{pmatrix}
        1 & x_n \\ 0 & 1
    \end{pmatrix} 
     \,.
     \label{eq:Iwasawa}
\end{align}
These coordinates are called the Iwasawa decomposition of $\SL(2,\R)$. It is convenient to parameterize the group element $g$ not by the numerical coordinates $(x_k,x_a,x_n)$, but by the matrices that these coordinates parameterize. Labeling these matrices $k,a,n$, respectively, we can think of these matrices as living in subgroups $K,A,N$ of $G$. For this reason, this splitting of $G$ is also sometimes the $KAN$ decomposition. Viewing $g$ as a matrix, this is just a $QR$ decomposition of $g$ into an orthonormal matrix $Q \equiv k$ and an upper triangular matrix $R \equiv an$, which we have further decomposed by splitting $R$ into a diagonal matrix $a$ and an upper triangular matrix $n$ with all $1$'s on the diagonal.  
These coordinates are well-defined because the QR decomposition of a matrix is unique, so every tuple $(k,a,n)$ determines a unique element $g(k,a,n)$.  The decomposition in \eqref{eq:Iwasawa}  is \emph{not} a group homomorphism between $K \times A \times N \to G$. Group multiplication of $K \times A \times N$ does not have a simple functional form in terms of group multiplication on $G$ because the factors $K,A,N$ do not commute.  Nevertheless, this splitting of $\SL(2,\R)$ is essential to understanding its representation theory. Furthermore, the Haar measure of $\SL(2,\R)$ is the product of the Haar measures
of these groups are simply related:
\begin{align}
    dg = dk \frac{da}{a} dn \,. \label{eqn:KANHaar}
\end{align}
By understanding each factor $K,A,N$ of $G$ separately and combining the results, we will be able to understand the entire group.

For a more general real, semi-simple Lie group, a similar decomposition still holds. In that case, $K$ is the maximal compact subgroup of $G$, perhaps $\SO(N)$, $\SU(N)$, or $\text{Sp}(N)$ for some $N$. Then, we can view the left coset space $G/K$ as a manifold equipped with a natural action of $G$: in particular, it is a symmetric space of $G$. Generally, $G/K$ is not a group because $K$ is not a normal subgroup, but it still has a geometric structure we can exploit. Let $x_0 \in G/K$ be the coset of $K$ itself, which will act as an ``origin'' for a natural coordinate system on $G/K$. Because $G$ is non-compact and $K$ is compact, $G/K$ will have some non-compact directions which we can think of as ``radial'' directions in $G/K$. 

Next, we consider the family of geodesics which travel from our base point $x_0$ to infinity. They are geodesics with respect to the metric of $G/K$ that is inherited from any left-invariant metric on $G$. Then, we define $A$ as the maximal abelian subgroup of $G$ which both acts as a pure scaling transformation on these geodesics and fixes the origin $x_0$. We can think of these radial geodesics as being the orbits of $A$ on $G/K$. The double quotient space $(G/K)/A$, then, can be thought of as the geometric space parameterizing the set of ``angular'' directions of $G/K$. $N$ acts transitively on $(G/K)/A$, and in fact, is defined to be the smallest subgroup of $G$ which does so. Essentially, $N$ being minimal means $N \cap A$ is trivial.

Putting the pieces together, we can work backwards to define a coordinate system for $G$ which generates the $KAN$ decomposition of a general semi-simple Lie group. Because $N$ is transitive on $(G/K)/A$ and is minimal, we can use an element of $N$ to label a particular radial geodesic $\gamma$ of $G/K$. Then, we can use an element of $A$ to label a particular point on this geodesic, so the pair $(a,n)$ labels an equivalence class of $G$ under the left action of $K$. Finally, an element of $K$ determines a particular representative $g(k,a,n)$ in this equivalence class. This decomposition is unique, so the tuple $(k,a,n)$  is a global coordinate system for $G$. In particular, the map
\begin{align}
    g: K \times A \times N &\to G \\
    (k,a,n) &\mapsto g(k,a,n)
\end{align}
is a smooth diffeomorphism, but it is not a group homomorphism. So while $K \times A \times N$ is not the same group as $G$, it is the same as $G$ when viewed as a manifold. Furthermore, the Haar measure in these coordinates splits into the product of Haar measures on $K,A,N$, which generalizes \eqref{eqn:KANHaar} to arbitrary non-compact groups.

As an example, consider $\SL(2,\R)$. In this case, $K=\SO(2)$, and $\SL(2,\R) / \SO(2)$ is the hyperbolic plane. Think of this space as the upper half plane with coordinates $z= x+iy$  and $y > 0$, a metric $ds^2 = y^{-2}(dx^2 + dy^2)$, and a measure $d^2z = y^{-1} dxdy$. The origin $x_0$ is the point $z=0$, $A$ is the group of dilatations $z \to a\cdot z$ which preserves the hyperbolic metric, and $N$ is the subgroup of translations $x \to x+n$.

\subsubsection{Cartan subgroups}

Now that we have decomposed $G = KAN$, we will use this splitting to understand the representation theory of non-compact groups. But first, we will review the representation theory of compact groups. Assume for the moment that $G$ is compact, so $G=K$ and $A = N = \{e\}$ where $e$ is the identity. Then, let $T_K$ be the maximal abelian subgroup of $K$, often called the maximal torus or the Cartan subgroup of $K$. The possible choices of maximal torus are related by conjugation, so they are isomorphic. For example, the maximal torus of $\SU(2)$ is $\text{U}(1)$, which can be thought of as the subgroup of rotations around a particular axis, say $\hat{z}$. This choice of axis is necessary but arbitrary, because any two axes are related by a rotation. So for compact groups, we will often speak of ``the'' maximal torus, when we really mean ``any'' maximal torus.

For a compact group, the Cartan subgroup $T_K$ plays an essential role in determining the unitary representations.
The reason is that when $G$ is compact, every element $g \in G$ is conjugate to an element of the maximal torus: for all $g \in G$, there exists an $h \in G$ and a $t \in T$ such that $g = h t h^{-1}$. So, by cyclicity of the trace, the character function $\chi_\pi(g) = \tr[\pi(g)]$ is uniquely determined by its values on the Cartan subgroup $T_K$. This is useful is because the character function $\chi_\pi(g)$ uniquely determines the entire representation $V_\pi$. To see this, suppose we knew the function $\chi_\pi(g)$ for any group element $g$. If we view the matrix $\pi(g)$ as a vector $ \ket{\pi(g)} \in E_\pi$ (the set of matrices acting on $V_\pi$, with the usual inner product $\tr[A^\dagger B]$), then we can think of
\begin{align}
    \chi_\pi(h^{-1} g) = \tr[\pi(h^{-1}g)]=\tr[\pi(h)^\dagger \pi(g)] = \braket{\pi(h)}{\pi(g)}_{E_\pi}\,,
\end{align}
with respect to the usual inner product for matrices on $E_\pi$ . Therefore, for fixed $g$, knowledge of the character function for arbitrary $h$ tells us the exact vector $\ket{\pi(g)} \in E_\pi$ because we know its overlap with an arbitrary vector in $E_\pi$. In other words, we know the matrix elements of the entire representation just by specifying the character function $\chi_\pi(g)$.

As we said above, the characters  of a compact group are determined by their value on the maximal torus $T_K$. Actually, the maximal torus $T_K$ has some redundancies: there are sometimes elements $t,t' \in T_K$ and a group element $g \in G$ such that $t' = g t g^{-1}$. To characterize this redundancy, we can define the Weyl group\footnote{$N(T_K)$ is called the normalizer of $T_K$ in $G$, and is not related to the $N$ of the $KAN$ decomposition of $G$.}
\begin{align}
    W &= N(T_K) / T_K \,,
    \\ N(T_K) &= \{g \in G | g t g^{-1} \in T_K \text{ for all } t \in T_K\} \,.
\end{align}
The Weyl group characterizes which elements of $T_K$ are conjugate to each other.\footnote{The Weyl group has an alternative definition as the group of reflections of the root system of the Lie algebra $\mathfrak{g}$ of $G$. These definitions are equivalent.} Therefore, the conjugacy classes of $G$ are actually characterized by the quotient $T = T_K / W$. Thus, the pair $(T_K,W)$ completely determines the (unitary) representation theory of $G$ when $G$ is compact.

When $G$ is non-compact, something similar is true, and will be useful for understanding the representation theory.  In this case there are multiple maximal tori which are not conjugate to each other;  so we must consider the collection of  maximal tori up to conjugation in $G$. For example, let $M_A$ be the centralizer of $A$ in $K$. In other words, $M_A$ is the maximal Abelian subgroup of $K$ which commutes with every element of $A$. Then $T_A = M_A A$ is  a maximal abelian subgroup of $G$: by definition of $A$, no element of $N$ commutes with $T_A$, and by definition of $M_A$, no element of $ K / M_A$ commutes with $T_A$. $T_A$ is called the maximally split torus of $G$. $T_K \equiv T_{\{e\}}$ is called the maximally compact torus of $G$. More generally, there are other maximal tori $T_B$ which are labeled by subgroups $B \subset A$, along with the centralizers $M_B \subset K$. A choice of $ T_B = M_B B$ is called a Cartan subgroup of $G$, and $\dim(B)$ is called the split rank of $T_B$, which measures the number of non-compact directions in $T_B$. As manifolds, all Cartan subgroups $T_B$ have the same dimension, regardless of a choice of $B$ \cite{knapp2002}.\footnote{One way to see this is by noting that all Cartan subgroups $T_B$ agree with each other if we complexify the group $G \to G_\C$, essentially because $e^x \in \R^+ \subset B$ and $e^{i\theta} \in \text{U}(1) \subset M_B$ both complexify to $e^{z} \in \C^\times$.} The dimension $\dim(T_B) = \dim(B) + \dim(M_B)$ of any Cartan subgroup is called the rank of $G$, and is equal to the number of nodes of the Dynkin diagram of $G$. 

We are almost ready to discuss the conjugacy classes of non-compact groups, and therefore the character functions that determine the representations of $G$. First, however, we must note a technical point. An element $g'\in G$ is said to be \emph{regular} if the centralizer of $g$  (the elements of $G$ that commute with $g$) has the smallest possible dimension (the rank of $G$) \cite{langSL2R,Haar}.\footnote{The centralizer of an element $g$ is the set of elements which commute with $g$.}  This is a statement about $g'$ being sufficiently ``generic''. If $G = \GL(n,\R)$, then the regular elements are the matrices with distinct eigenvalues, because the centralizer of a fixed matrix with repeated eigenvalues includes rotations of the repeated eigenspaces, and these additional rotations would not fix a generic matrix. The set of regular elements is denoted $G'$, and is dense in $G$ (just as in the case $G=\GL(n,\R)$). 

Every regular element of $G'$ is conjugate to an element in some Cartan subgroup $T_B$ of $G$. Further Harish-Chandra showed \cite{HarishChandra1976} that we can  define the character of the representation on the dense subset $G' \subset G$, and extend it to all of $G$ by a completion. 
Thus, up to details involving the Weyl groups $W(T_B)$ of $G$, we can understand the representation theory of $G$ by first defining the character functions on all of the maximal Cartan subgroups $T_B$, and then extend these character functions to all of $G$. Each Cartan subgroup has its own Weyl group $W(T_B)$ which parameterizes the redundancy of conjugacy classes in $T_B$. A full understanding of the representation theory of $G$, then, requires constructing every possible Cartan subgroup $T_B$, the associated Weyl group $W(T_B) = N(T_B) / T_B$, as well as how the different pairs $(T_B, W(T_B))$ are related to each other (i.e., what pairs $(T_B, W(T_B))$ are related by conjugation in $G$). In general, this can be a difficult task, but in principle these are the subsets of $G$ which determine its conjugacy classes, and therefore its unitary representation theory.

\subsubsection{Character distributions} \label{sec:sl2Rexample}

Now that we understand the set of conjugacy classes of $G$, we are ready to discuss the character of a representation. The definition of the character function $\chi_\pi(g)$ is more subtle when $G$ is non-compact.  For example, $\tr[\pi(e)] = \dim(V_\pi) = \infty$, so we cannot define the character directly in terms of the trace of representations.\footnote{Notice that the identity is not a regular element of $G$. This is not a coincidence: the character function $\chi_\pi(g)$ we define below is generally singular on $G \setminus G'$.} Nevertheless, we can still define the character function as follows.
Let $f(g)$ be a compactly supported test function. Given a unitary irreducible representation $\pi \in \widehat{G}$, we  define the operator
\begin{align}
    \pi(f) = \int dg \, f(g) \pi(g^{-1}) \,. \label{eqn:fouriertransform}
\end{align}
We can think of the Fourier transform of $f$ as the operator-valued map on $\widehat{G}$ which sends
\begin{align}
    \widehat{f}: \pi \mapsto \pi(f) \,. \label{eqn:fouriertransform2}
\end{align}
Indeed, this is one way to think about the Fourier transform of $\R$: it is the map $\widehat{f}$ which sends a momentum $k$ to the 1 $\times$ 1 matrix $\widehat{f}(k)$, or $k(f)$ in the above notation. The inverse in $\pi(g^{-1})$ is there to match the  conventions for the minus sign in the exponential of the Fourier transform of $\R$. This is the definition of the Fourier transform which generalizes to  arbitrary transformable groups. Sometimes, we will just refer to $\pi(f)$ as the Fourier transform of $f$.

The operator $\pi(f)$ is a trace class \cite{Haar} (this is where the assumption that $G$ is type I comes in), and so we can define the distribution
\begin{align}
    \chi_\pi(f) = \tr[\pi(f)] \,.
\end{align}
It turns out that there exists a locally integrable function $\chi_\pi(g)$ such that this distribution can be calculated by the integral
\begin{align}
    \chi_\pi(f) = \int_G dg f(g) \chi_\pi(g^{-1}) \,. \label{eqn:charactercompact}
\end{align}
This holds for any compactly supported $f(g)$ \cite{Haar}. Because compactly supported functions are dense in $L^2(G)$, \eqref{eqn:charactercompact} can be extended to hold for all $L^2$ functions as well. This function $\chi_\pi(g)$ is called the global character of the irrep $\pi$, and is an $L^1$ function. When $G$ is compact, we can freely think of $\chi_\pi(g) = \tr[\pi(g)]$, as suggested by commuting the trace and the integral. When $G$ is non-compact, the reason we cannot swap the integral and the trace to reach the same conclusion is because of conditional convergence issues which do not let us swap these sums. Despite this subtlety, the locally integrable function $\chi_\pi(g)$ can be thought of as a renormalized version of the naive definition $\tr[\pi(g)]$. That this function always exists when $G$ is a transformable group is a deep theorem due to Harish-Chandra \cite{HarishChandra1952,HarishChandra1954complex,HarishChandra1976}.

The characters $\chi_\pi(g)$ are class functions, which means that they are constant on conjugacy classes of $g$: $\chi_\pi(g) = \chi_\pi(hgh^{-1})$ for any $h \in G$. Thus, for regular elements $g'$, we can restrict the character functions to the collection of Cartan subgroups of $G$ without any loss of generality. Harish-Chandra's theorem says that this is sufficient to determine the entire representation.
When $G$ is compact, the maximal torus $T_K$ can be thought of as containing $\text{rank}(G)$ copies of $\text{U}(1)$, so the representations will be labeled by $\text{rank}(G)$ integers. On the other hand, the group $\R^r$ will have representations labeled by a continuous family of $r$ real numbers. More generally, Cartan subgroups $T_B = M_B B$ will contain some compact directions $M_B$ and some non-compact directions $B$, and so we should expect the representations of $G$ to depend on both continuous and discrete parameters. But the precise form that these parameters take depends on the group, as well as on the details of the Weyl group $W(T_B)$. 

We then define the unitary dual $\widehat{G}$ of $G$ to be the set of parameters $\pi$ which label the distinct, irreducible, unitary representations of $G$. Alternatively, the points $\pi \in \widehat{G}$ can be thought of as labeling character distributions $\chi_\pi$ themselves. Finally, we note that $\widehat{G}$ inherits a natural topology from character distributions. 
A sequence of representations $\pi_n$ is said to converge to another representation $\pi$ if their character distributions converge for any compactly supported $f$:
\begin{align}
    \lim_{n \to \infty} \chi_{\pi_n}(f) = \chi_{\pi}(f) \,.
\end{align}
Because $f$ was  arbitrary, this is the same as demanding that the matrix elements of the $\pi_n$ representation converge to the matrix elements of the $\pi$ representation. More intuitively, two representations $\pi,\omega$ are ``close'' in $\widehat{G}$ if all of their matrix elements are close. The topology on $\widehat{G}$ that is induced by this definition of convergence is called the Fell topology \cite{Dixmier1977}. It is the topology which makes $\widehat{G}$ a discrete series of points when $G$ is compact, and $\widehat{\SL}(2,\R)$ contain discrete and continuous families of representations, rather than a union of uncountably many disjoint points.

We conclude this section by writing down the characters of $\SL(2,\R)$ which appear in the Plancherel formula for $L^2(\SL(2,\R))$. There are two families of tempered representations of $\SL(2,\R)$. The first is the \emph{discrete series} $D_n$, which are labeled by an integer $n \neq 0$. The second family is the \emph{principal series} $P^\pm_{\lambda}$, which is labeled by a continuous parameter $\lambda > 0$ and a sign $\pm$.

\paragraph{The discrete series $D_n$: }

For an element of the maximal compact torus $T_K = \text{U}(1)$, and letting $k_\theta \in T_K$ denote the rotation matrix by an angle $\theta$, the character of the discrete series is
\begin{align}
    \chi_{n}(k_\theta) = -\text{sign}(n) \frac{e^{in\theta} }{e^{i\theta} - e^{-i\theta}}\,.
\end{align}
We can interpret this as saying that the principal series has a non-zero charge under the action of $\text{U}(1) \subset \SL(2,\R)$. This is important in Sec.~\ref{sec:gravity}.

For an element of the split maximal Cartan subgroup $T_A = M_A A = \pm \R^+$, and an element
\begin{align}
    \pm a_t = \begin{pmatrix}
        \pm e^t & 0 \\ 0 & \pm e^{-t}
    \end{pmatrix}\,,
\end{align}
the character of the discrete series is
\begin{align}
    \chi_n(\pm a_t) = (-1)^{1+|n|} \frac{e^{-|nt|}}{e^{t} - e^{-t}}\,.
\end{align}

\paragraph{The principal series $P_\lambda^\pm$: }

The character of the principal series is
\begin{align}
    \chi_{\lambda,+}(\pm a_t) &= \frac{e^{\lambda t} + e^{-\lambda t}}{|e^{t} - e^{-t}|}\,, \\
    \chi_{\lambda,-}(\pm a_t) &= \pm \frac{e^{\lambda t} + e^{-\lambda t}}{|e^{t} - e^{-t}|}\,.
\end{align}
$\chi_{\lambda,+}$ vanishes on the maximal compact torus of $\SL(2,\R)$. We can interpret this as saying the representations in the principal series have no $\text{U}(1)$ charge with respect to any $\text{U}(1)$ subgroup of $\SL(2,\R)$. This is important in Sec.~\ref{sec:gravity}.

\subsection{The Plancherel Formula}

Above, we defined the Fourier transform of a function $f \in L^2(G)$ as the transformation $f \mapsto \widehat{f}$, where $\widehat{f}$ is defined in \eqref{eqn:fouriertransform2}. We also need to define an inverse Fourier transform $\widehat{f} \mapsto f$, which requires us to integrate over $\widehat{G}$. To integrate over $\widehat{G}$ we need a {\it Radon measure} that is compatible  with the Fell topology. The requirements on this measure are that we should be able to: (1) take approximate integrals over $\widehat{G}$ using compact subsets (inner regularity), and (2) extend local results to the full space (outer regularity).
The necessity of a Radon measure stems from practical requirements: we need to be able to approximate integrals over $\widehat{G}$ using compact subsets (inner regularity) and extend local results to the full space (outer regularity). These properties ensure that physical observables computed via integration on $\widehat{G}$ are stable under approximation and that infinite-dimensional spaces of representations can be handled systematically.  These regularity properties also allow us to exchange limits and integrals, which will be essential for applications like computing traces, matrix elements, and spectral decompositions. 

While Radon measures are well-behaved, they are not unique. For example, if $dk$ is the Lebesgue measure for $\R$ and $f(k)$ is a positive $L^1$ function, then $f(k) dk$ is also a Radon measure for $\R$. Nevertheless, there is a particular Radon measure \cite{Dixmier1977,Haar}, called the \emph{Plancherel measure} $d\mu(\pi)$, that is uniquely defined by the inverse Fourier transform
\begin{align}
    f(g) = \int_{\widehat{G}} d\mu(\pi) \tr_{V_\pi}[\pi(g)\pi(f)] \,.\label{eqn:plancherelinversion}
\end{align}
The Plancherel measure $d\mu(\pi)$ weighs different representations non-uniformly according to their contribution to the regular representation of $G$. Physically, it tells us how much each irreducible sector contributes to the total Hilbert space. Crucially, for the unitary dual $\widehat{G}$ of a semi-simple non-compact group, no translation-invariant, finite measure exists that satisfies the regularity conditions: there is no natural ``uniform Radon measure'' on $\widehat{G}$. The Plancherel measure circumvents this obstacle by having a non-trivial, representation-dependent density that reflects the geometric structure of $G$ itself. Examples of the Plancherel measure includes $\mu(j)$ defined above for $\SU(2)$, or $dk$ for $\R$. See Sec.~\ref{sec:themodel} for further discussion about the Plancherel measure.

Note that the Plancherel measure does not have support on all of $\widehat{G}$: there are open sets of $\widehat{G}$ to which it assigns zero measure. This is a feature, not a bug. The representations outside of the support of the Plancherel measure do not have square-normalizable matrix elements,\footnote{A representation is in the support of the Plancherel measure if and only if its matrix elements $\pi_{ij}(g)$ are in $L^{2+\epsilon}(G)$ for any $\epsilon > 0$. These are the so-called ``tempered'' representations. In physics jargon, this means that the support of the Plancherel measure only includes representations whose matrix elements are normalizable after introducing an IR regulator the group integral, and removing the regulator at the end. } so they do not appear in the Fourier expansion of an $L^2$ function. 

The inverse Fourier transform \eqref{eqn:plancherelinversion}, first proven by Harish-Chandra \cite{HarishChandra1952,HarishChandra1954complex,HarishChandra1976}, has many striking implications. First, by integrating both sides of \eqref{eqn:plancherelinversion} with respect to another $L^2$ function $f'(g)$, we  see that
\begin{align}
    \int_G dg\, f'(g)^* f(g) = \int_{\widehat{G}} d\mu(\pi) \tr_{V_\pi}[\pi(f')^{\dagger}\pi(f)] \,. \label{eqn:fourierisunitary}
\end{align}
The left side is the inner product of $f,f'$ as $L^2$ functions. $\tr_{V_\pi}[\pi(f')^{\dagger}\pi(f)]$ is the Hilbert-Schmidt inner product of the operators $\pi(f),\pi(f')$, which can be thought of as vectors in the Hilbert space $E_\pi = V_\pi \otimes V_\pi^*$. Thus, because $f,f'$ are arbitrary, we have an equivalence of Hilbert spaces
\begin{align}
    L^2(G) = \int_{\widehat{G}} d\mu(\pi) V_\pi \otimes V_\pi^* \,.
\end{align}
This is called the Plancherel decomposition of $L^2(G)$. It is clear that \eqref{eqn:fouriertransform} is a linear map from $L^2(G)$ to the right side of the Plancherel decomposition above. \eqref{eqn:fourierisunitary} additionally implies  that the map is \emph{unitary}. Thus, the inner products of $f(g)$ and $\pi(f)$ agree. The Plancherel measure is the unique measure which ensures that the Fourier transform is unitary.

We can further understand the non-Abelian Fourier transform by setting $g=e$ in the inverse  transform \eqref{eqn:plancherelinversion}. We can see
\begin{align}
    f(e) &= \int_{\widehat{G}} d\mu(\pi) \tr_{V_\pi}[\pi(f)]
    \\&= \int_{\widehat{G}} d\mu(\pi) \chi_\pi(f)
    \\&= \int_{\widehat{G}} d\mu(\pi) \int_G dg f(g) \chi_\pi(g^{-1}) \,.
\end{align}
This formula is equivalent to the equality of distributions
\begin{align}
    \delta(g) =  \int_{\widehat{G}} d\mu(\pi)  \chi_\pi(g^{-1}) \,. \label{eqn:deltafn}
\end{align}
This is the generalization of the familiar Fourier transform of the delta function of $\R$. Indeed, for $G=\R$, the Fourier transform of $\delta_a(x) = \delta(x-a)$ is $e^{-ika}$. For a more general transformable group, using \eqref{eqn:fouriertransform}, we can see that if we define $\delta_h(g) = \delta(h^{-1}g)$, then
\begin{align}
    \pi(\delta_h) = \pi(h^{-1}) \,,
\end{align}
which is the natural generalization. The Plancherel measure is essential for this equality to hold. 

We conclude this discussion by giving some examples of the Plancherel measure.

\paragraph{$\SL(2,\C):$ } The tempered representations of $\SL(2,C)$ are labeled by an integer $n$ and a real number $\nu$. The Plancherel measure of these representations are $d\mu(n,\nu) = (n^2 + \nu^2) d\nu$.

\paragraph{$\SL(2,\R)$: } There are two families of tempered representations of $\SL(2,\R)$. The first is the \emph{discrete series} $D_n^\pm$, which are labeled by an integer $n \neq 0$. The Plancherel measure of the discrete series is $d\mu(n) = |n| $. The second family is the \emph{principal series} $P^\pm_{\lambda}$, which is labeled by a continuous parameter $\lambda > 0$ and a sign $\pm$. The Plancherel measure of the principal series depends on the sign: $d\mu(\lambda,+) = \frac{1}{2}\lambda \tanh(\pi \lambda/2) d\lambda$, and $d\mu(\lambda,-) = \frac{1}{2}\lambda \,\text{cotanh}(\pi \lambda/2) d\lambda$. 


\subsection{Intertwiners}

So far, we have discussed the mathematics required to understand $L^2(G)$, which in our context, is the pre-Hilbert space associated with a single leg of the tensor network $\Lambda$. The electric and magnetic constraints couple these legs together, so we also need to understand the structure of operators which have support on multiple legs.

The magnetic constraints are simple to understand in the group basis, in which they are diagonal; we will not discuss them further here. In contrast, the electric constraints $A_v[1]$ (see Sec.~\ref{sec:themodel}) are not diagonal in the group basis, so their physical effect is not as immediate. However, the electric constraints are easier to understand in the representation basis in which they are diagonal.

For simplicity, we will focus on the example of the reduced lattice $\Lambda_r$ with $n$ boundary legs and no matter legs (see Sec.~\ref{sec:Hphys} for the definition of the reduced lattice). There is no loss of generality in this assumption, because we can repeat the same analysis at each bulk vertex $v$ of a more general graph $\Lambda$ since the electric constraints at each vertex commute. Furthermore, the inclusion of matter legs is straightforward by using the substitution \eqref{eqn:Aconstraintwithmatter}.

For notational simplicity, define
\begin{align}
    \ket{\vec{\pi},\vec{a} \vec{b}} &= \ket{\pi_1, a_1 b_1} \cdots \ket{\pi_n, a_n b_n}\,, \\
    \ket{\vec{g}} &= \ket{g_1} \cdots \ket{g_n} \,, \\
    \vec{\pi}(\vec{g})_{L} &= \pi_1(g_1) \otimes \cdots \otimes \pi_n(g_n)\,,\\
    \vec{\pi}(\vec{g})_{R} &= \pi_1(g_1)^\dagger \otimes \cdots \otimes \pi_n(g_n)^\dagger\,.
\end{align}
Here, $ \ket{\vec{\pi},\vec{a} \vec{b}}$ is a basis for the pre-Hilbert space $\Ha(\Lambda_r)$ in which every leg is in the representation basis. $\vec{\pi}(\vec{g})_L$ and $\vec{\pi}(\vec{g})_R$ are operators which act on $\Ha(\Lambda_r)$ by the $\pi_i$ group action on the $V_{\pi_i}$ and $V_{\pi_i}^*$ subspaces of the $i$th boundary leg respectively, and the identity elsewhere. The $L$ and $R$ subscripts stand for left and right multiplication, which matches the understanding of $\ket{\pi,ab}$ as a matrix. If we write $\vec{\pi}(g)_{L,R}$ without the arrow on the group element, we mean the same operator but with $g_1 = g_2 = \cdots = g_n = g$. In terms of representation theory, $\vec{\pi}(\vec{g})_{L,R}$ is an irreducible representation of $G^n$, because there are $n$ group elements that go into its definition. In contrast, $\vec{\pi}(g)_{L,R}$ is a representation of $G$, because there is only one group element which defines these operators. As a $G$ representation, $\vec{\pi}(g)_{L,R}$ is generally reducible. For example, if $\pi_F$ is the fundamental representation of $G$ and $\pi_{\overline{F}}$ is the anti-fundamental representation, then $\pi_F(g) \otimes \pi_{\overline{F}}(g)$ will decompose into a direct sum of the trivial and adjoint representations. This perspective will be important below.

We orient the legs of $\Lambda_r$ to be inflowing, so the $a_i$ index is ``on the boundary'' and the $b_i$ index is ``in the bulk''. By Fourier transforming the group-basis definition presented in Sec.~\ref{sec:electricops} (see also Fig.~\ref{fig:Adef}), a straightforward calculation shows that a gauge transformation of $\Ha(\Lambda_r)$ in the representation basis acts as
\begin{align}
    A_v(h) \ket{\vec{\pi}, \vec{a} \vec{b}} = \vec{\pi}(h)_R \ket{\vec{\pi}, \vec{a} \vec{b}} = \sum_{\vec{c}} \bra{\vec{c}\,} \vec{\pi}(h)_R |\vec{b} \rangle\ket{\vec{\pi}, \vec{a} \vec{c}\,} \,.
\end{align}
Graphically, this follows because the gauge transformations act on the bulk vertex, which transforms the $\vec{b}$ index and leaves the $\vec{a}$ index free.

Now consider the constraint operator $\Pi_A = A_v[1] = \int dh A_v(h)$. For the moment, assume $G$ is compact, so $\Pi_A$ is a  projection operator on $\Ha(\Lambda_r)$, rather than a map between Hilbert spaces with different inner products (as explained in Sec.~\ref{sec:themodel}). If we define the operator
\begin{align}
    \vec{\pi}[1]_R = \int dh \,\vec{\pi}(h)_R\,,
\end{align}
then 
\begin{align}
    \Pi_A \ket{\vec{\pi}, \vec{a} \vec{b}} = \vec{\pi}[1]_R \ket{\vec{\pi}, \vec{a} \vec{b}} \,.
\end{align}
Thus, the constraint operator can be understood as applying the operator $ \vec{\pi}[1]_R$ on the $V_{\vec{\pi}} \otimes V_{\vec{\pi}}^*$ subspace of $\Ha(\Lambda_r)$, subspace by subspace.

The operator $\vec{\pi}[1]_R$  is an \emph{intertwiner} from the $\vec{\pi}_R$ representation of $G$ to the trivial representation. An intertwiner is a map $I$ which interpolates between representations of $G$. More precisely, let $\pi(g)$ and $\omega(g)$ be the matrices for representations of $G$, possibly reducible, which act on the vector spaces $V_\pi$ and $V_\omega$, respectively. Then a map $I_{\pi,\omega}: V_\omega \to V_\pi$ is an intertwiner if and only if for any $g \in G$, 
\begin{align}
    I_{\pi,\omega} \,\omega(g) = \pi(g) I_{\pi,\omega} \,.
\end{align}
If $V_\pi, V_\omega$ are finite dimensional representations, we can think of the intertwiner $I_{\pi,\omega}$ as a rectangular matrix which interpolates between these representations. The intertwiner condition is linear in $I$, so if $I_1$ and $I_2$ are two intertwiners between the same representations of $G$, then so is $I_1 + c \cdot I_2$ for any constant $c \in \C$. Thus, the collection of intertwiners 
\begin{align}
    \mathcal{I}_{\pi,\omega} = \{I_{\pi,\omega} \,|\, \forall g \in G \,,\,I_{\pi,\omega} \omega(g) = \pi(g) I_{\pi,\omega}\}
\end{align}
forms a vector space. One way to think of Schur's lemma is in terms of intertwiners. If $\pi$ and $\omega$ are irreducible, then Schur's lemma says
\begin{align}
    \dim(\mathcal{I}_{\pi,\omega}) = \begin{cases}
        1 & \pi \cong \omega \,,\\
        0 & \text{else}\,.
    \end{cases}
\end{align}
If the $\pi$ and $\omega$ representations are isomorphic, then the one dimensional vector space $\mathcal{I}_{\pi,\pi}$ is spanned by the identity operator $\Id_{V_\pi}$. 

Intertwiners can be thought of as generalizations of Clebsch-Gordan coefficients. For the group $G=\SU(2)$, the Clebsch-Gordan coefficients determine the overlap of spins $(j_1, m_1)$ and $(j_2, m_2)$ with another total spin $(j,m)$ of the combined system. We can think of this as defining the matrix elements of a linear map $C_{j_1 j_2}^j: V_{j_1} \otimes V_{j_2} \to V_{j}$. This map $C_{j_1 j_2}^j$ is an intertwiner for $\SU(2)$. 

\subsubsection{Quadratic forms}

The operator $\vec{\pi}[1]_R$ is an intertwiner between the $V_{\vec{\pi}}^*$ representation of $G$ and the trivial representation. To see this, we compute
\begin{align}
    \vec{\pi}[1]_R \vec{\pi}(g)_R = \int dh \, \vec{\pi}(h)_R \vec{\pi}(g)_R = \int dh \, \vec{\pi}(hg)_R = \int dh \, \vec{\pi}(h)_R =  \Id_{V^*_{\vec{\pi}}} \vec{\pi}[1]_R \,.
\end{align}
and notice that $\Id_{V^*_{\vec{\pi}}}$ is the matrix for the action of $g$ in the trivial representation. 

Actually, $\vec{\pi}[1]_R$ does not just act as the identity: it projects $V_{\vec{\pi}}^*$ onto the vector subspace spanned by all the copies of the trivial representation within $V_{\vec{\pi}}^*$. This follows from Schur's lemma; if we decompose $V_{\vec{\pi}}^*$ into a direct sum of irreducible $G$ representations $\omega$, then there are no intertwiners from $\omega$ to the trivial representation unless $\omega$ is already trivial. This is clearest  when $G$ is compact: in this case, choosing the Haar measure with $\text{Vol}(G) = 1$, we can think of $\vec{\pi}[1]_R$ as a literal projection operator from $V_{\vec{\pi}}^*$ onto the subspace spanned by the copies of the trivial representation within $V_{\vec{\pi}}^*$. The dimension of this subspace is $\tr[\vec{\pi}[1]_R]$, and this is precisely the subspace of gauge invariant states within $V_{\vec{\pi}}^*$. 

When $G$ is non-compact, this is still essentially true, although the vector subspace spanned by the copies of the trivial representation is not a Hilbert subspace. That is because the trivial representation is not normalizable within the $\Ha(\Lambda_r)$ inner product. However, we can circumvent this normalizability problem by instead viewing $\vec{\pi}[1]_R$ not as projector onto a subspace of $V_{\vec{\pi}}^*$, but as quadratic form. A quadratic form is a map $Q(\ket{\psi},\ket{\sigma}): V_{\vec{\pi}}^* \otimes V_{\vec{\pi}}^* \to \C$ which takes two vectors of $V_{\vec{\pi}}^*$ as input and outputs a complex number, and is (anti-)linear in the (first) second slot of $Q$. This is a useful perspective  because the normalizablity issues of $\vec{\pi}[1]_R$ arise when we square it, which we must be allowed to do if $\vec{\pi}[1]_R$ is viewed as an operator. In contrast, if we view $\vec{\pi}[1]_R$ as the quadratic form
\begin{align}
    Q(\ket{\psi},\ket{\sigma}) =  \bra{\psi} \vec{\pi}[1]_R \ket{\sigma} \,,
\end{align}
then we do not need to worry about the volume divergence of $\vec{\pi}[1]_R $, because quadratic forms are not operators and can not be ``squared''. 

A quadratic form is similar to an inner product of vector spaces, with an important difference: inner products must be non-degenerate. In contrast, $Q(\cdot, \ket{\sigma}) = 0$ if $\ket{\sigma}$ has no support on the vector subspace of $V_{\vec{\pi}}^*$ spanned by the trivial representations of $G$. To make $Q$ into an inner product on $V_{\vec{\pi}}^*$, we must quotient by these null states. 

So far, this discussion has taken place within a single subspace $V_{\vec{\pi}}^*$ of $\Ha(\Lambda_r)$. As we saw above, the electric constraint $\Pi_A$ acts as $\vec{\pi}[1]_R$ on each subspace of this form: therefore, $\Pi_A$ is a quadratic form. This quadratic form has support on the gauge invariant vector subspace of $\Ha(\Lambda_r)$. To make this vector subspace into a Hilbert space, we quotient out the null states of $\Pi_A$ and use this quadratic form as an inner product. This is precisely the procedure we adopted in Sec.~\ref{sec:themodel} to define the physical Hilbert space. 

\subsubsection{Multipartite entanglement from intertwiners}

Given an intertwiner $I: V_{\vec{\pi}} \to V_{\vec{\omega}}$, we can raise the indices associated with $V_{\vec{\pi}}$ (i.e., flip the bras into kets) and define a vector $\ket{I} \in V_{\vec{\pi}}^* \otimes V_{\vec{\omega}}$. In the case of $\vec{\pi}[1]_R: V_{\vec{\pi}} \to \C$, this defines an invariant vector $\ket{\vec{\pi}_R} \in V_{\vec{\pi}}$. This name comes from the fact that for any $g \in G$,
\begin{align}
    \vec{\pi}(g)_R \ket{\vec{\pi}_R} = \ket{\vec{\pi}_R} \,. \label{eqn:invariantvector}
\end{align}
Note that this implies that we can use a generalized ``transpose trick'' on an invariant vector to trade group multiplication on some tensor factors to group multiplication on the other factors. In other words, we can multiply both sides of \eqref{eqn:invariantvector} by $\pi_1(g^{-1}) \otimes \Id_{\pi_2} \otimes \Id_{\pi_n}$ to see that
\begin{align}
    \Id_{\pi_1} \otimes \pi_2(g) \otimes \cdots \otimes \pi_n(g) \ket{\vec{\pi}_R} = \pi_1(g^{-1}) \otimes \Id_{\pi_2} \otimes \cdots \otimes \Id_{\pi_n}\ket{\vec{\pi}_R} \,. \label{eqn:invariantvector2}
\end{align}
This implies that $\ket{\vec{\pi}_R}$ is multiparty entangled in a product basis of $V_{\vec{\pi}}$. To see that it is entangled, note that \eqref{eqn:invariantvector2} would be impossible if $\ket{\vec{\pi}_R}$ was unentangled across $V_{\pi_1}$ and the rest of the Hilbert space. The multiparty nature of the entanglement is implied by the fact that \eqref{eqn:invariantvector2} holds no matter the bipartition of $V_{\vec{\pi}}$ we choose, not just the bipartition into $V_{\pi_1}$ and its complement. 

To be more concrete, consider the example of the identity map $\Id_\pi : V_\pi \to V_\pi$, where $V_\pi$ is an irreducible representation. Then the associated vector $\ket{\Id_\pi}$ is
\begin{align}
    \ket{\Id_\pi} = \sum_{m} \ket{\pi^*,n} \ket{\pi,n}
\end{align}
where $\ket{\pi^*,n}$ and $\ket{\pi,n}$ are orthonormal bases for $V_\pi^*$ and $V_\pi$, respectively. Clearly, $\ket{\Id}$ is highly entangled in this basis. 

To see the multiparty entanglement, we need to consider an invariant vector with at least three tensor factors, such as $\ket{\pi_1,\pi_2,\pi_3} \in V_{\pi_1} \otimes V_{\pi_2} \otimes V_{\pi_3}$, where we take each $V_\pi$ to be irreducible. Then in terms of the $3j$ symbols of $G$ \cite{Derome,DeromeSharp} (which are proportional to the Clebsch-Gordan coefficients), 
\begin{align}
    \ket{\pi_1,\pi_2,\pi_3} = \sum_{m_1, m_2,m_3} \begin{pmatrix}
        \pi_1 & \pi_2 & \pi_3 \\
        m_1 & m_2 & m_3 
    \end{pmatrix} \ket{\pi_1,m_1}\ket{\pi_2,m_2}\ket{\pi_3,m_3}\,.
\end{align}
This is the \emph{definition} of the $3j$ symbol. In the notation of Sec.~\ref{sec:Hphys}, the state $\ket{\pi_1,\pi_2,\pi_3}$ is an example of a state in the gauge invariant subspace $\Pi_A[V_{\vec{\pi}}^*]$. The fact that $\ket{\pi_1,\pi_2,\pi_3}$ is multiparty entangled follows from the orthogonality relation of the $3j$ symbols \cite{Derome,DeromeSharp}
\begin{align}
    \int d\mu(\pi_1) \sum_{m_1} \begin{pmatrix}
        \pi_1 & \pi_2 & \pi_3 \\
        m_1 & m_2 & m_3 
    \end{pmatrix} \begin{pmatrix}
        \pi_1 & \pi_2 & \pi_3 \\
        m_1 & n_2 & n_3 
    \end{pmatrix} = \delta_{m_2, n_2} \delta_{m_3,n_3} \,.
\end{align}
This relation implies the reduced density matrix $\rho_{\pi_2 \pi_3} = \tr_{\pi_1}[\ketbra{\pi_1,\pi_2,\pi_3}]$ on $V_{\pi_2} \otimes V_{\pi_3}$ will be maximally mixed. Because this holds regardless of the choice of tensor factor $V_\pi$, the state $\ket{\pi_1,\pi_2,\pi_3}$ is multiparty entangled. Along with the generalization for intertwiners with $n$ legs, this is the entanglement structure which we use in Sec.~\ref{sec:Hphys} to analyze the physical Hilbert space.

\subsection{A series of useful equations}

We end this appendix by collecting a useful set of equations that we either proved or argued for above.

\begin{align}
    \Id_{L^2(G)} &= \int_G dg \ketbra{g} = \int_{\widehat{G}} d\mu(\pi)\sum_{a,b} \ketbra{\pi,ab} \\
    \delta(g)&= \int d\mu(\pi) \chi_\pi(g^{-1}) \label{eqn:groupdeltacharacter}
    \\ \braket{\pi,ab}{\omega,mn} &= \delta(\pi,\omega) \delta_{am} \delta_{bn} \\
    \chi_\pi(g^{-1}) &= \chi_\pi(g)^* \label{eqn:charmanip1} \\
    \chi_{\pi \oplus \omega}(g) &= \chi_\pi(g) + \chi_\pi(g) \\
    \chi_{\pi \otimes \omega}(g) &= \chi_\pi(g) \cdot \chi_\pi(g) \label{eqn:charmanip3} 
    \\ \int_G dg \chi_\omega(g^{-1}) \chi_\pi(g h) &= \delta(\pi,\omega) \chi_\pi(h) \,.
    \label{eqn:characterdeltafn}
\end{align}

We can prove the last equation, which we have not yet shown, as follows. First, note that we can interpret this integral as the matrix element
\begin{align}
    \int_G dg \chi_\omega(g^{-1}) \chi_\pi(g h) = \bra{\chi_\omega} \pi(h^{-1})_R \ket{\chi_\pi}
\end{align}
where $\pi(h^{-1})_R$ is the right multiplication operator on $V_\pi \otimes V_\pi^* \subset L^2(G)$. This right multiplication maps $V_\pi \otimes V_\pi^*$ to itself, which in particular is orthogonal to $V_\omega \otimes V_\omega^*$ for $\pi \neq \omega$. So this integral must vanish unless $\pi = \omega$. To determine its value in this case, we can use \eqref{eqn:groupdeltacharacter} and integrate\footnote{The convolution of $L^1$ functions is $L^1$, so we can swap the integrals freely.}
\begin{align}
    \int_{\widehat{G}} d\mu(\omega)\int_G dg \chi_\omega(g^{-1}) \chi_\pi(g h)
    &= \int_G dg \int_{\widehat{G}} d\mu(\omega) \chi_\omega(g^{-1}) \chi_\pi(g h)
    \\&= \int_G dg\, \delta(g) \chi_\pi(g h)
    \\&= \chi_\pi(h) \,.
\end{align}
Thus, as an equality of distributions, we have proven \eqref{eqn:characterdeltafn}.

\section{Topological Tensor Network Moves} \label{sec:move12}

In this appendix, we will define the topological moves 1 and 2, as well as the isometries $\Delta_{1}$ and $\Delta_2$ which map physical states from one Hilbert space to another. For finite groups, these moves are equivalent to those defined in Appendix A of \cite{akers2024multipartite}. So we need only prove that those moves are maps between normalizable states in our new inner products.

\subsection{Move 1} \label{sec:move1}

\begin{figure}
    \centering
    \begin{tikzpicture}
        \def\sep{3}
        \def\nn{7}
        \def\scale{2}
        \def\hh{0.75}
        \node at (-\sep,0) {\begin{tikzpicture}[scale=\scale]
        
        \filldraw (0,0) circle (0.025);
        \foreach \xx in {1,...,3}{
        \draw[thick,->-=0.6] (0,0) -- ({cos(360*(\xx-1)/\nn)},{sin(360*(\xx-1)/\nn)});
        }
        \foreach \xx in {4,...,\nn}{
        \draw[thick,->-=0.6] ({cos(360*(\xx-1)/\nn)},{sin(360*(\xx-1)/\nn)}) -- (0,0);
        }
        \end{tikzpicture}};
        \node[scale=2] at (0,0) {$\Leftrightarrow $};
        \node at (\sep,0) {\begin{tikzpicture}[scale=\scale]
        \filldraw (0,0) circle (0.025);
        \filldraw (\hh,0) circle (0.025);
        \node[anchor=south] at (0,0) {$v$};
        \node[anchor=north] at (\hh,0) {$v'$};
        
        \foreach \xx in {1,...,3}{
        \draw[thick,->-=0.6] (\hh,0) -- ({\hh+cos(360*(\xx-1)/\nn)},{sin(360*(\xx-1)/\nn)});
        }
        \foreach \xx in {4,...,\nn}{
        \draw[thick,->-=0.6] ({cos(360*(\xx-1)/\nn)},{sin(360*(\xx-1)/\nn)}) -- (0,0);
        }
        \draw[thick,->-=0.5] (\hh,0) -- (0,0);
        \end{tikzpicture}};
       
    \end{tikzpicture}
    \caption{An example of move 1, which can be performed in either direction to add or remove a bulk vertex/leg from the graph $\Lambda$.}
    \label{fig:move1apx}
\end{figure}
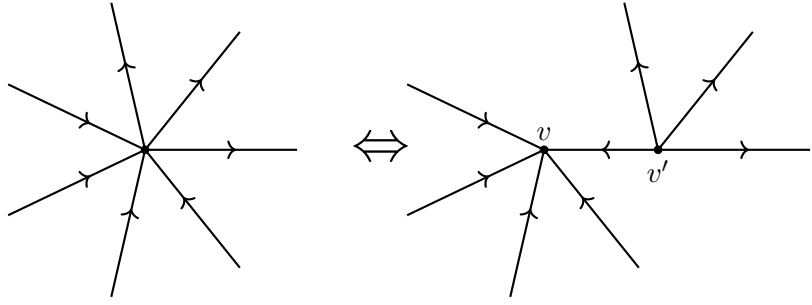

Let $\Delta_1: \Pi_A \Pi_B \Ha(\Lambda) \to \Pi_A \Pi_B \Ha(\Lambda')$ be the map which adds a vertex as in Fig.~\ref{fig:move1apx}. It is convenient to introduce the compact notation
\begin{align}
    \psi(\vec{g}) &= \psi(g_1, \cdots,g_{|L|})\,, \label{eqn:compactnotation1}
    \\ \ket{\vec{g}} &= \ket{g_1,\cdots,g_{|L|}}\,,
    \\d[\vec{g}] &= dg_1 \cdots dg_{|L|}\,. \label{eqn:compactnotation3}
\end{align}
Let $\ket{\Psi}\rangle$ be a state in the physical Hilbert space $\Pi_A \Pi_B\Ha(\Lambda)$, with representative
\begin{align}
    \ket{\Psi} = \int d[\vec{g}] \psi(\vec{g}) \ket{\vec{g}}\,.
\end{align}
We would like to map this state to a new state in the physical Hilbert space $\Pi_A \Pi_B \Ha(\Lambda')$, where $\Lambda$ and $\Lambda'$ are related by move 1, so $\Lambda'$ has one additional leg compared to $\Lambda$. Let $v,v'$ be the two vertices of the new leg of $\Lambda'$, and $[\Pi_A]_v, [\Pi_A]_{v'}$ be the Gauss law constraints for these vertices. For notational simplicity, we take all the legs of $v$ to be inflowing, and all the legs of $v'$ to be outflowing, but any orientation of legs is allowed. When $G$ is a compact group, this map can be defined as
\begin{align}
    \Delta_1\ket{\Psi} &= [\Pi_A]_v[\Pi_A]_{v'} \ket{e} \ket{\Psi} \,.
\end{align}
Here, $\ket{e}$ is the state of the new leg of $\Lambda'$, and $\ket{\Psi}$ is the state on the legs of $\Lambda'$ which descend directly from $\Lambda$. More concretely, we can expand the action of $[\Pi_A]_v,[\Pi_A]_{v'}$ to see that
\begin{align}
    \Delta_1\ket{\Psi} &= \int d[h,k,\vec{g}] \,  \psi(\vec{g}) \, \ket{hk^{-1}} \ket{h\cdot\vec{g} \cdot k^{-1}} \,.
\end{align}
The notation $h\cdot\vec{g} \cdot k^{-1}$ means that we left multiply by $h$ on the legs attached to the $v$ vertex, and right multiply by $k^{-1}$ on the legs attached to the $v'$ vertex. For other choices of orientations of legs feeding into $v,v'$, we must instead apply the appropriate group action depending on the orientations of the legs.
The integration over $h,k$ implements the Gauss constraint on the two vertices of this new leg.

When $G$ is a more general transformable group, we saw in Sec.~\ref{sec:themodel} that the Gauss constraint $\Pi_A$ must instead be moved to the definition of the inner product. In this case, we can define the isometry of move 1 to be
\begin{align}
    \Delta_1\ket{\Psi}\rangle = [\ket{e} \ket{\Psi} \sim \ket{e} \ket{\Psi} + \ket{\chi}] \equiv \ket{e,\Psi}\rangle
\end{align}
where $\ket{\chi}$ is a null state of $\Pi_A \Pi_B \Ha(\Lambda')$. We can think of this representative $\ket{e} \ket{\Psi}$ as a particular gauge choice for the gauge invariant state $\ket{e,\Psi}\rangle$. Although the leg $\ket{e}$ in this representative seems unentangled from $\ket{\Psi}$, the entanglement is  intrinsic to the definition of the inner product of $\Pi_A \Pi_B \Ha(\Lambda')$. In other words, the entanglement comes from the quotient by null states.

To see that $\Delta_1$ is an isometry, we will show that $\Delta_1^\dagger \Delta_1$ acts as the identity on any state of $\Pi_A \Pi_B \Ha(\Lambda)$. For simplicity, we focus on the Gauss constraint for the vertices $v,v'$, and suppress the Gauss constraint on the other vertices of $\Lambda$. Then, we can see that
\begin{align}
    \langle \bra{\psi} \Delta_1^\dagger \Delta_1 \ket{\sigma}\rangle 
    &= \bra{e}\langle\bra{\psi} [\Pi_A]_{v}  [\Pi_A]_{v'} \ket{e}\ket{\sigma}\rangle 
    \\&= \int d[h,k,\vec{g},\vec{\ell}] \psi^*(\vec{g}) \sigma(\vec{\ell}) \braket{e,\vec{g}}{hk^{-1},h\cdot\vec{\ell} \cdot k^{-1}}
    \\&= \int d[\vec{g},h] \psi^*(\vec{g}) \sigma(h^{-1} \cdot\vec{g}\cdot h)
    \\&= \langle \braket{\psi}{\sigma}\rangle\,.
\end{align}
The last line follows because this is the definition of the inner product on $\Ha(\Lambda)$. $\Delta_1$ therefore embeds $\Pi_A \Pi_B \Ha(\Lambda)$ into $\Pi_A \Pi_B \Ha(\Lambda')$ isometrically.

The inverse move $\Delta_1^\dagger: \Pi_A \Pi_B \Ha(\Lambda') \to \Pi_A \Pi_B \Ha(\Lambda)$ is not an isometry, but is a one-to-one identification of physical states. This is to be expected: in general, $\Delta_1\Delta_1^\dagger $ is a projector on $\Ha(\Lambda')$, which  does not preserve the norm of a state. This map is defined by reversing the definition of $\Delta_1$: if $\ell$ is the leg which we wish to remove from $\Lambda' \to \Lambda$, then $\Delta_1^\dagger \ket{\Psi}\rangle$ is defined to be the equivalence class
\begin{align}
    \Delta_1^\dagger \ket{\Psi}\rangle = [(\bra{e}_\ell \otimes \Id_{\Lambda})\ket{\Psi} ]
\end{align}
where the equivalence class is again taken to be up to the null states of $\Pi_A \Pi_B \Ha(\Lambda)$. Note that this definition is consistent with our calculation of $\Delta_1^\dagger \Delta_1$. We can think of this inverse move as having first  ``gauge fixed'' the state $\ket{\Psi}\rangle$ on $\Pi_A \Pi_B \Ha(\Lambda')$ to have $e$ in the first slot, and declaring that this gauge fixed state is the physical state on $\Pi_A \Pi_B \Ha(\Lambda)$ that we are interested in. If we wish for the resulting state to be properly normalized in $\Pi_A \Pi_B \Ha(\Lambda)$, we must simply demand that it is in the image of $\Delta_1$.

\subsection{Move 2}

\begin{figure}
    \centering
    \begin{tikzpicture}
        \def\sep{4}
        \def\nn{6}
        \def\rr{1.3}
        \def\scale{2}
        \def\hh{0.75}
        \node at (-\sep,0) {\begin{tikzpicture}[scale=\scale]
        
        \foreach \xx in {1,...,\nn}{
        \filldraw ({cos(360*(\xx-1)/\nn)},{sin(360*(\xx-1)/\nn)}) circle (0.025);
        \draw[thick,->-=0.6] ({cos(360*(\xx-1)/\nn)},{sin(360*(\xx-1)/\nn)}) -- ({\rr*cos(360*(\xx-1)/\nn)},{\rr*sin(360*(\xx-1)/\nn)});
        \draw[thick,->-=0.6] ({cos(360*(\xx-1)/\nn)},{sin(360*(\xx-1)/\nn)}) -- ({cos(360*\xx/\nn)},{sin(360*\xx/\nn)});
        }
        \end{tikzpicture}};
        \node[scale=2] at (0,0) {$\Leftrightarrow $};
        \node at (\sep,0)  {\begin{tikzpicture}[scale=\scale]

        \draw[thick,->-=0.5] (1,0) -- ({cos(360*(2)/\nn)},{sin(360*(2)/\nn)});
        \node[anchor=north east] at ({0.5*(1+cos(360*(2)/\nn))},{0.5*sin(360*(2)/\nn)}) {$\ell$};
        
        \foreach \xx in {1,...,\nn}{
        \filldraw ({cos(360*(\xx-1)/\nn)},{sin(360*(\xx-1)/\nn)}) circle (0.025);
        \draw[thick,->-=0.6] ({cos(360*(\xx-1)/\nn)},{sin(360*(\xx-1)/\nn)}) -- ({\rr*cos(360*(\xx-1)/\nn)},{\rr*sin(360*(\xx-1)/\nn)});
        \draw[thick,->-=0.6] ({cos(360*(\xx-1)/\nn)},{sin(360*(\xx-1)/\nn)}) -- ({cos(360*\xx/\nn)},{sin(360*\xx/\nn)});
        }
        \end{tikzpicture}};
       
    \end{tikzpicture}
    \caption{An example of move 2, which can be performed in either direction to add or remove a leg/plaquette from the graph $\Lambda$.}
    \label{fig:move2apx}
\end{figure}
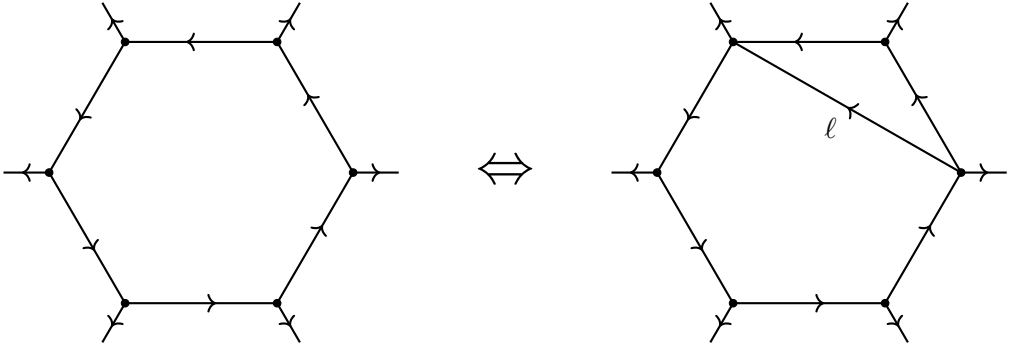

Let $\Delta_2: \Pi_A \Pi_B \Ha(\Lambda) \to \Pi_A \Pi_B \Ha(\Lambda')$ be the move which adds a plaquette as in Fig.~\ref{fig:move2apx}. Let $\ell$ be the new leg of $\Lambda'$ compared to $\Lambda$, and let $[\Pi_B]_p,[\Pi_B]_{p'}$ be the magnetic constraint operators for the plaquettes whose boundaries containing $\ell$. Without loss of generality, we take the orientation of $\ell$ to be aligned with the orientation of $\partial p$ (i.e., it is pointing counter-clockwise) and anti-aligned with the orientation of $\partial p'$ (i.e., clockwise). This is just a convention, but the orientation of $\ell$ on $\partial p$ and $\partial p'$ will always be opposite, which is clear from inspection of Fig.~\ref{fig:move2apx}. Using the same notation as in Sec.~\ref{sec:move1} and \eqref{eqn:finiteGintegral}, when $G$ is discrete, we can write move 2 as
\begin{align}
    \Delta_2\ket{\Psi} = [\Pi_B]_p [\Pi_B]_{p'} \left(\int dg \ket{g}\right) \ket{\Psi}\,.
\end{align}
Here, $\ket{g}$ is a state on $L^2(G)_\ell$, and $\ket{\Psi}$ has support on all the legs of $\Lambda'$ which descend directly from $\Lambda$. If we expand the definitions of $[\Pi_B]_p,[\Pi_B]_{p'}$, we can see that
\begin{align}
     \Delta_2\ket{\Psi} = \int_G d[h,\vec{g}] \delta(h g_{\partial p})  \delta(h^{-1}g_{\partial p'})\psi(\vec{g}) \ket{h}\ket{\vec{g}}\,.
\end{align}
Here, $g_{\partial p}$ is the product of group elements starting from the outward flowing end of $\ell$, traversing the boundary of $p$, and ending at the other vertex of $\ell$. $g_{\partial p'}$ is defined similarly, but orientation reversed because of our convention for the orientation of $\ell$. This explains why the $\partial p'$ delta function has an $h^{-1}$ instead of an $h$.

When $G$ is a more general transformable group, we saw that $[\Pi_B]_p,[\Pi_B]_{p'}$ must instead be included as part of the definition of the inner product. Thus, we should instead define the map $\Delta_2$ by
\begin{align}
    \Delta_2 \ket{\Psi}\rangle = \left[\left(\int dg \ket{g}\right) \ket{\Psi} \sim \left(\int dg \ket{g}\right) \ket{\Psi} + \ket{\chi}\right] \equiv \ket{1,\Psi}\rangle
\end{align}
as an equivalence class up to null states $\ket{\chi}$ of the $\Pi_A\Pi_B \Ha(\Lambda')$ inner product. The resulting state satisfies the magnetic constraint because all the states $\ket{g}$ in the superposition $\int dg \ket{g}$ that do not satisfy the magnetic constraints of the new plaquette are projected out by the null state quotient. 

This definition of $\Delta_2$ is an isometry. To see this, we focus on the magnetic constraints at $p$ and $p'$, and compute 
\begin{align}
    \langle \bra{\psi} \Delta_2^\dagger \Delta_2 \ket{\sigma}\rangle 
    &= \int d[h,k] \bra{h} \langle\bra{\psi} [\Pi_B]_p [\Pi_B]_{p'} \ket{k} \ket{\sigma}\rangle \,,
    \\&= \int d[h,k,\vec{g},\vec{m}] \psi^*(\vec{g}) \sigma(\vec{m}) \delta(h g_{\partial p})\delta(h^{-1} g_{\partial p'})  \braket{h,\vec{g}}{k,\vec{m}}\,,
    \\&= \int d[\vec{g}] \psi^*(\vec{g}) \sigma(\vec{g}) \delta(g_{\partial p} g_{\partial p'}) \,,
    \\&= \langle \braket{\psi}{\sigma}\rangle \,.
\end{align}
In the third line, we used the $h,k,\vec{m}$ integrals to simplify integral. Recognizing the remaining integral as the definition of the $\Pi_A \Pi_B \Ha(\Lambda)$ inner product, we are done.

Similar to move 1, the inverse of this move, $\Delta_2^\dagger: \Pi_A \Pi_B \Ha(\Lambda') \to \Pi_A \Pi_B \Ha(\Lambda)$, is not an isometry (unless we restrict to the image of $\Delta_2$), but it is a one-to-one map between physical states. For completeness, this map is given by
\begin{align}
    \Delta_2^\dagger \ket{\Psi}\rangle = \left[\left(\int dg\bra{g}_\ell \otimes \Id_{\Lambda}\right)\ket{\Psi} \right]
\end{align}
where the equivalence class is with respect to null states of $\Pi_A \Pi_B \Ha(\Lambda)$.

\section{The quantum double algebra for transformable groups} \label{sec:DA}

In this appendix, we will describe the algebraic structure of the electric and magnetic operators. In our application to gravity, the electric and magnetic operators are a tool for defining the constraints. But in condensed matter theory, one instead views $\Ha(\Lambda)$ as a physical system with Hamiltonian
\begin{align}
    H = -\sum_v A_v[1] -\sum_p B_{(v,p)}(e)\,.
\end{align}
From this perspective, the physical Hilbert space $\Ha_{phys}(\Sigma)$ is the space of ground states of $\Ha(\Lambda)$, and the gauge symmetry and topological behavior of $\Ha_{phys}(\Sigma)$ is emergent in the IR limit of $\Ha(\Lambda)$. So understanding how the electric and magnetic operators act on all of $\Ha(\Lambda)$ may be of interest. We can show directly from the definitions that 
the electric operators $A_v(g)$ and the magnetic operators $B_{(v,p)}(h)$ satisfy a quantum double algebra
\begin{align}
    A_v(g)^\dagger &= A_v(g^{-1})\,, \label{eqn:qda1}\\
    A_v(g) A_v(h) &= A_v(gh) \,,\\
    B_{(v,p)}(g)^\dagger &= B_{(v,p)}(g)\,, \\
    B_{(v,p)}(g) B_{(v,p)}(h) &= \delta(g^{-1} h) B_{(v,p)}(h) \,,\\
    A_v(g) B_{(v,p)}(h) &= B_{(v,p)}(ghg^{-1}) A_v(g) \label{eqn:qda5} \,.
\end{align}
Note that the last relation implies that $[\Pi_A,\Pi_B]=0$. When $G$ is a finite group, the above relations are well-defined because the delta functions are finite, and the electric and magnetic operators are bounded, so they are well-defined operators on the Hilbert space.
But for a more general transformable group, in order to ensure the electric and magnetic operators are bounded, we must instead define the smeared operators
\begin{align}
    A_v[f_1] = \int dg\, f_1(g) A_v(g) && B_{(v,p)}[f_\infty] = \int dg\, f_\infty(g) B_{(v,p)}(g)
    \label{eq:smearedAB}
\end{align}
where $f_1 \in L^1(G)$ and $f_\infty \in L^\infty(G)$, where recall again that $L^\infty$ is space of bounded functions.
In Sec.~\ref{sec:themodel}, our demonstration that the inner product is finite for bounded $L^1$ functions is equivalent to a proof that operators smeared as in \eqref{eq:smearedAB} are bounded. This is in contrast with $A_v(g),B_{(v,p)}(h)$ themselves. Because products of bounded operators with compatible domains are bounded, we can define the quantum double algebra as the completion of the union of the algebras generated by these smeared operators \eqref{eq:smearedAB}. For the rest of this appendix, we will assume that the argument of $A_v[f]$ is an $L^1$ function and the argument of $ B_{(v,p)}[f]$ is an $L^\infty$ function, unless otherwise stated.

To understand this generalized quantum double algebra, is enlightening to smear both sides of \eqref{eqn:qda1}--\eqref{eqn:qda5} with $L^1$ and $L^\infty$ smearing functions, as in \eqref{eq:smearedAB}. Doing so, we find that
\begin{align}
    A_v[f]^\dagger &= A_v(\overline{f}) \label{eqn:Ainvolution} \\
    A_v[f] A_v[f'] &= A_v[f \ast f']\label{eqn:Aproduct} \\
    B_{(v,p)}[f]^\dagger &= B_{(v,p)}[f^*]\label{eqn:Binvolution}  \\
    B_{(v,p)}[f] B_{(v,p)}[f'] &= B_{(v,p)}[f \cdot f']\label{eqn:Bproduct} \\
    A_v(g) B_{(v,p)}[f'] &=  B_{(v,p)}[\Ad_g(f')]  A_v(g) \label{eqn:ABcommutation}\,.
\end{align}
We presented the last relation in terms of a specific group element $g$ for simplicity, but by smearing both sides with an $L^1$ function $f(g)$, there is no loss of generality in this description of \eqref{eqn:ABcommutation}.
In the above relations, 
\begin{align}
    \overline{f}(g) &= f^*(g^{-1})\,, \label{eqn:L1inv}
    \\f^*(g) &\text{ is the complex conjugate of } f(g)\,, \label{eqn:Linfinv}
    \\ (f \ast f') (g) &= \int dh f(h) f'(h^{-1} g)\,, \label{eqn:L1mult}
    \\ (f \cdot f')(g) &= f(g) f'(g) \,, \label{eqn:Linfmult}
    \\ \Ad_g(f)(h) &= f(g^{-1} h g)\,. \label{eqn:adjoint}
\end{align}
The last equation defines the adjoint action of $G$ on $f$.

For the electric operators, this is interesting because if we define a norm on the function $f \in L^1(G)$ by
\begin{align}
    ||f||_* = \sup_{\pi \in \widehat{G}} ||\pi(f)|| = ||A_v[f]||_\infty \,,
\end{align}
then the completion of $L^1(G)$ with respect to this norm, equipped with involution \eqref{eqn:L1inv} and multiplication \eqref{eqn:L1mult}, is called the group $C^*$ algebra of $G$. This $C^*$ algebra has many useful and interesting mathematical properties (see \cite{Buerschaper:2010vwd,lin2015,Dixmier1977,Haar}). $A_v$ can then be interpreted as a $C^*$ algebra homomorphism from the group $C^*$ algebra into the algebra of bounded operators acting the Hilbert space $\Ha(\Lambda)$. One could then complete this $C^*$ algebra into the group von Neumann algebra $W^1(G)$ by taking a double commutant.

For the magnetic operators, bounded functions with involution \eqref{eqn:Linfinv} and multiplication \eqref{eqn:Linfmult} also form a commutative von Neumann algebra $W^\infty(G)$. $B_{(v,p)}$ can be interpreted as a homomorphism from $W^\infty(G)$ into the von Neumann algebra of bounded operators on $\Ha(\Lambda)$. 

Note that the composition of two bounded operators with compatible domains is bounded, so products of $A_v[f], B_{(v,p)}[f']$ are also bounded. The final commutation relation \eqref{eqn:qda5} links the representations of the group von Neumann algebra $W^1(G)$ and $W^\infty(G)$. The quantum double algebra is then the completion $(W^1(G) \vee W^\infty(G))''$, subject to the commutation relation \eqref{eqn:ABcommutation}.\footnote{More precisely, it is the representation of this algebra on $\Ha(\Lambda)$.}
This mathematical structure is the crossed product algebra $L^\infty(G) \rtimes_\alpha G$ \cite{takesaki2003theory}, where the automorphism $\alpha$ of the crossed product is the adjoint action, as specified by the commutation relation \eqref{eqn:ABcommutation}. This mathematical structure is well-studied, and has had many interesting recent applications to quantum gravity \cite{Witten2022, Leutheusser2023a, Leutheusser2023b, Chandrasekaran:2022cip,Klinger:2023auu,DeVuyst:2024fxc,Ali:2024jkx,Jensen2023}. 

\paragraph{Lattice Yang-Mills: }In lattice Yang-Mills theory, physical states are required to be gauge invariant, but there is no constraint requiring the flux around a plaquette to vanish. Thus, the relevant Hilbert space for lattice Yang-Mills theory is $\Pi_A \Ha(\Lambda)$, not $\Pi_A \Pi_B \Ha(\Lambda)$. Therefore, the electric operators $A_v[f]$ will continue to act trivially on physical states, but the magnetic operators $B_{(v,p)}[f]$ can act more generally and still be physical operators. A magnetic operator $B_{(v,p)}[f]$ is physical if it commutes with gauge transformations $A_v(g)$ for any $g \in G$. Inspecting \eqref{eqn:ABcommutation} and \eqref{eqn:adjoint}, this requires $f(h) = f(ghg^{-1})$ for any $g \in G$. Thus, the gauge invariant magnetic operators are always smeared by class functions, which by definition are constant on the conjugacy classes of $G$. The subalgebra $W_c^\infty(G) $ of $W^\infty(G)$ spanned by these operators is also a von Neumann algebra.

\bibliographystyle{JHEP}
\bibliography{biblio}

\end{document}